\documentclass[12pt]{article}
\usepackage{amsmath,amssymb,bm,graphicx,hyperref,physics, cleveref, subfig, slashed, stmaryrd, tensor, float}
\usepackage[backend=biber,natbib,style=numeric-comp,sorting=none,maxbibnames=99,doi=false,isbn=false,url=false]{biblatex}
\usepackage{tikz}
\usetikzlibrary{arrows.meta, decorations.markings}

\definecolor{bisque}{rgb}{1.0, 0.89, 0.77}
\definecolor{forestgreen(web)}{rgb}{0.13, 0.55, 0.13}

\def\be{\begin{equation}}
\def\ee{\end{equation}}
\def\bea{\begin{eqnarray}}
\def\eea{\end{eqnarray}}
\def\ie{\begin{equation}\begin{aligned}}
\def\fe{\end{aligned}\end{equation}}
\numberwithin{equation}{section}

\renewcommand{\title}[1]{\vbox{\center\LARGE{#1}}\vspace{5mm}}
\renewcommand{\author}[1]{\vbox{\center#1}\vspace{5mm}}
\newcommand{\address}[1]{\vbox{\center\em#1}}
\newcommand{\email}[1]{\vbox{\center\tt#1}\vspace{5mm}}

\newtheorem{theorem}{Theorem}

\begin{document}

\unitlength = .8mm

\hypersetup{pageanchor=false}
\begin{titlepage}

\begin{center}

\hfill \\
\hfill \\
\vskip 1cm

\title{Bosonic String Amplitudes From Ribbon Graphs}

\author{
Sam Christian\textsuperscript{1},
Ben Mazel\textsuperscript{1},
Yuchen Wang\textsuperscript{1},
Xi Yin\textsuperscript{1,2},
Yutai Zhang\textsuperscript{1}
}

\address{
\textsuperscript{1}Jefferson Physical Laboratory, Harvard University,
Cambridge, MA 02138 USA\\
\textsuperscript{2}OpenAI
}

\email{
    samchristian@fas.harvard.edu, bmazel@g.harvard.edu, yuchen\_wang@fas.harvard.edu, xiyin@fas.harvard.edu, yutaizhang@fas.harvard.edu}

\end{center}

\abstract{We develop a method for numerically computing higher genus CFT correlators and string amplitudes based on Kontsevich's ribbon graph parametrization of the moduli space $\mathcal{M}_{g,n}$ of punctured Riemann surfaces. Using this method, we compute the compact boson partition function up to genus 6 as well as the moduli space integrand for bosonic string vacuum amplitudes up to genus 3. Our results pass a number of nontrivial numerical checks. In particular, the moduli space integrand of bosonic string amplitudes reproduce the Igusa cusp forms at genera 2 and 3.}

\vfill

\end{titlepage}
\hypersetup{pageanchor=true}

\eject

\begingroup
\hypersetup{linkcolor=black}

\tableofcontents

\endgroup

\section{Introduction}

In the on shell formalism of string theory, string amplitudes are computed by integrating CFT correlators over the moduli space of punctured Riemann surfaces \cite{Polyakov:1981}. At genus $g$, an $n$ point amplitude of the bosonic string takes the form
\ie 
\mathcal{A}_{g,n} = \mathcal{N}_{g,n} \int_{\mathcal{M}_{g,n}} \left\langle  e^{\mathcal{B}} \prod_{i=1}^n \mathcal{V}_i\right\rangle,
\fe
where $\mathcal{M}_{g,n}$ is the moduli space of genus $g$ Riemann surfaces with $n$ punctures, the angled brackets stand for correlators of the worldsheet CFT, $\mathcal{V}_i$ are vertex operators that correspond to the asymptotic one particle states, $\mathcal{N}_{g,n}$ is a normalization constant that is fixed unambiguously by unitarity, and $\mathcal{B}$ is a Grassmann even 1-form on $\mathcal{M}_{g,n}$ that is linear in the $b$-ghosts. To compute the string amplitude, the conventional on-shell prescription is to first compute the worldsheet CFT correlator on a general Riemann surface and then integrate over the moduli space $\mathcal{M}_{g,n}$. A similar expression exists for the superstring, where one also needs to integrate over the Grassmann odd moduli and sum over the spin structures on the Riemann surface.

A remarkable number of amplitudes have been computed at genus 1, where the integral over moduli space is a two dimensional integral over the fundamental domain of $PSL(2,\mathbb{Z})$. For genus $g\geq 2$,CFT correlation functions on a given Riemann surface can either be computed analytically for free theories in terms of the period matrix \cite{Verlinde:1986kw,DHokerPhong:1988,DHokerPhong:1989Chiral} or numerically in terms of plumbing parameters and CFT data for general worldsheet theories \cite{Cho-Collier-Yin}. Constructing the corresponding string amplitude additionally requires a globally defined moduli space integration form and integrating it over $\mathcal{M}_{g,n}$, whose real dimension is $6g-6+2n$.

Analytic results for genus two have been obtained for certain amplitudes where the modular integration reduces to the Siegel volume of the moduli space or where the amplitude can be evaluated entirely in a degeneration limit and the resulting computation factorizes as the product of two genus one moduli space integrals  \cite{DHokerGutperlePhong:2005,DHokerGreenPiolineRusso:2015}. For genus $g\geq 2$, analytic results have been  determined recursively through holomorphic anomaly equations, geometric recursion, or matrix model topological recursion for special worldsheet theories \cite{Witten:1991Intersection,BCOV:1993,Mirzakhani:2007,EynardOrantin:2007,CollierEtAl:2024}, or obtained using zero mode counting in the pure spinor formalism, with the remaining genus three modular integration reducing to a Siegel volume integral \cite{Berkovits:2004Multiloop,Berkovits:2006HigherDerivative,GomezMafra:2013}. Beyond these special cases, direct evaluations of higher genus string amplitudes by numerically integrating over $\mathcal{M}_{g,n}$ remain largely unexplored.

One reason for this problem is that the modular integrands are mostly computed in terms of the period matrix or the plumbing parameters. The period matrix is defined by choosing a symplectic basis $\{\alpha^I, \beta_I\}_{I=1}^g$ of one dimensional cycles on the Riemann surface \cite{farkas1980riemann,StringNotes}, whose intersection numbers satisfy
\ie
\alpha^I \cdot \alpha^J=\beta_I \cdot \beta_J=0, \quad \alpha^I \cdot \beta_J=\tensor{\delta}{^I_J} .
\fe
One can then choose a basis of holomorphic 1-forms $\omega_I$ that are normalized on the $\alpha$ cycles, namely
\ie
\int_{\alpha^I}\omega_J = \delta^I_J,
\fe
and the period matrix of the Riemann surface is defined by
\ie
\Omega_{IJ} \equiv \int_{\beta_I} \omega_J.
\fe
The period matrix $\Omega$ is a symmetric matrix with positive definite imaginary part, and two period matrices related by a $Sp(2g,\mathbb{Z})$ change of basis parametrize the same Riemann surface. At genera 1 and 2, the elements of the period matrix define local complex coordinates everywhere on the moduli spaces $\mathcal{M}_{1,0}$ and $\mathcal{M}_{2,0}$, and one can utilize the $Sp(2g,\mathbb{Z})$ modular invariance to construct moduli integrands of string amplitudes in Minkowski spacetime in terms of $\Omega$ \cite{Rauch:1959}. For genus three, $\Omega_{IJ}$ provide a set of nondegenerate local coordinates except for hyperelliptic Riemann surfaces, where the coordinates become degenerate. But the set of hyperelliptic Riemann surfaces is of measure zero and therefore $\Omega_{IJ}$ are still a suitable parameterization for the purposes of modular integration provided the string integrand is locally integrable on the hyperelliptic locus. However, the entries $\Omega_{IJ}$ cease to provide local coordinates anywhere on $\mathcal{M}_{g,0}$ for $g\geq 4$.

%Let us denote the local coordinates on $\mathcal{M}_{3,0}$ by $t^k$.
%
%For hyperelliptic Riemann surfaces of genus 3, the variations $\delta\Omega_{IJ}$ under an infinitesimal change $\delta t^k$ of complex structure are subject to a linear relation, which leads to a vanishing Jacobian $\det \frac{\partial \Omega_{IJ}}{\partial t^k}$; hence, $\Omega_{IJ}$ do not give local coordinates at the hyperelliptic locus within $\mathcal{M}_{3,0}$ \cite{Rauch:1959}. For genus $g\geq 4$, the period matrix has $\frac{1}{2}g(g+1)$ independent complex entries, which exceeds the complex dimension $\dim_{\mathbb{C}} \mathcal{M}_{g,0}=3g-3$ of the moduli space. The entries of the period matrix obey nontrivial constraints.  $\Omega_{IJ}$ cannot be regarded as independent coordinates on $\mathcal{M}_{g,0}$, and one cannot simply integrate the moduli integrand over the fundamental region of $Sp(2g,\mathbb{Z})$ within the Siegel half space.

A complementary description is given by constructing Riemann surfaces by gluing together 3-punctured Riemann spheres \cite{Polchinski:1988Factorization}. More specifically, given a pair of punctures on the Riemann spheres, let $z_1, z_2$ be the local coordinates around the two punctures; the plumbing map then identifies a pair of annular regions in those coordinates on the two spheres using
\ie
z_1 z_2 = q,\quad |q|\leq 1.
\fe
For each pair of punctures plumbed this way, the plumbing map introduces 1 complex modulus given by the plumbing parameter $q$. For $g\geq 2$, a genus $g$ Riemann surface can be constructed by plumbing $(3g-3)$ pairs of punctures, and the moduli of the constructed Riemann surface can be described by the $(3g-3)$ plumbing parameters. On backgrounds of string theory where the conformal data of the worldsheet CFT are known, the moduli integrand for string amplitudes can be computed in terms of the plumbing parameters using the Zamolodchikov recursion relation of Virasoro conformal blocks \cite{Zamolodchikov}, which has been constructed for arbitrary plumbing channels \cite{Cho-Collier-Yin}. However, a Riemann surface can admit multiple different pair of pants decompositions, which correspond to different ways of forming the same Riemann surface using the plumbing fixture. The plumbing parameters define overlapping coordinate patches on the moduli space $\mathcal{M}_{g,0}$. Integrating over $\mathcal{M}_{g,0}$ therefore requires a partition of unity, whose construction depends on the transition maps between overlapping patches. In practice, computing these transition maps %review: add "requires the precise relation between coordinates in the different plumbing channels, and"
is difficult.

As an attempt to solve this problem, we utilize Kontsevich's ribbon graph parametrization of the moduli space $\mathcal{M}_{g,n}$ \cite{Kontsevich:1992ti}. The details of this construction will be reviewed in Section~\ref{sec:Kontsevich}, and we sketch the main ideas here. For the case where $n\geq 1$ and $2-2g-n<0$, one can associate to a punctured Riemann surface, together with a positive perimeter around each puncture, a genus $g$ $n$-faced metric ribbon graph using the Strebel differential\cite{Strebel:1967,Strebel:1984}. The faces of the ribbon graph correspond to the punctures on the Riemann surface. Conversely, the Riemann surface can be reconstructed from the ribbon graph by taking a semi-infinite cylinder for each face and gluing together boundary segments of the cylinders according to the ribbon graph edges. This procedure gives an isomorphism
\ie
\mathcal{M}_{g,n}\times \mathbb{R}_+^n \xrightarrow{\sim} \mathcal{M}_{g,n}^{\text{comb}},
\fe
where $\mathcal{M}_{g,n}^{\text{comb}}$ is the moduli space of metric ribbon graphs of genus $g$ with $n$ faces. The ribbon graphs provide a cell decomposition of moduli space \cite{Penner}. Each top dimensional cell is labeled by the topology of a ribbon graph with three edges at each vertex, the positive edge lengths $\{\ell_a\}$ of the ribbon graph give coordinates inside each cell, and neighboring cells meet when one or more edge lengths shrink to zero, and trivalent vertices merge into vertices of higher valence. The moduli space integral can then be decomposed into sums of integrals over the edge lengths in each cell, where the edge lengths are subject to a simple constraint that the edge lengths around each face add up to a fixed number. For the case where $n=1$, the moduli integral is a sum of integrals over simplexes. We will focus on this case.

To compute the moduli integrand in the ribbon graph parameters, we utilize the fact that a Riemann surface can be reconstructed by gluing together boundary segments of semi-infinite cylinders. According to the Strebel construction, there exists a conformal frame where a 1-punctured Riemann surface at arbitrary genus is described as a unit disk with pairs of boundary segments identified. We call this conformal frame the disk frame.

In this paper, we develop two complementary numerical algorithms utilizing the disk frame. The first algorithm numerically computes the holomorphic data, including a basis of holomorphic 1-forms and the period matrix, from the ribbon graph parameters. This is achieved by discretizing the disk frame boundary circle and imposing matching conditions on the holomorphic 1-forms across every pair of identified points. The same data can then be used to compute Abel-Jacobi maps, Riemann theta functions, prime forms, and the other objects that enter free field and ghost correlators. This construction gives a numerical map from the ribbon graph parameters to the period matrix, which allows formulae written in the period matrix frame to be evaluated in terms of ribbon graph parameters.

The second algorithm computes free CFT correlators directly in terms of the ribbon graph data by explicit evaluation of the path integral. In the disk frame, a path integral on the Riemann surface can be written as a functional integral over the field variables on the boundary circle of the unit disk, subject to the identifications between glued segments. In this way, the complicated topology of the surface reduces to a set of simple boundary identifications. For free CFTs, this path integral is Gaussian, and the functional integration can be directly performed after discretizing the boundary circle.

We test these algorithms against analytic results at several genera. At genus 1, after accounting for the Weyl transformation between the disk frame and the flat torus frame, we reproduce the expected moduli dependence of the free boson partition function, the string integration measure, and one point functions of simple operators. At higher genus, we check the compact boson partition function against its period matrix representation at genera 2 through 6, which serves as a test of the algorithm for the free boson partition function. We also check that the combined matter and ghost correlators on genus 2 surfaces and nonhyperelliptic genus 3 surfaces reproduce the expected dependence on the Igusa cusp forms $\chi_{10}$ and $\chi_{18}$, respectively \cite{DHokerPhong:1988,Igusa:1962,Igusa:1967}. This serves as a test of the algorithm for the holomorphic data.

The rest of this paper is organized as follows. Section~\ref{geometrysection} reviews Kontsevich's ribbon graph parametrization of the moduli space $\mathcal{M}_{g,n}$ and introduces the algorithm of numerically computing the holomorphic data from the ribbon graph parameters. Section~\ref{bosonsection} introduces the algorithm of computing free boson parameters in terms of the ribbon graph data. Section~\ref{bcsection} determines the $b$ ghost insertions associated with the ribbon graph moduli and introduces the algorithm we use to compute $bc$ ghost correlators. We test our algorithm by comparing the results of CFT correlators and moduli integrands against known analytic results in Section~\ref{numericssection}. In Section~\ref{discussionsection}, we conclude with a summary and discuss the limitations of the current algorithm and future directions. The code used to generate our numerical results is available in an online repository.\footnote{\url{https://github.com/Sam-2727/string-amplitudes}}

\section{Riemann surfaces from ribbon graphs}

\label{geometrysection}

\subsection{Kontsevich's isomorphism}
\label{sec:Kontsevich}

In this section, we review Kontsevich's parametrization of the moduli space $\mathcal{M}_{g,n}$ of genus $g$ Riemann surfaces with $n$ punctures by metric ribbon graphs \cite{Kontsevich:1992ti}. A ribbon graph is defined as a collection of vertices, edges connecting the vertices, and a cyclic ordering of edges associated with each vertex. The faces of the ribbon graph are the disjoint cycles of vertices and edges and each face corresponds to one puncture of the Riemann surface. A metric ribbon graph is a ribbon graph where each edge is associated with a positive length $\ell_i$.

Specifically, Kontsevich's isomorphism identifies $\mathcal{M}_{g,n}$ with the combinatorial moduli space $\mathcal{M}_{g,n}^{\text{comb}}$ of metric ribbon graphs:
\ie
\mathcal{M}_{g,n}\times \mathbb{R}_+^n \xrightarrow{\sim} \mathcal{M}_{g,n}^{\text{comb}}.
\fe
Here, each point of $\mathcal{M}_{g,n}^{\text{comb}}$ corresponds to a metric ribbon graph of genus $g$ with $n$ faces, and the $n$ positive numbers in $\mathbb{R}_+^n$ parametrize the total edge length around each face.

This correspondence between $\mathcal{M}_{g,n}$ and $\mathcal{M}^{\text{comb}}_{g,n}$ can be constructed using a certain meromorphic quadratic differential, the Strebel differential, on the punctured Riemann surface. For a meromorphic quadratic differential $\omega=\varphi(z)dz^2$, a horizontal trajectory is defined to be a curve $\gamma:t\mapsto z(t)$ satisfying
\ie
\varphi(z(t))\left(\frac{\partial z(t)}{\partial t}\right)^2\in \mathbb{R}_{+}.
\fe

The Strebel differential $\omega$ is defined using the following theorem \cite{Strebel:1967, Strebel:1984}.
\begin{theorem}
Let $\Sigma$ be a genus $g$ Riemann surface with $n$ punctures $\{z_1,\cdots, z_n\}$, where $n\geq 1$ and $2-2g-n<0$. Given a positive number $p_i\in \mathbb{R}_+$ for each puncture, there exists a unique quadratic differential $\omega = \varphi(z)dz^2$ that satisfies
\begin{enumerate}
\item $\omega$ is holomorphic on $\Sigma\backslash\{z_1,\cdots, z_n\}$, and $\omega$ has a double pole at each $z_i$.
\item The union of all noncompact horizontal trajectories of $\omega$ has measure zero.
\item Each compact horizontal trajectory of $\omega$ is a simple loop around one of the punctures $z_i$. Horizontal trajectories $\gamma_i$ around puncture $z_i$ satisfy
\ie
\oint_{\gamma_i}\sqrt{\varphi(z)}\, dz =2\pi p_i.
\label{lengthconst}
\fe
\end{enumerate}
\end{theorem}

% All figures are temporary. We should redraw them in tikZ...
\begin{figure}
    \begin{center}
        \includegraphics[scale=0.3]{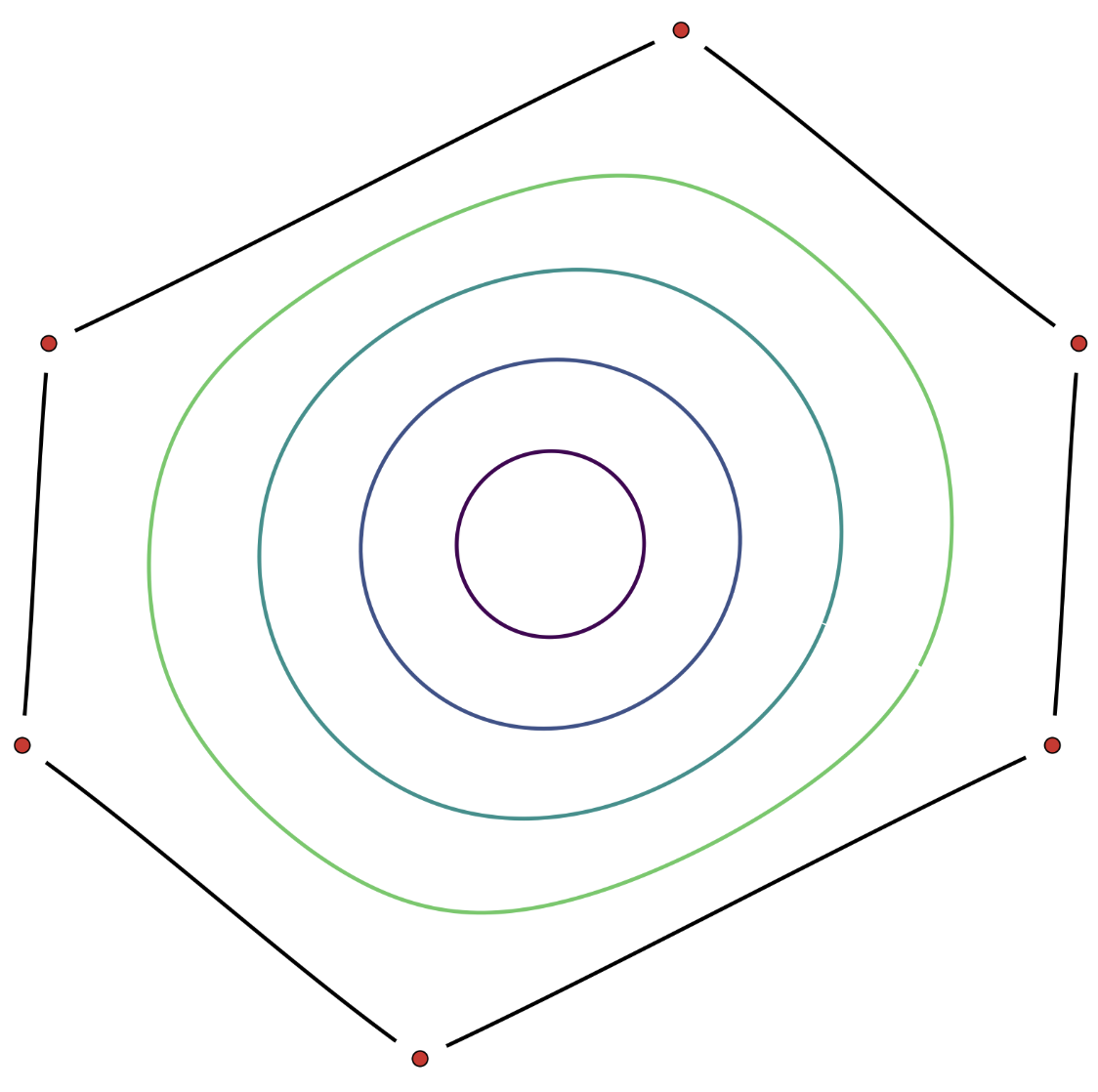}
    \end{center}
    \caption{Horizontal trajectories of the Strebel differential $\omega$. The red points correspond to the zeros of $\omega$, the black lines are the noncompact horizontal trajectories, and the colored lines are the compact horizontal trajectories.}
    \label{celldecompSigma}
\end{figure}

It can be proven that each noncompact horizontal trajectory of $\omega$ is critical, namely, that it is bounded by two zeros of $\omega$. The union of all noncompact horizontal trajectories then forms a graph $\Gamma$ with $n$ faces, whose vertices are the zeros of $\omega$ and whose edges are the noncompact horizontal trajectories. The Riemann surface $\Sigma$ can be divided into $n$ domains $D_i$, each swept out by the compact horizontal trajectories around $z_i$, as illustrated in figure \ref{celldecompSigma}. This gives a cell decomposition of $\Sigma$: the cells of dimensions 2, 1, and 0 are the domains $D_i$, edges, and vertices of the graph $\Gamma$, respectively. If we define the conformally flat metric
\ie ds^2=|\varphi(z)|dzd\bar{z},
\label{flatcylmetric}\fe
then the condition \eqref{lengthconst} tells us that all compact horizontal trajectories around $z_i$ have the same length $2\pi p_i$, and the domains $D_i$ have the geometry of a flat semi infinite cylinder or, equivalently, a disk. If we define the disk coordinate $w_i$ on the domain $D_i$ by
\ie
w_i(z) = \exp\left(\frac{i}{p_i}\int^z \sqrt{\omega}\right),
\label{diskcoord}
\fe
then the compact horizontal trajectories will have constant radius $|w_i|$, and the puncture $z_i$ will be located at $w_i=0$. The metric \eqref{flatcylmetric} will take the form
\ie
ds^2 = p_i^2 |w_i|^{-2} dw_i d\bar{w}_i.
\label{diskmetric}
\fe

\begin{figure}
    \begin{center}
        \includegraphics[width=0.7\textwidth]{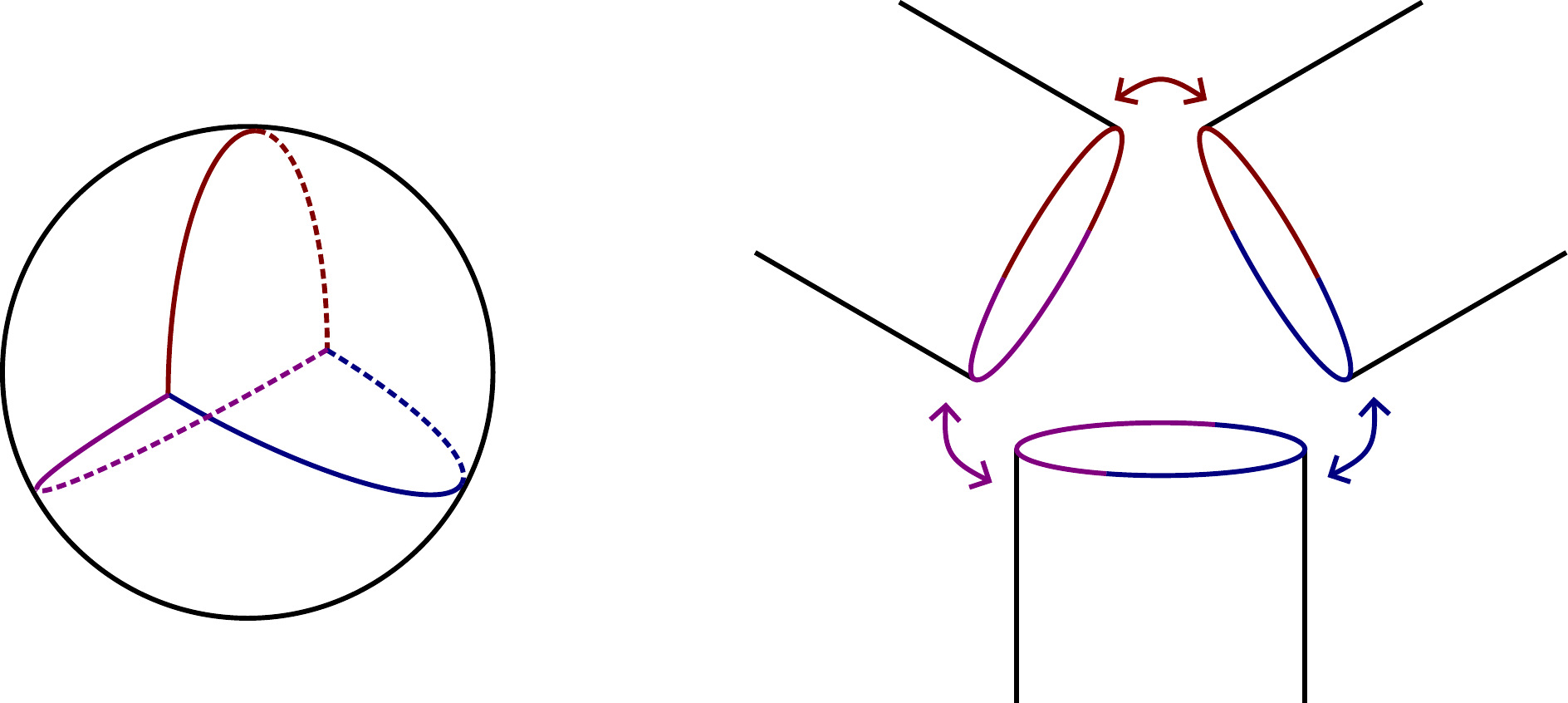}
    \end{center}
    \caption{Reconstructing a sphere with 3 punctures by gluing 3 semi infinite cylinders. The left picture shows the ribbon graph $\Gamma$.}
    \label{gluingcylinders}
\end{figure}

The map $\mathcal{M}_{g,n}\times \mathbb{R}_+^n \rightarrow \mathcal{M}_{g,n}^{\text{comb}}$ is constructed as follows. Since the edge lengths and the cyclic order of edges at each vertex are preserved under biholomorphism, the graph $\Gamma$ carries the structure of a metric ribbon graph. The map is then defined by sending $(\Sigma, \{p_i\})$ to the metric ribbon graph $\Gamma$ constructed in this way. Conversely, from a metric ribbon graph $\Gamma$, the Riemann surface $\Sigma$ can be reconstructed by gluing a semi infinite flat cylinder along each face of $\Gamma$, where the circumference of the cylinder is equal to the total edge length around that face, as shown in figure \ref{gluingcylinders}. This process defines the inverse map $\mathcal{M}_{g,n}^{\text{comb}}\rightarrow \mathcal{M}_{g,n}\times \mathbb{R}_+^n$.

The combinatorial moduli space $\mathcal{M}_{g,n}^{\text{comb}}$ comes with a natural cell decomposition in which each cell is labeled by the topology, or combinatorial type, of the ribbon graph. In particular, the top dimensional cells correspond to trivalent graphs, namely graphs with only cubic vertices. In this case, each zero of the Strebel differential has degree $1$. The full combinatorial moduli space $\mathcal{M}_{g,n}^{\text{comb}}$ is obtained by gluing these cells together along boundaries where one or more edge lengths degenerate to zero. For a trivalent graph with $n$ faces, its dual graph defines a triangulation of the Riemann surface $\Sigma$ with $n$ vertices, and the number of top dimensional cells in $\mathcal{M}_{g,n}^{\text{comb}}$ can be counted by enumerating such triangulations on $\Sigma$. 

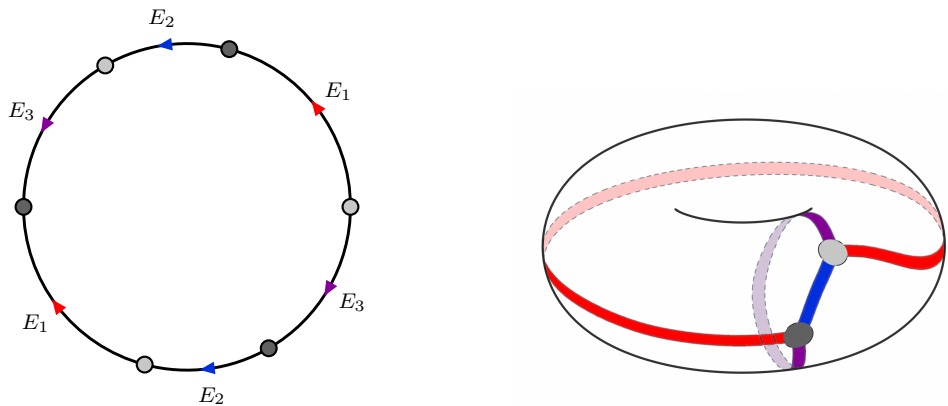
\begin{figure}
	\definecolor{edge1}{HTML}{FE0000}
	\definecolor{edge2}{HTML}{002FDD}
	\definecolor{edge3}{HTML}{850096}
	\definecolor{vertexlight}{HTML}{C7C7C7}
	\definecolor{vertexdark}{HTML}{616161}
    \begin{center}
   		\subfloat[]{
   			\begin{tikzpicture}[
   				scale=0.6,
   				line cap=round,
   				boundary edge/.style={draw=black, line width=1.1pt},
   				edge label/.style={font=\scriptsize, fill=white, inner sep=0.8pt},
   				endpoint light/.style={circle, draw=black, fill=vertexlight, line width=0.8pt, inner sep=2.1pt},
   				endpoint dark/.style={circle, draw=black, fill=vertexdark, line width=0.8pt, inner sep=2.1pt}
   				]
   				\def\r{3.6}
   				
   				\draw[boundary edge] (0.00000:\r) arc[start angle=0.00000, end angle=75.00000, radius=\r];
   				\begin{scope}[shift={(37.50000:\r)}, rotate=127.50000]
   					\fill[edge1] (0.19,0) -- (-0.14,0.14) -- (-0.14,-0.14) -- cycle;
   				\end{scope}
   				\node[edge label] at (37.50000:4.20) {$E_{1}$};
   				\draw[boundary edge] (75.00000:\r) arc[start angle=75.00000, end angle=120.00000, radius=\r];
   				\begin{scope}[shift={(97.50000:\r)}, rotate=187.50000]
   					\fill[edge2] (0.19,0) -- (-0.14,0.14) -- (-0.14,-0.14) -- cycle;
   				\end{scope}
   				\node[edge label] at (97.50000:4.20) {$E_{2}$};
   				\draw[boundary edge] (120.00000:\r) arc[start angle=120.00000, end angle=180.00000, radius=\r];
   				\begin{scope}[shift={(150.00000:\r)}, rotate=240.00000]
   					\fill[edge3] (0.19,0) -- (-0.14,0.14) -- (-0.14,-0.14) -- cycle;
   				\end{scope}
   				\node[edge label] at (150.00000:4.20) {$E_{3}$};
   				\draw[boundary edge] (180.00000:\r) arc[start angle=180.00000, end angle=255.00000, radius=\r];
   				\begin{scope}[shift={(217.50000:\r)}, rotate=127.50000]
   					\fill[edge1] (0.19,0) -- (-0.14,0.14) -- (-0.14,-0.14) -- cycle;
   				\end{scope}
   				\node[edge label] at (217.50000:4.20) {$E_{1}$};
   				\draw[boundary edge] (255.00000:\r) arc[start angle=255.00000, end angle=300.00000, radius=\r];
   				\begin{scope}[shift={(277.50000:\r)}, rotate=187.50000]
   					\fill[edge2] (0.19,0) -- (-0.14,0.14) -- (-0.14,-0.14) -- cycle;
   				\end{scope}
   				\node[edge label] at (277.50000:4.20) {$E_{2}$};
   				\draw[boundary edge] (300.00000:\r) arc[start angle=300.00000, end angle=360.00000, radius=\r];
   				\begin{scope}[shift={(330.00000:\r)}, rotate=240.00000]
   					\fill[edge3] (0.19,0) -- (-0.14,0.14) -- (-0.14,-0.14) -- cycle;
   				\end{scope}
   				\node[edge label] at (330.00000:4.20) {$E_{3}$};
   				
   				\node[endpoint light] at (0.00000:\r) {};
   				\node[endpoint dark] at (75.00000:\r) {};
   				\node[endpoint light] at (120.00000:\r) {};
   				\node[endpoint dark] at (180.00000:\r) {};
   				\node[endpoint light] at (255.00000:\r) {};
   				\node[endpoint dark] at (300.00000:\r) {};
   			\end{tikzpicture}
   		}
   		\qquad\qquad
   		\subfloat[]{
   			\includegraphics[width=0.35\linewidth]{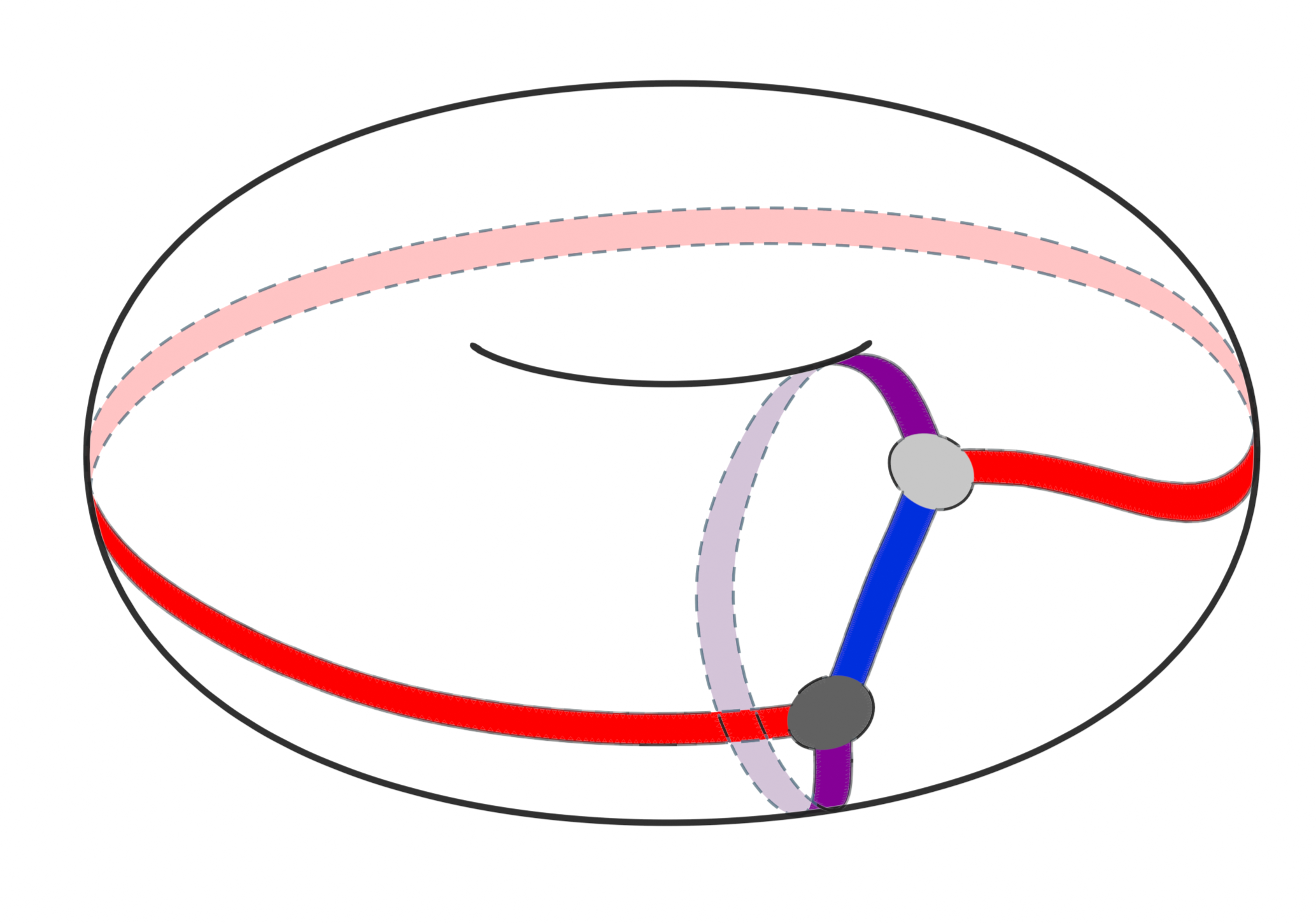} 
   		}
   	\end{center}        
    \caption{The disk frame representation of the ribbon graph that gives rise to the torus with 1 puncture, together with the corresponding genus 1 ribbon graph embedded on the surface of a torus. Note that each trivalent vertex (indicated by light and dark gray) has three representatives in the disk frame. Edges are identified in the direction of the colored arrows.}
    \label{gluingdisk}
\end{figure}

\begin{figure}
	\definecolor{edge1}{HTML}{FE0000}
	\definecolor{edge2}{HTML}{009ADB}
	\definecolor{edge3}{HTML}{00DC00}
	\definecolor{edge4}{HTML}{00BA00}
	\definecolor{edge5}{HTML}{850096}
	\definecolor{edge6}{HTML}{FF7500}
	\definecolor{edge7}{HTML}{CC0000}
	\definecolor{edge8}{HTML}{007ADD}
	\definecolor{edge9}{HTML}{FFB500}
	\definecolor{edge10}{HTML}{00A668}
	\definecolor{edge11}{HTML}{002FDD}
	\definecolor{edge12}{HTML}{E0F300}
	\definecolor{edge13}{HTML}{00FF00}
	\definecolor{edge14}{HTML}{00A8AF}
	\definecolor{edge15}{HTML}{58009F}
	\definecolor{vertexgray}{HTML}{999999}
	\begin{center}
		\subfloat[]{
			\begin{tikzpicture}[
				scale=0.8,
				line cap=round,
				boundary edge/.style={draw=black, line width=1.1pt},
				edge label/.style={font=\scriptsize, fill=white, inner sep=0.8pt},
				endpoint filled/.style={
					circle, draw=black, fill=vertexgray, line width=0.8pt, inner sep=1.8pt
				}
				]
				\def\r{5.2}
				
				\draw[boundary edge] (0.00000:\r) arc[start angle=0.00000, end angle=12.00000, radius=\r];
				\begin{scope}[shift={(6.00000:\r)}, rotate=96.00000]
					\fill[edge1] (0.16,0) -- (-0.11,0.105) -- (-0.11,-0.105) -- cycle;
				\end{scope}
				\node[edge label] at (6.00000:5.75) {$E_{1}$};
				\draw[boundary edge] (12.00000:\r) arc[start angle=12.00000, end angle=24.00000, radius=\r];
				\begin{scope}[shift={(18.00000:\r)}, rotate=108.00000]
					\fill[edge4] (0.16,0) -- (-0.11,0.105) -- (-0.11,-0.105) -- cycle;
				\end{scope}
				\node[edge label] at (18.00000:5.75) {$E_{4}$};
				\draw[boundary edge] (24.00000:\r) arc[start angle=24.00000, end angle=36.00000, radius=\r];
				\begin{scope}[shift={(30.00000:\r)}, rotate=120.00000]
					\fill[edge5] (0.16,0) -- (-0.11,0.105) -- (-0.11,-0.105) -- cycle;
				\end{scope}
				\node[edge label] at (30.00000:5.75) {$E_{5}$};
				\draw[boundary edge] (36.00000:\r) arc[start angle=36.00000, end angle=48.00000, radius=\r];
				\begin{scope}[shift={(42.00000:\r)}, rotate=132.00000]
					\fill[edge9] (0.16,0) -- (-0.11,0.105) -- (-0.11,-0.105) -- cycle;
				\end{scope}
				\node[edge label] at (42.00000:5.75) {$E_{9}$};
				\draw[boundary edge] (48.00000:\r) arc[start angle=48.00000, end angle=60.00000, radius=\r];
				\begin{scope}[shift={(54.00000:\r)}, rotate=144.00000]
					\fill[edge7] (0.16,0) -- (-0.11,0.105) -- (-0.11,-0.105) -- cycle;
				\end{scope}
				\node[edge label] at (54.00000:5.75) {$E_{7}$};
				\draw[boundary edge] (60.00000:\r) arc[start angle=60.00000, end angle=72.00000, radius=\r];
				\begin{scope}[shift={(66.00000:\r)}, rotate=156.00000]
					\fill[edge13] (0.16,0) -- (-0.11,0.105) -- (-0.11,-0.105) -- cycle;
				\end{scope}
				\node[edge label] at (66.00000:5.75) {$E_{13}$};
				\draw[boundary edge] (72.00000:\r) arc[start angle=72.00000, end angle=84.00000, radius=\r];
				\begin{scope}[shift={(78.00000:\r)}, rotate=168.00000]
					\fill[edge15] (0.16,0) -- (-0.11,0.105) -- (-0.11,-0.105) -- cycle;
				\end{scope}
				\node[edge label] at (78.00000:5.75) {$E_{15}$};
				\draw[boundary edge] (84.00000:\r) arc[start angle=84.00000, end angle=96.00000, radius=\r];
				\begin{scope}[shift={(90.00000:\r)}, rotate=180.00000]
					\fill[edge12] (0.16,0) -- (-0.11,0.105) -- (-0.11,-0.105) -- cycle;
				\end{scope}
				\node[edge label] at (90.00000:5.75) {$E_{12}$};
				\draw[boundary edge] (96.00000:\r) arc[start angle=96.00000, end angle=108.00000, radius=\r];
				\begin{scope}[shift={(102.00000:\r)}, rotate=192.00000]
					\fill[edge10] (0.16,0) -- (-0.11,0.105) -- (-0.11,-0.105) -- cycle;
				\end{scope}
				\node[edge label] at (102.00000:5.75) {$E_{10}$};
				\draw[boundary edge] (108.00000:\r) arc[start angle=108.00000, end angle=120.00000, radius=\r];
				\begin{scope}[shift={(114.00000:\r)}, rotate=24.00000]
					\fill[edge13] (0.16,0) -- (-0.11,0.105) -- (-0.11,-0.105) -- cycle;
				\end{scope}
				\node[edge label] at (114.00000:5.75) {$E_{13}$};
				\draw[boundary edge] (120.00000:\r) arc[start angle=120.00000, end angle=132.00000, radius=\r];
				\begin{scope}[shift={(126.00000:\r)}, rotate=216.00000]
					\fill[edge14] (0.16,0) -- (-0.11,0.105) -- (-0.11,-0.105) -- cycle;
				\end{scope}
				\node[edge label] at (126.00000:5.75) {$E_{14}$};
				\draw[boundary edge] (132.00000:\r) arc[start angle=132.00000, end angle=144.00000, radius=\r];
				\begin{scope}[shift={(138.00000:\r)}, rotate=48.00000]
					\fill[edge15] (0.16,0) -- (-0.11,0.105) -- (-0.11,-0.105) -- cycle;
				\end{scope}
				\node[edge label] at (138.00000:5.75) {$E_{15}$};
				\draw[boundary edge] (144.00000:\r) arc[start angle=144.00000, end angle=156.00000, radius=\r];
				\begin{scope}[shift={(150.00000:\r)}, rotate=60.00000]
					\fill[edge10] (0.16,0) -- (-0.11,0.105) -- (-0.11,-0.105) -- cycle;
				\end{scope}
				\node[edge label] at (150.00000:5.75) {$E_{10}$};
				\draw[boundary edge] (156.00000:\r) arc[start angle=156.00000, end angle=168.00000, radius=\r];
				\begin{scope}[shift={(162.00000:\r)}, rotate=252.00000]
					\fill[edge11] (0.16,0) -- (-0.11,0.105) -- (-0.11,-0.105) -- cycle;
				\end{scope}
				\node[edge label] at (162.00000:5.75) {$E_{11}$};
				\draw[boundary edge] (168.00000:\r) arc[start angle=168.00000, end angle=180.00000, radius=\r];
				\begin{scope}[shift={(174.00000:\r)}, rotate=264.00000]
					\fill[edge6] (0.16,0) -- (-0.11,0.105) -- (-0.11,-0.105) -- cycle;
				\end{scope}
				\node[edge label] at (174.00000:5.75) {$E_{6}$};
				\draw[boundary edge] (180.00000:\r) arc[start angle=180.00000, end angle=192.00000, radius=\r];
				\begin{scope}[shift={(186.00000:\r)}, rotate=96.00000]
					\fill[edge4] (0.16,0) -- (-0.11,0.105) -- (-0.11,-0.105) -- cycle;
				\end{scope}
				\node[edge label] at (186.00000:5.75) {$E_{4}$};
				\draw[boundary edge] (192.00000:\r) arc[start angle=192.00000, end angle=204.00000, radius=\r];
				\begin{scope}[shift={(198.00000:\r)}, rotate=288.00000]
					\fill[edge8] (0.16,0) -- (-0.11,0.105) -- (-0.11,-0.105) -- cycle;
				\end{scope}
				\node[edge label] at (198.00000:5.75) {$E_{8}$};
				\draw[boundary edge] (204.00000:\r) arc[start angle=204.00000, end angle=216.00000, radius=\r];
				\begin{scope}[shift={(210.00000:\r)}, rotate=120.00000]
					\fill[edge9] (0.16,0) -- (-0.11,0.105) -- (-0.11,-0.105) -- cycle;
				\end{scope}
				\node[edge label] at (210.00000:5.75) {$E_{9}$};
				\draw[boundary edge] (216.00000:\r) arc[start angle=216.00000, end angle=228.00000, radius=\r];
				\begin{scope}[shift={(222.00000:\r)}, rotate=312.00000]
					\fill[edge2] (0.16,0) -- (-0.11,0.105) -- (-0.11,-0.105) -- cycle;
				\end{scope}
				\node[edge label] at (222.00000:5.75) {$E_{2}$};
				\draw[boundary edge] (228.00000:\r) arc[start angle=228.00000, end angle=240.00000, radius=\r];
				\begin{scope}[shift={(234.00000:\r)}, rotate=324.00000]
					\fill[edge3] (0.16,0) -- (-0.11,0.105) -- (-0.11,-0.105) -- cycle;
				\end{scope}
				\node[edge label] at (234.00000:5.75) {$E_{3}$};
				\draw[boundary edge] (240.00000:\r) arc[start angle=240.00000, end angle=252.00000, radius=\r];
				\begin{scope}[shift={(246.00000:\r)}, rotate=156.00000]
					\fill[edge11] (0.16,0) -- (-0.11,0.105) -- (-0.11,-0.105) -- cycle;
				\end{scope}
				\node[edge label] at (246.00000:5.75) {$E_{11}$};
				\draw[boundary edge] (252.00000:\r) arc[start angle=252.00000, end angle=264.00000, radius=\r];
				\begin{scope}[shift={(258.00000:\r)}, rotate=168.00000]
					\fill[edge12] (0.16,0) -- (-0.11,0.105) -- (-0.11,-0.105) -- cycle;
				\end{scope}
				\node[edge label] at (258.00000:5.75) {$E_{12}$};
				\draw[boundary edge] (264.00000:\r) arc[start angle=264.00000, end angle=276.00000, radius=\r];
				\begin{scope}[shift={(270.00000:\r)}, rotate=180.00000]
					\fill[edge14] (0.16,0) -- (-0.11,0.105) -- (-0.11,-0.105) -- cycle;
				\end{scope}
				\node[edge label] at (270.00000:5.75) {$E_{14}$};
				\draw[boundary edge] (276.00000:\r) arc[start angle=276.00000, end angle=288.00000, radius=\r];
				\begin{scope}[shift={(282.00000:\r)}, rotate=192.00000]
					\fill[edge7] (0.16,0) -- (-0.11,0.105) -- (-0.11,-0.105) -- cycle;
				\end{scope}
				\node[edge label] at (282.00000:5.75) {$E_{7}$};
				\draw[boundary edge] (288.00000:\r) arc[start angle=288.00000, end angle=300.00000, radius=\r];
				\begin{scope}[shift={(294.00000:\r)}, rotate=204.00000]
					\fill[edge8] (0.16,0) -- (-0.11,0.105) -- (-0.11,-0.105) -- cycle;
				\end{scope}
				\node[edge label] at (294.00000:5.75) {$E_{8}$};
				\draw[boundary edge] (300.00000:\r) arc[start angle=300.00000, end angle=312.00000, radius=\r];
				\begin{scope}[shift={(306.00000:\r)}, rotate=216.00000]
					\fill[edge1] (0.16,0) -- (-0.11,0.105) -- (-0.11,-0.105) -- cycle;
				\end{scope}
				\node[edge label] at (306.00000:5.75) {$E_{1}$};
				\draw[boundary edge] (312.00000:\r) arc[start angle=312.00000, end angle=324.00000, radius=\r];
				\begin{scope}[shift={(318.00000:\r)}, rotate=228.00000]
					\fill[edge2] (0.16,0) -- (-0.11,0.105) -- (-0.11,-0.105) -- cycle;
				\end{scope}
				\node[edge label] at (318.00000:5.75) {$E_{2}$};
				\draw[boundary edge] (324.00000:\r) arc[start angle=324.00000, end angle=336.00000, radius=\r];
				\begin{scope}[shift={(330.00000:\r)}, rotate=240.00000]
					\fill[edge5] (0.16,0) -- (-0.11,0.105) -- (-0.11,-0.105) -- cycle;
				\end{scope}
				\node[edge label] at (330.00000:5.75) {$E_{5}$};
				\draw[boundary edge] (336.00000:\r) arc[start angle=336.00000, end angle=348.00000, radius=\r];
				\begin{scope}[shift={(342.00000:\r)}, rotate=252.00000]
					\fill[edge6] (0.16,0) -- (-0.11,0.105) -- (-0.11,-0.105) -- cycle;
				\end{scope}
				\node[edge label] at (342.00000:5.75) {$E_{6}$};
				\draw[boundary edge] (348.00000:\r) arc[start angle=348.00000, end angle=360.00000, radius=\r];
				\begin{scope}[shift={(354.00000:\r)}, rotate=264.00000]
					\fill[edge3] (0.16,0) -- (-0.11,0.105) -- (-0.11,-0.105) -- cycle;
				\end{scope}
				\node[edge label] at (354.00000:5.75) {$E_{3}$};
				
				\node[endpoint filled] at (0.00000:\r) {};
				\node[endpoint filled] at (12.00000:\r) {};
				\node[endpoint filled] at (24.00000:\r) {};
				\node[endpoint filled] at (36.00000:\r) {};
				\node[endpoint filled] at (48.00000:\r) {};
				\node[endpoint filled] at (60.00000:\r) {};
				\node[endpoint filled] at (72.00000:\r) {};
				\node[endpoint filled] at (84.00000:\r) {};
				\node[endpoint filled] at (96.00000:\r) {};
				\node[endpoint filled] at (108.00000:\r) {};
				\node[endpoint filled] at (120.00000:\r) {};
				\node[endpoint filled] at (132.00000:\r) {};
				\node[endpoint filled] at (144.00000:\r) {};
				\node[endpoint filled] at (156.00000:\r) {};
				\node[endpoint filled] at (168.00000:\r) {};
				\node[endpoint filled] at (180.00000:\r) {};
				\node[endpoint filled] at (192.00000:\r) {};
				\node[endpoint filled] at (204.00000:\r) {};
				\node[endpoint filled] at (216.00000:\r) {};
				\node[endpoint filled] at (228.00000:\r) {};
				\node[endpoint filled] at (240.00000:\r) {};
				\node[endpoint filled] at (252.00000:\r) {};
				\node[endpoint filled] at (264.00000:\r) {};
				\node[endpoint filled] at (276.00000:\r) {};
				\node[endpoint filled] at (288.00000:\r) {};
				\node[endpoint filled] at (300.00000:\r) {};
				\node[endpoint filled] at (312.00000:\r) {};
				\node[endpoint filled] at (324.00000:\r) {};
				\node[endpoint filled] at (336.00000:\r) {};
				\node[endpoint filled] at (348.00000:\r) {};
			\end{tikzpicture}
		}\\[0.5em]
		\subfloat[]{
			\includegraphics[width=0.75\linewidth]{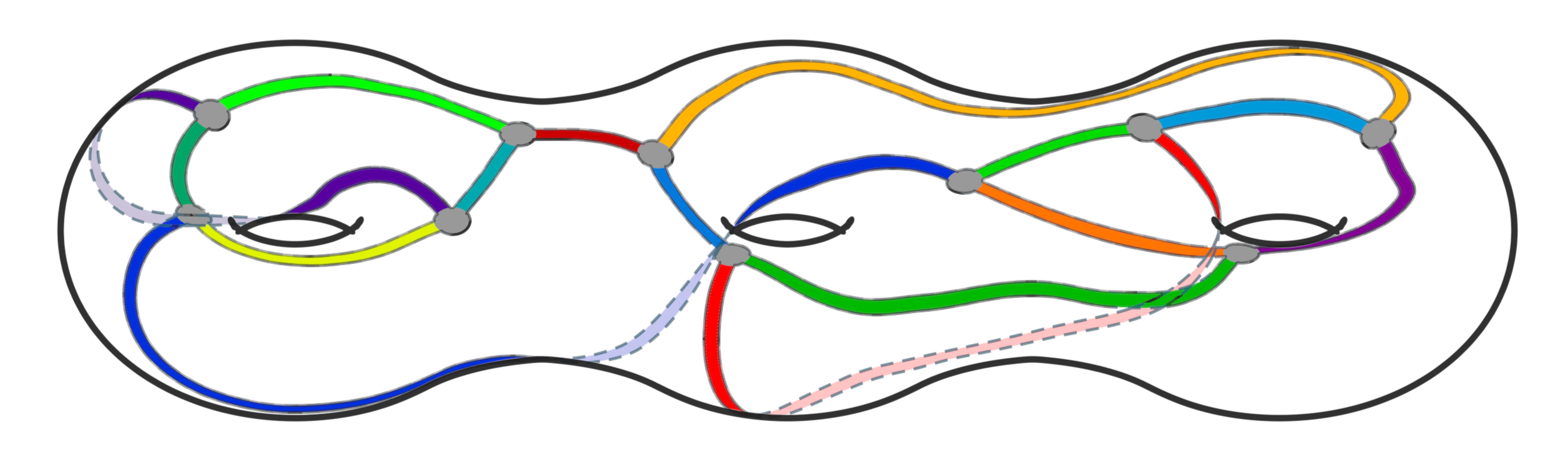} 
		}
	\end{center}
	\caption{Disk frame visualization of a genus 3 ribbon graph and the reconstructed genus 3 Riemann surface with the ribbon graph embedded in it. The edges are shown with all equal lengths for visual clarity. To reconstruct the genus 3 Riemann surface, one glues together the edges with the same color in opposite orientation.}
	\label{genus3_Example}
\end{figure}
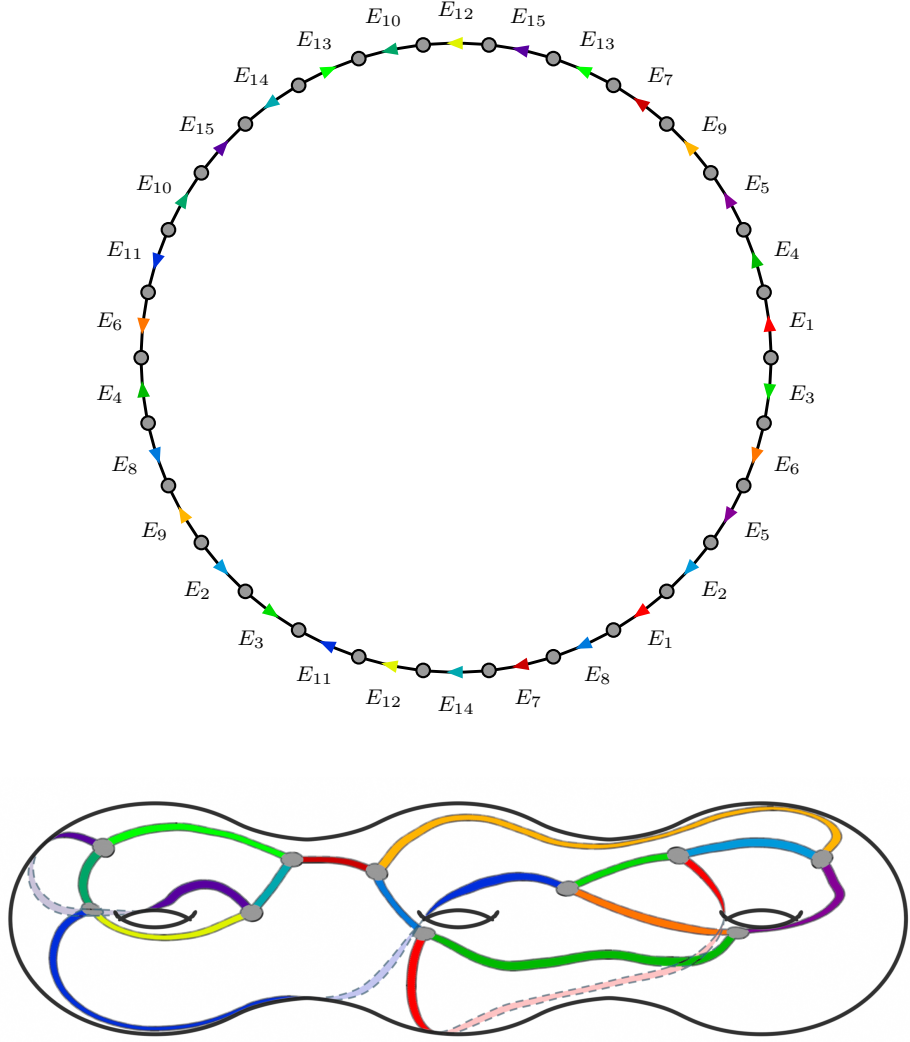

In the rest of this paper, we focus on the case of Riemann surfaces with a single puncture. In this case, the Riemann surface $\Sigma$ can be reconstructed from the ribbon graph by identifying segments on the boundary of a flat unit disk, with the puncture located at its center. We call this conformal frame the disk frame. The example of reconstructing a torus with 1 puncture by gluing boundary segments of the disk frame is illustrated in figure~\ref{gluingdisk}. A more nontrivial example of reconstructing a 1-punctured genus 3 Riemann surface is shown in figure \ref{genus3_Example}.

A generic metric ribbon graph in $\mathcal{M}_{g,n}^{\text{comb}}$ has only cubic vertices. The number of vertices $V$ and edges $E$ in the ribbon graph $\Gamma$ therefore satisfy $3V=2E$. Combining this with Euler's formula $V-E+F=2-2g$, one finds
\ie E=6g-3,\quad V=4g-2.\fe 
Thus, reconstructing $\Sigma$ requires gluing $2E=6(2g-1)$ segments along the boundary of the disk frame, and different ways of gluing correspond to different topologies of the ribbon graph. The number of one vertex triangulations of a genus $g$ surface was determined by Bacher and Vdovina \cite{BACHER200213}, and the corresponding numbers of ribbon graph topologies are listed in table \ref{topologynumber}. 

\begin{table}
\begin{center}
\begin{tabular}{cccc}
\hline
Genus & Number of edges & Number of vertices & Number of ribbon graph topologies \\ \hline
1     & 3 & 2 & 1                                 \\
2     & 9 & 6 & 9                                 \\
3     & 15 & 10 & 1726                              \\
4     & 21 & 14 & 1349005                           \\
5     & 27 & 18 & 2169056374                        \\ \hline
\end{tabular}
\end{center}
\caption{Number of edges, vertices, and topologies of genus $g$ ribbon graphs with a single face.}
\label{topologynumber}
\end{table}

\subsection{Holomorphic data from ribbon graphs: Genus 1}

Given a metric ribbon graph $\Gamma$, one can numerically compute a basis of holomorphic 1-forms on $\Sigma$. For simplicity, we first present the algorithm for computing the holomorphic 1-form $f(z)dz$ on a genus 1 surface.

The torus with 1 puncture can be reconstructed by gluing 3 pairs of boundary segments on the unit disk. For genus 1, there is only one ribbon graph topology, and the corresponding gluing map is the same as the one illustrated in figure \ref{gluingdisk}. If we parametrize the unit circle by the angular coordinate $\sigma \in (0,2\pi)$, the gluing map reads
\ie
&\sigma \sim \pi + \theta_1 - \sigma, \quad 0<\sigma<\theta_1,\\
&\sigma \sim \pi + \theta_1 + \theta_2 - \sigma, \quad \theta_1 <\sigma<\theta_2,\\
&\sigma \sim 2\pi + \theta_2 - \sigma,\quad \theta_2 < \sigma < \pi.
\label{gluinggenus1}
\fe
The angles $\theta_{1}, \theta_2-\theta_1$, and $\pi-\theta_2$ are the angular sizes of the three segments. In other words, they are the edge lengths of the ribbon graph if we take the radius $p$ to be $1$. The moduli space $\mathcal{M}_{1,1}$ can be parametrized by the two angles $\theta_1$ and $\theta_2$.

Let us denote the disk coordinate by $z$, where the unit circle is given by $|z|=1$. The holomorphic 1-form $f(z)dz$ is regular for $|z|<1$, while $f(z)$ diverges at the Strebel vertices. The vertices are located at the zeroes $z=\pm 1$, $\pm e^{i\theta_1}$, and $\pm e^{i\theta_2}$ of the Strebel differential $\omega = \varphi(z)dz^2$. In the flat torus coordinate $u$, where the holomorphic 1-form is simply $du$, the disk coordinate $z$ is related to $u$ by equation \eqref{diskcoord}, which implies
\ie
f^{-1}(z) = \frac{dz}{du} = iz\sqrt{\varphi(u)}.
\fe
%review: should we write (f(z))^{-1} for clarity?
Since the ribbon graph $\Gamma$ has only cubic vertices, each zero $u_i$ of $\omega$ is of degree one, \textit{i.e.}, $\varphi(u)\sim (u-u_i)$ near $u_i$, and hence $f(z)$ diverges like $(u-u_i)^{-1/2}$. Translating back to the disk coordinate, one finds the following divergent behavior of $f(z)$ around the zeros $z_i$:
\ie
f(z) \sim (z-z_i)^{-1/3}.
\fe
Based on this divergent behavior, one can make the following ansatz with appropriate poles at the six representatives of the Strebel vertices:
\ie\label{fzoneform}
f(z) dz &= \frac{\sum_{n=1}^\infty a_n z^{n-1}}{(1-z^2)^{\frac13} (1-z^2 e^{-2 i \theta_1})^{\frac13} (1-z^2 e^{-2i \theta_2})^{\frac13}} dz.
\fe

The coefficients $a_n$ can be determined numerically as follows. We discretize the unit circle into $L$ points, where $L$ is an even number. The discretized points $z(k)$ are located at
\ie\label{zkposition}
z(k) = e^{\frac{2\pi i}{L}(k-\frac{1}{2})},\quad k=1,\cdots, L,
\fe
and we make the identifications
\ie
&z(k) \sim z(\tfrac{L}{2}+\ell_1+1-k),\quad 1\leq k \leq \ell_1,\\
&z(k) \sim z(\tfrac{L}{2}+2\ell_1+\ell_2+1-k),\quad \ell_1+1 \leq k \leq \ell_1+\ell_2,\\
&z(k) \sim z(L+\ell_1+\ell_2+1-k),\quad \ell_1+\ell_2+1\leq k \leq \tfrac{L}{2},\\
\fe
as a discretized version of the gluing map \eqref{gluinggenus1}. We will denote the point identified with $z(k)$ by $z(k')$. The relative segment lengths %review: I think 2\pi should be included
$2\pi\tfrac{\ell_1}{L}, 2\pi\tfrac{\ell_2}{L}$ can be identified with the moduli $\theta_1, \theta_2-\theta_1$, respectively. The Strebel vertices $\pm 1,\pm e^{i\theta_1}, \pm e^{i\theta_2}$ are located halfway between the discretized points at the edges of identified sections of the boundary.

In order to solve \eqref{fzoneform}, we can truncate the Fourier coefficients $a_n$ by restricting to the lowest $\tfrac{L}{2}$ modes, normalize with $a_1=1$, and solve for $a_2,\cdots, a_{L/2}$ by imposing the condition that $f(z)dz$ matches across identified discretized points:
\ie
{}&\frac{\sum_{n=1}^{\frac{L}{2}} a_n e^{in (2 \pi \frac{k-\frac 12}{L})} }{(1-e^{4i\pi \frac{k-\frac12}{L}})^{\frac13} (1- e^{4i \pi \frac{k-\frac12-\ell_1}{L} })^{\frac13} (1- e^{4i\pi \frac{k-\frac12-\ell_1 -\ell_2}{L} })^{\frac13}}\\
&\qquad\qquad = - \frac{\sum_{n=1}^{\frac{L}{2}} a_n e^{in (2 \pi \frac{k'-\frac 12}{L})} }{(1-e^{4i\pi \frac{k'-\frac12}{L}})^{\frac13} (1- e^{4i \pi \frac{k'-\frac12-\ell_1}{L} })^{\frac13} (1- e^{4i\pi \frac{k'-\frac12-\ell_1 -\ell_2}{L} })^{\frac13}}, ~~~~1 \leq k \leq \tfrac{L}{2},
\fe
where the relative sign is due to flipping the orientation of $dz$. 

\begin{figure}
    \begin{center}
        \begin{tikzpicture}[
            >=Latex,
            line cap=round,
            boundary/.style={line width=1.1pt},
            cycle/.style={
                line width=1.3pt,
                postaction={
                    decorate,
                    decoration={
                        markings,
                        mark=at position 0.5 with {\arrow{Latex[length=2.4mm]}}
                    }
                }
            },
            arc label/.style={font=\small, fill=white, inner sep=1.3pt},
            endpoint filled/.style={circle, fill=black, inner sep=1.8pt},
            endpoint open/.style={circle, draw=black, fill=white, line width=0.8pt, inner sep=1.8pt},
        ]
        \def\r{2.2}

        \coordinate (p0) at (0:\r);
        \coordinate (p30) at (30:\r);
        \coordinate (p100) at (100:\r);
        \coordinate (p180) at (180:\r);
        \coordinate (p210) at (210:\r);
        \coordinate (p280) at (280:\r);

        \draw[boundary] (0:\r) arc[start angle=0, end angle=30, radius=\r]
            node[midway, right=4pt, arc label] {$\ell_1$};
        \draw[boundary] (30:\r) arc[start angle=30, end angle=100, radius=\r]
            node[midway, above=4pt, arc label] {$\ell_2$};
        \draw[boundary] (100:\r) arc[start angle=100, end angle=180, radius=\r]
            node[midway, above left=2pt, arc label] {$\frac{L}{2}-\ell_1-\ell_2$};
        \draw[boundary] (180:\r) arc[start angle=180, end angle=210, radius=\r]
            node[midway, left=4pt, arc label] {$\ell_1$};
        \draw[boundary] (210:\r) arc[start angle=210, end angle=280, radius=\r]
            node[midway, below=4pt, arc label] {$\ell_2$};
        \draw[boundary] (280:\r) arc[start angle=280, end angle=360, radius=\r]
            node[midway, below right=2pt, arc label] {$\frac{L}{2}-\ell_1-\ell_2$};

        \coordinate (betaCenter) at ({-2.1734*\r},{1.0135*\r});
        \coordinate (alphaCenter) at ({0.6638*\r},{-2.4773*\r});

        \begin{scope}[shift={(betaCenter)}]
            \draw[cycle, red!80!black]
                (310.82:{2*\r}) arc[start angle=310.82, end angle=359.18, radius={2*\r}]
                node[midway, above left=1pt, arc label] {$\beta$};
        \end{scope}
        \begin{scope}[shift={(alphaCenter)}]
            \draw[cycle, blue!80!black]
                (127.73:{2.5*\r}) arc[start angle=127.73, end angle=82.27, radius={2.5*\r}]
                node[midway, below right=1pt, arc label] {$\alpha$};
        \end{scope}

        \node[endpoint filled] at (p30) {};
        \node[endpoint open] at (p100) {};
        \node[endpoint filled] at (p180) {};
        \node[endpoint open] at (p210) {};
        \node[endpoint filled] at (p280) {};
        \node[endpoint open] at (p0) {};
        \end{tikzpicture}
    \end{center}
    \caption{Symplectic basis of cycles on the torus.}
    \label{torus1cycles}
\end{figure}
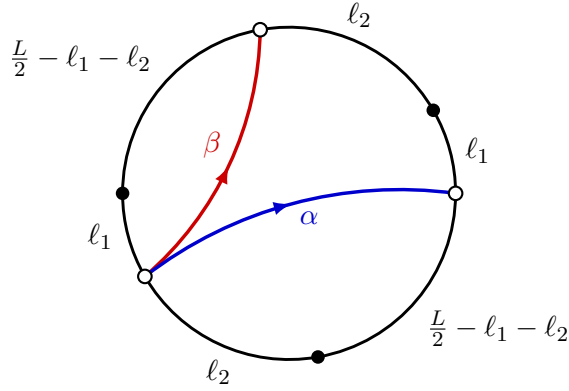

Once we have solved for the holomorphic 1-form, we can numerically compute the usual complex modulus $\tau$ of the torus by integrating it against a symplectic basis $\alpha, \beta$ of cycles
\ie
\tau = \frac{\int_\beta f(z)dz}{\int_\alpha f(z)dz}.
\fe
In the disk frame, both of these cycles are given by chords that connect the Strebel vertices, as illustrated in figure \ref{torus1cycles}. 

\subsection{Holomorphic data from ribbon graphs: Higher genus}

\label{holodata}

The algorithm above generalizes to the case $g\geq 2$. The genus $g$ Riemann surface can be reconstructed by gluing $12g-6$ boundary segments on the disk. Upon discretizing the unit circle into $L$ sites, the moduli space $\mathcal{M}_{g,1}$ can be parametrized by the $6g-4$ independent edge lengths.

Going counterclockwise from $z=1$, we label the boundary segments $S_i$ from $i=1$ to $i=12g-6$. We label the edges by $E_a$ with $a=1,\cdots, 6g-3$, where each edge $E_a$ pairs two segments $S_{i_1(a)}$ and $S_{i_2(a)}$ and glues them with opposite orientation. The number of discretized sites contained in $E_a$ is denoted by $\ell_a$. We always take $i_1(a)<i_2(a)$. Conversely, we denote the edge associated with segment $i$ by $a(i)$. 

The map $a(i)$ depends on the ribbon graph topology. Given a ribbon graph topology, the corresponding top dimensional cell in $\mathcal{M}_{g,1}^{\text{comb}}$ can be parametrized by the relative edge lengths $\tfrac{\ell_a}{L}$. In this language, the gluing map on the unit circle can be written as
\ie\label{gluinggenusg}
z(k)\sim z\left(\sum_{j\leq i_1(a)}\ell_{a(j)}+\sum_{j< i_2(a)}\ell_{a(j)}-k+1\right),\quad \sum_{j<i_1(a)}\ell_{a(j)}+1\leq k \leq \sum_{j\leq i_1(a)}\ell_{a(j)},
\fe
where the discretized points $z(k)$ are given by \eqref{zkposition}. The Strebel vertices are located at
\ie
z_i = \exp\left(\frac{2\pi i}{L}\sum_{j<i}\ell_{a(j)}\right).
\fe
One additional complication is that for higher genus the independent boundary points are not simply $z(k)$ with $1\leq k \leq L/2$, since the segments in the same half circle can be glued together. This is evident in figure \ref{genus3_Example}. We denote the set of independent variables by $\mathcal{K}=\cup_{a=1}^{6g-3}\{k : k\in S_{i_1(a)}\}$.

On a genus $g$ surface, there are $g$ linearly independent holomorphic 1-forms. We start with the following ansatz for each holomorphic 1-form
\ie \label{fzdz}
f(z)dz= \frac{\sum_{n=1}^\infty a_n z^{n-1}}{\prod_{i=1}^{12g-6}\left(1-z\exp\left({-\frac{2\pi i}{L}\sum_{j<i}\ell_{a(j)}}\right)\right)^{\frac{1}{3}}}dz,
\fe
and solve for the coefficients $a_1,\cdots, a_{L/2}$ by matching the values of $f(z)dz$ on the identified boundary points. This produces a system of linear equations for the coefficients $a_n$ of the form
\ie
&\frac{\sum_{n=1}^{\frac{L}{2}} a_n z(k)^{n}}{\prod_{i=1}^{12g-6}\left(1-z(k)\exp\left({-\frac{2\pi i}{L}\sum_{j<i}\ell_{a(j)}}\right)\right)^{\frac{1}{3}}}\\&\qquad \qquad \qquad = - \frac{\sum_{n=1}^{\frac{L}{2}} a_n z(k')^{n}}{\prod_{i=1}^{12g-6}\left(1-z(k')\exp\left({-\frac{2\pi i}{L}\sum_{j<i}\ell_{a(j)}}\right)\right)^{\frac{1}{3}}},\quad k \in \mathcal{K}.
\fe

This system of linear equations can be written in the form $Ba=0$, where $a=(a_1,\cdots, a_{L/2})\in \mathbb{C}^{L/2}$ is a vector in the coefficient space, and $B$ is a $\tfrac{L}{2}\times \tfrac{L}{2}$ matrix defined by
\ie
B_{k,n} &= \frac{z(k)^{n}}{\prod_{i=1}^{12g-6}\left(1-z(k)\exp\left({-\frac{2\pi i}{L}\sum_{j<i}\ell_{a(j)}}\right)\right)^{\frac{1}{3}}}\\
&\qquad \qquad \qquad + \frac{z(k')^{n}}{\prod_{i=1}^{12g-6}\left(1-z(k')\exp\left({-\frac{2\pi i}{L}\sum_{j<i}\ell_{a(j)}}\right)\right)^{\frac{1}{3}}}.
\fe
Since there are $g$ linearly independent holomorphic 1-forms, the matrix $B$ has a $g$ dimensional kernel. To obtain a numerically stable basis of holomorphic 1-forms, we first find the $g$ singular vectors $v_I$ of $B$ with the smallest singular values, and choose $g$ coefficients $a_{n_1},\cdots, a_{n_g}$ such that the matrix $H_{IJ}=(v_I)_{n_J}$ has the largest possible smallest singular value. We then normalize the holomorphic 1-form $\omega_J=f^{(J)}(z)dz$ by imposing $a^{(J)}_{n_I} = \delta_{IJ}$ for $I=1,\cdots,g$. The remaining coefficients $a^{(J)}_n$ are then obtained by a Hermitian block solve.

For each edge $E_a$, we define a chord connecting the midpoints of $S_{i_1(a)}$ and $S_{i_2(a)}$. One can then choose a symplectic basis $\{\alpha^I, \beta_I\}_{I=1}^g$ of cycles among these chords, with intersection numbers satisfying
\ie
\alpha^I\cdot \alpha^J = \beta_I\cdot \beta_J = 0,\quad \alpha^I\cdot \beta_J= \delta^{I}_J.
\fe
The period matrix $\Omega$ of the Riemann surface can then be computed as
\ie \label{periodmatrix}
\Omega=\mathcal{A}^{-1}\mathcal{B},\quad \mathcal{A}_{I}{}^J=\oint_{\alpha^J}\omega_I,\quad \mathcal{B}_{IJ}=\oint_{\beta_J} \omega_I.
\fe
Further geometric quantities on the Riemann surface, such as the Riemann constant vector and the prime form, can be computed using the holomorphic 1-forms and the period matrix. These quantities are used in the computation of $bc$ ghost correlators; the details are presented in section \ref{bcsection}.

\section{Free boson CFT from ribbon graphs}

\label{bosonsection}

\subsection{The free boson partition function}
\label{sec:freeBosonPartitionFunction}

The partition function of the free boson CFT on a Riemann surface of arbitrary genus can be computed in the ribbon graph formalism by performing the path integral on the reconstructed surface. For simplicity, we again consider the genus 1 case first. On a torus parametrized by the angle moduli $(\theta_1, \theta_2)$, the free boson partition function can be written as a path integral over the disk, with delta functions implementing the gluing of boundary segments:
\ie
Z_{\text{bos}}^{(1)}(\theta_1, \theta_2)&=\int [\mathcal{D}X] \exp(-S[X]) \prod_{0<\sigma<\theta_1} \delta(X(1,\sigma)-X(1,\pi+\theta_1-\sigma))
\\
&~~~\times \prod_{\theta_1<\sigma< \theta_2} \delta(X(1,\sigma)-X(1,\pi+\theta_1+\theta_2 - \sigma))  \prod_{\theta_2<\sigma< \pi} \delta(X(1,\sigma)-X(1,2\pi+\theta_2 - \sigma)),
\fe
where we have parametrized the boson field $X(r,\sigma)$ in polar coordinates $z=re^{i\sigma}$. Our convention for the free boson action $S[X]$ is such that
\ie
S[X] = \frac{1}{2\pi \alpha'} \int_\Sigma d^2z~ \partial X \bar{\partial} X.
\fe

We also know that the path integral inside the disk is given by the wavefunctional $\Psi_1[X(\sigma)]$ of the identity operator
\ie\label{psionefx}
\Psi_1[X(\sigma)] \propto \exp\left( - \frac{1}{\alpha'}\sum_{m\geq 1} m X_m X_{-m} \right),
\fe
where $X_m$ is the Fourier mode of $X(\sigma)$ indexed by $m$. Therefore,
\ie\label{zthetas}
Z_{\text{bos}}^{(1)}(\theta_1, \theta_2) &= \int [DX(\sigma)]_{0<\sigma<2\pi} \Psi_1[X(\sigma)] \prod_{0<\sigma<\theta_1} \delta(X(\sigma)-X(\pi+\theta_1-\sigma))
\\
&~~~\times \prod_{\theta_1<\sigma< \theta_2} \delta(X(\sigma)-X(\pi+\theta_1+\theta_2 - \sigma))  \prod_{\theta_2<\sigma< \pi} \delta(X(\sigma)-X(2\pi+\theta_2 - \sigma)).
\fe
Since the identity wavefunctional is Gaussian in $X$, the torus partition function $Z(\theta_1, \theta_2)$ can be computed by a Gaussian integral. This relation generalizes straightforwardly to higher genus and to operators other than the identity, and the final integral remains Gaussian. 

To compute the partition function numerically, we again discretize the unit circle into $L$ sites, with a variable $X(k)$ on each site $k=1,\cdots,L$. The wavefunctional \eqref{psionefx} can be approximated by
\ie \label{discrPsi1}
\Psi_1[X]&\propto\exp\left[- \frac{1}{\pi \alpha' L}\left(\sum_{m=1}^{L/2-1}\sin \frac{\pi m}{L}\sum_{k,\ell=1}^LX(k)X(\ell)e^{2\pi i \frac{m}{L}(k-\ell)}\right.\right.\\[-2mm]
&\hspace{5.0cm}\left.\left.+ \frac{1}{2}\sum_{k,\ell=1}^LX(k)X(\ell)(-1)^{k-\ell}\right)\right]\\
&=\exp\left[- \frac{1}{2\pi\alpha' L}\sum_{k,r=1}^LX(k)X(k+r)\frac{\sin \frac{\pi}{L}}{\cos \frac{2\pi r}{L}-\cos \frac{\pi}{L}}\right].
\fe
Note that we have summed only over the modes with $m$ from 1 to $L/2$. This is because there are only $L$ independent discrete Fourier modes on a circle with $L$ sites. The delta functions on the boundary can be imposed by making the identifications
\ie\label{pairthree}
& X(\tfrac{L}{2} + \ell_1+1 - k) = X(k),~~~~ 1\leq k \leq \ell_1,
\\
& X(\tfrac{L}{2} + 2 \ell_1 + \ell_2 +1 - k) = X(k),~~~~ \ell_1+1 \leq k \leq \ell_1 + \ell_2,
\\
& X(L+\ell_1 + \ell_2+1 - k) = X(k),~~~~ \ell_1+\ell_2+1 \leq k \leq \tfrac{L}{2}.
\fe
The partition function \eqref{zthetas} is then discretized into
\ie
Z_{\text{bos}}^{(1)}(L;\ell_1, \ell_2) = \int \prod_{k=1}^{L} dX(k)\prod_{k=1}^{L/2} \delta(X(k)-X(k'))\,\Psi_1[X].
\fe

After replacing each $X(k)$ with $\tfrac{L}{2}+1\leq k\leq L$ by the corresponding identified variable in \eqref{discrPsi1}, the wavefunctional $\Psi_1$ can be written as
\ie
\Psi_1[X]\propto \exp\left[ - \frac{1}{\pi}\sum_{k,\ell=1}^{L/2} \mathbf{A}_{k\ell}(L;\ell_1, \ell_2) X(k) X(\ell) \right],
\fe
where the matrix $\mathbf{A}$ is given by
\ie
\label{Amatrix}
\mathbf{A}_{k\ell}(L;\ell_1, \ell_2) = \frac{1}{2 \alpha' L}&\left[\frac{\sin \frac{\pi}{L}}{\cos \frac{2\pi (k-\ell)}{L}-\cos \frac{\pi}{L}}+\frac{\sin \frac{\pi}{L}}{\cos \frac{2\pi (k-\ell')}{L}-\cos \frac{\pi}{L}}\right.\\
&\qquad\left.+\frac{\sin \frac{\pi}{L}}{\cos \frac{2\pi (k'-\ell)}{L}-\cos \frac{\pi}{L}}+\frac{\sin \frac{\pi}{L}}{\cos \frac{2\pi (k'-\ell')}{L}-\cos \frac{\pi}{L}}\right],
\fe
where $k'$ is the partner site of $k$ determined by the identifications \eqref{pairthree}.

The eigenvalues of the matrix $\mathbf{A}_{k\ell}$ are all positive except for a single zero. This zero mode corresponds to the constant mode of $X(\sigma)$ on the unit circle. The integration over $X(k)$, $1\leq k\leq \frac{L}{2}$, then produces, up to a moduli independent constant normalization factor, the determinant
\ie\label{detprime}
Z_{\text{bos}}^{(1)}(L;\ell_1, \ell_2) \propto \frac{1}{\sqrt{\det \mathbf{A}'(L;\ell_1, \ell_2)}},
\fe
where the prime on $\mathbf{A}$ denotes the removal of the zero mode $\sum_k X(k)$. This can be done by imposing the condition
\ie X(\tfrac{L}{2})=-\sum_{k=1}^{\tfrac{L}{2}-1}X(k),
\fe 
and the $(\tfrac{L}{2}-1)\times (\tfrac{L}{2}-1)$ matrix $\mathbf{A}'$ in the subspace orthogonal to the zero mode is constructed as
\ie\label{Aprime}
\mathbf{A}_{k\ell}'= \mathbf{A}_{k\ell}-\mathbf{A}_{k,\frac{L}{2}}-\mathbf{A}_{\frac{L}{2},\ell}+\mathbf{A}_{\frac{L}{2}, \frac{L}{2}},\quad k,\ell = 1,\cdots, \tfrac{L}{2}-1.
\fe

This relation can be easily generalized to higher genus. For a genus $g$ surface, the discretized expression for the partition function reads
\ie \label{detprimeg}
Z_{\text{bos}}^{(g)}(L;\ell_a) &=\int \prod_{k=1}^{L} dX(k)\prod_{k\in \mathcal{K}} \delta(X(k)-X(k'))\,\Psi_1[X]\propto \frac{1}{\sqrt{\det \mathbf{A}'(L;\ell_a)}},
\fe
where the matrix $\mathbf{A}(L;\ell_a)$ is defined by \eqref{Amatrix}, with the identification map replaced by \eqref{gluinggenusg}. The matrix $\mathbf{A}'(L;\ell_a)$ is obtained from $\mathbf{A}(L;\ell_a)$ by imposing the condition $\sum_{k\in\mathcal{K}}X(k)=0$, as in \eqref{Aprime}.

We conclude this section with a few remarks. First, the algorithm outlined above can be easily generalized to one point functions, where one simply replaces the wavefunctional $\Psi_1[X]$ of the identity operator with the wavefunctional $\Psi_\mathcal{O}[X]$ of the corresponding operator. The discretized expression for the one point function $\langle \mathcal{O}\rangle$ on the Riemann surface reads
\ie \label{1ptfuncgeneral}
\langle \mathcal{O}(0)\rangle_{\Sigma(\ell_a)}= \frac{1}{Z_{\text{bos}}^{(g)}(L;\ell_a)}\int \prod_{k=1}^{L} dX(k)\prod_{k\in \mathcal{K}} \delta(X(k)-X(k'))\,\Psi_\mathcal{O}[X].
\fe
In the ribbon graph construction, the operator is always inserted at the origin of the disc coordinate system, and we denote $\Psi_{\mathcal{O}}[X]$ as the wavefunctional of the operator $\mathcal{O}$ inserted at the origin of the disc coordinate system. When $\mathcal{O}$ has nonzero conformal dimension, the correlation function will depend on the local coordinate system where $\mathcal{O}$ is inserted.

Second, the quantity $Z_{\text{bos}}^{(g)}(L;\ell_a)$ computed in this way is subject to a Weyl anomaly, and the corresponding Weyl anomaly factor generally depends on the moduli $\frac{\ell_a}{L}$, or equivalently for genera one and two, on the period matrix $\Omega$. In particular, the genus 1 partition function $Z_\text{bos}^{(1)}(L;\ell_1, \ell_2)$ and the analytic expression obtained in the flat torus frame are related by a nontrivial Weyl transformation between the disk and flat torus frames. However, the ratio of the partition functions of two different theories with the same central charge at the same moduli, or compute the one point function $\langle \mathcal{O}\rangle_{\Sigma(\ell_a)}$ of a nontrivial operator, and such observables are free of the Weyl anomaly factor.

Third, the quantity $Z^{(1)}_{\text{bos}}(L;\ell_1, \ell_2)$ is also divergent as a function of the total length $L$ in the continuum limit $L\rightarrow \infty$. There are two sources of this divergence. One comes from the discretization scheme and the normalization of the Gaussian measure, and scales like $\exp(\text{const}\cdot L)$. This piece is moduli independent. The other comes from the divergence of the Weyl anomaly factor, which scales like a constant power of $L$. This divergent behavior can be estimated directly as follows. 

Around a cubic vertex of the ribbon graph, the metric \eqref{flatcylmetric} has the behavior $ds^2 =e^{2\omega} dz d\bar z \sim |z| dz d\bar z$, and so the Weyl transformation $\omega$ that relates the disk metric to the locally flat metric behaves like $\omega \sim \frac{1}{2} \log|z|$. Near the puncture, on the other hand, we have $\omega \sim - \log|z|$. For a theory with central charge $c=1$, the Weyl anomaly factor in going from the locally flat metric to the ribbon graph metric on a genus $g$ Riemann surface is given by $\exp({S_\text{Weyl}})$ with
\ie\label{sweyl}
S_\text{Weyl} = \frac{1}{12\pi} \int_{\Sigma} d^2z \,\partial\omega \bar\partial\omega \sim \frac{2g-1}{24}\log \frac{1}{\delta} +\frac{1}{12} \log \frac{1}{\epsilon},
\fe
where the asymptotic relation indicates only the divergent part. To regularize the divergence, we impose the cutoffs $|z-z_i|>\delta$ around the $(4g-2)$ Strebel vertices and $|z|>\epsilon$ around the puncture. From this, one can see that the coefficient of the logarithmic divergence is independent of the moduli $\Omega$.

If we want to have a finite answer when taking the $L\rightarrow \infty$ limit, we need to further renormalize the action by adding counterterms. We define the quantity $Z^{(g)}_{\text{bos}}(L;\ell_a)$ to be
\ie
Z_{\text{bos}}^{(g)}(L;\ell_a) = \frac{(2\pi)^{-g}}{\sqrt{\det \mathbf{A}'(L;\ell_a)}}.
\fe
The large $L$ behavior of this partition function takes the form
\ie\label{largeL}
\log Z^{(g)}_{\text{bos}}(L;\ell_a) \sim c(\Omega) + \gamma L + \alpha(g) \log L,
\fe
where $c(\Omega)$ is the moduli dependent finite part. One can compute $c(\Omega)$, $\gamma$, and $\alpha(g)$ by fitting the large $L$ behavior of $Z^{(g)}_{\text{bos}}(L;\ell_a)$ with the relative lengths $\tfrac{\ell_a}{L}$ held fixed. As argued above, $\gamma$ and $\alpha$ are moduli independent coefficients. Moreover, as the behavior \eqref{sweyl} suggests, the coefficient $\alpha(g)$ depends linearly on the genus $g$, and one can write
\ie \alpha(g)=\alpha_0+\alpha_1 g.\fe
In order to arrive at a finite partition function in the $L\rightarrow \infty$ limit, we choose the renormalization scheme by adding the following counterterms to the free boson action:
\ie
S_\text{CT} = \gamma L + (\alpha_0+\alpha_1)\log L- \frac{\alpha_1}{8\pi}\log L\int_\Sigma \mathrm{d}^2\sigma \sqrt{g}\, R(g).
\fe
These counterterms will cancel the divergences in \eqref{largeL}, and the renormalized partition function in this scheme is simply given by
\ie \label{renZ}
Z_\text{ren}^{(g)}(\Omega) = \exp c(\Omega).
\fe
Of course, the renormalization scheme here is not unique, and one can add an arbitrary finite part to the counterterms to define a different renormalization scheme.

\subsection{The Weyl anomaly at genus 1}
\label{sec:WeylAnomalyGenus1}

At genus 1, one can relate the renormalized partition function obtained in the disk frame to the partition function in the flat torus frame by explicitly computing the finite moduli dependent part of the Weyl anomaly.

Let $u\in\mathbb{C}/(\mathbb{Z}+\tau\mathbb{Z})$ be the flat torus coordinate. It can be constructed from the holomorphic 1-form $\omega = f(z)dz$ by
\ie \label{ufromz}
u= \frac{1}{\mathcal{A}}\int^z_{z_0} f(z)dz,
\fe
where $z_0$ is a reference point. The denominator $\mathcal{A}$ is the $\alpha$ cycle period
\ie
\mathcal{A}=\oint_\alpha f(z)dz,
\fe
and the normalization by $\mathcal{A}$ is chosen so that the period of $u$ around the $\alpha$ cycle is equal to 1. In the flat torus frame, the metric on the Riemann surface $\Sigma$ is given by
\ie
ds^2 = du d\bar{u},
\fe
while the metric in the disk frame is given by \eqref{diskmetric} as
\ie
ds^2=\frac{dzd\bar{z}}{|z|^2} = \frac{|\mathcal{A}|^2}{|z f(z)|^2}dud\bar{u}.
\fe
Hence the Weyl transformation $g_{\mu\nu}\rightarrow e^{2\omega} g_{\mu\nu}$ from the disk frame to the flat torus frame is
\ie
\omega(z, \bar{z}) = -\log |z f(z)| + \log |\mathcal{A}|.
\fe

To evaluate the Weyl anomaly, we need the local behavior of $f(z)$ near the Strebel vertices. From \eqref{fzdz}, for each Strebel vertex $z_i$ there exists a number $\nu_i$ such that $f(z)$ behaves like
\ie f(z)\sim \nu_i \left(1-\frac{z}{z_i}\right)^{-\frac{1}{3}}
\fe 
near $z_i$. The coefficients $\nu_i$ can be computed directly from the coefficients $a_n$ in \eqref{fzdz}, and they depend on the moduli $\tau,\, \bar{\tau}$ (or equivalently, $\tfrac{\ell_a}{L}$).

The Weyl anomaly factor \eqref{sweyl} is divergent, so we regulate it by imposing the cutoffs $|z-z_i|>\delta$ near the Strebel vertices and $|z|>\epsilon$ near the puncture. After imposing these cutoffs, one finds
\ie \label{genus1Weyl}
S_\text{Weyl}(\tau, \bar{\tau}) = \frac{\log |\nu_1|+\log|\nu_2|}{24}+\frac{1}{24}\log \frac{1}{\delta} +\frac{1}{12} \log \frac{1}{\epsilon}.
\fe
The first term is the finite moduli dependent part, while the last two terms are cutoff dependent divergences. These divergent terms are removed in the renormalized free boson partition function \eqref{renZ}. Therefore, the torus partition function in the flat frame is given by
\ie
Z_\text{flat}^{(1)}(\tau, \bar{\tau}) = \exp\left[-S_\text{Weyl}(\tau, \bar{\tau})\right]Z_\text{ren}^{(1)}(\tau, \bar{\tau})
\fe
in a definite renormalization scheme. It may differ from the standard result
\ie
\label{eq:Zflat}
Z_\text{flat}^{(1)}(\tau, \bar{\tau})=\tau_2^{-\frac12} |\eta(\tau)|^{-2}
\fe
obtained from canonical quantization by a moduli independent multiplicative constant. We will test this relation numerically, and find agreement, in section \ref{numericssection}.

\subsection{The compact boson partition function}

The algorithm above generalizes to the compact boson CFT. Consider the genus $g$ partition function of a compact boson $X$ satisfying $X\sim X+2\pi R$. On the unit circle, a configuration of $X$ can be parametrized by the winding number $p\in\mathbb{Z}$ and the Fourier modes $X_m$:
\ie
X_b(\sigma)=p \sigma R+\sum_{m=-\infty}^\infty X_m e^{im\sigma}.
\fe
For any configuration on the disk with nonzero winding number on the boundary, the classical action is logarithmically divergent. Consequently, the wavefunctional $\Psi_1[X_b]=\int [\mathcal{D}X]_{X_b}e^{-S[X]}$ vanishes. Therefore, $\Psi_1[X_b]$ is identical to the wavefunctional of a noncompact boson in the zero winding sector and vanishes in sectors with nontrivial winding:
\ie
\Psi_1[p,X_m]\propto \delta_{p,0}\exp\left[-\frac{1}{\alpha'}\sum_{m\geq 1}mX_m X_{-m}\right].
\fe
This wavefunctional can then be discretized as in \eqref{discrPsi1}.

\begin{figure}
    \begin{center}
        \begin{tikzpicture}[
            >=Latex,
            line cap=round,
            boundary/.style={line width=1.1pt},
            boundary arrow/.style={
                boundary,
                postaction={
                    decorate,
                    decoration={
                        markings,
                        mark=at position 0.5 with {
                            \draw[line width=1.1pt, -{Latex[length=2.2mm]}] (-2.8pt,0) -- (2.8pt,0);
                        }
                    }
                }
            },
            internal cycle/.style={
                line width=0.8pt,
                blue!80!black,
                postaction={
                    decorate,
                    decoration={
                        markings,
                        mark=at position 0.5 with {
                            \draw[blue!80!black, line width=0.8pt, -{Latex[length=1.9mm]}] (-2.2pt,0) -- (2.2pt,0);
                        }
                    }
                }
            },
            arc label/.style={font=\small, fill=white, inner sep=1.3pt},
            vertex label/.style={font=\scriptsize, inner sep=1pt},
            endpoint filled/.style={circle, fill=black, inner sep=1.8pt},
            endpoint open/.style={circle, draw=black, fill=white, line width=0.8pt, inner sep=1.8pt},
        ]
        \def\r{2.2}
        \def\smallr{0.25*\r}

        \coordinate (p0) at (0:\r);
        \coordinate (p30) at (30:\r);
        \coordinate (p100) at (100:\r);
        \coordinate (p180) at (180:\r);
        \coordinate (p210) at (210:\r);
        \coordinate (p280) at (280:\r);

        \begin{scope}
            \clip (0,0) circle[radius=\r];
            \begin{scope}[shift={(p100)}]
                \draw[internal cycle] (362.82:\smallr) arc[start angle=362.82, end angle=197.18, radius=\smallr];
            \end{scope}
            \begin{scope}[shift={(p210)}]
                \draw[internal cycle] (472.82:\smallr) arc[start angle=472.82, end angle=307.18, radius=\smallr];
            \end{scope}
            \begin{scope}[shift={(p0)}]
                \draw[internal cycle] (262.82:\smallr) arc[start angle=262.82, end angle=97.18, radius=\smallr];
            \end{scope}
        \end{scope}

        \draw[boundary arrow] (0:\r) arc[start angle=0, end angle=30, radius=\r]
            node[midway, right=4pt, arc label] {$E_1$};
        \draw[boundary arrow] (30:\r) arc[start angle=30, end angle=100, radius=\r]
            node[midway, above=4pt, arc label] {$E_2$};
        \draw[boundary arrow] (100:\r) arc[start angle=100, end angle=180, radius=\r]
            node[midway, above left=2pt, arc label] {$E_3$};
        \draw[boundary arrow] (210:\r) arc[start angle=210, end angle=180, radius=\r]
            node[midway, left=4pt, arc label] {$E_1$};
        \draw[boundary arrow] (280:\r) arc[start angle=280, end angle=210, radius=\r]
            node[midway, below=4pt, arc label] {$E_2$};
        \draw[boundary arrow] (360:\r) arc[start angle=360, end angle=280, radius=\r]
            node[midway, below right=2pt, arc label] {$E_3$};

        \node[endpoint filled] at (p30) {};
        \node[endpoint open] at (p100) {};
        \node[endpoint filled] at (p180) {};
        \node[endpoint open] at (p210) {};
        \node[endpoint filled] at (p280) {};
        \node[endpoint open] at (p0) {};

        \node[vertex label] at (30:{\r+0.32}) {$V_2$};
        \node[vertex label] at (100:{\r+0.32}) {$V_1$};
        \node[vertex label] at (180:{\r+0.32}) {$V_2$};
        \node[vertex label] at (210:{\r+0.32}) {$V_1$};
        \node[vertex label] at (280:{\r+0.32}) {$V_2$};
        \node[vertex label] at (0:{\r+0.32}) {$V_1$};
        \end{tikzpicture}
    \end{center}
    \caption{The no winding condition $s_1-s_2+s_3=0$ around the Strebel vertex $V_1$ for a genus 1 Riemann surface. The contractible cycle around $V_1$ is shown in blue.}
    \label{nowinding}
\end{figure}
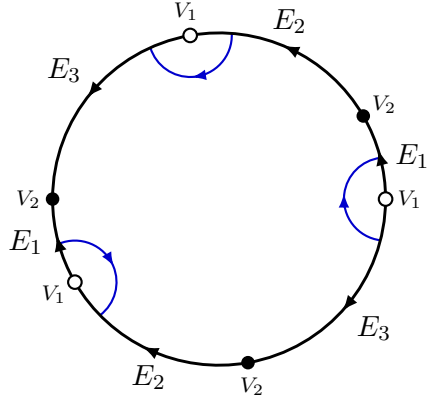

For the compact boson, the gluing map of boundary fields $X(k)$ is modified to allow an integer shift $s_a \in \mathbb{Z}$ for each edge $E_a$. The most general gluing map reads
\ie
~X\left(\sum_{j\leq i_1(a)}\ell_{a(j)}+\sum_{j<i_2(a)}\ell_{a(j)}-k+1\right)=X(k)+2\pi R s_a, \\
\hspace{1.2cm}\sum_{j<i_1(a)}\ell_{a(j)}+1\leq k\leq\sum_{j\leq i_1(a)}\ell_{a(j)},
\fe
which includes $6g-3$ integer shifts. On a genus $g$ surface, there are only $2g$ independent cycles, which means that there are $4g-3$ linear relations among the $6g-3$ shifts. These relations come from the fact that a field configuration with finite action should have no winding around each contractible cycle. Specifically, the three shifts associated with going around each trivalent Strebel vertex should add up to zero, as illustrated in figure \ref{nowinding}.

The discretized compact boson partition function is then obtained by summing over all physical shifts, the sets of ${s_a}$ that obey the linear relations in equation \ref{nowindingeq}:
\ie \label{compactPI}
Z_{\text{comp}}^{(g)}(R;L;\ell_a)= \sum_{\{s_a\}} \int \prod_{k=1}^{L} dX(k)\prod_{a=1}^{6g-3} \prod_{k\in S_{i_1(a)}} \delta\left(X(k')-X(k)-2\pi R s_a\right)\,\Psi_1[X],
\fe
We then directly compute this Gaussian path integral. The details of our algorithm used to determine the physical shifts $\{s_a\}$, together with the details of this Gaussian integral, are included in Appendix \ref{compactappendix}.

\section{String integration measure from ghosts}

\label{bcsection}

In order to compute string amplitudes in our formalism, we also need correlators of the $bc$ system. More specifically, we need the correlator $\langle \prod_{a=1}^{6g-4}\mathcal{B}_{\ell_a}\rangle$ that enters the integrand of string amplitudes, where $\mathcal{B}_{\ell_a}$ is the ghost-valued 1-form insertion corresponding to the ribbon graph modulus $\ell_a$. In this section, we first pin down the form of the $b$ ghost insertion $\mathcal{B}_{\ell_a}$, and then present the algorithm we use for general $bc$ correlators.

In the disk frame of section \ref{sec:Kontsevich}, the $b$ ghost insertion $\mathcal{B}_{\ell_a}$ related to the modulus $\ell_a$ can be determined by the following. Generally, one divides up a Riemann surface into domains $D_\mathtt{i}$, and associates each domain $D_\mathtt{j}$ with a local coordinate chart $z_\mathtt{j}$. The complex structure of the Riemann surface is then stored in the biholomorphic transition functions $z_{\mathtt{i}} = f_{\mathtt{ij}}(z_\mathtt{j};t)$. The $b$ ghost insertion for a general modulus $t^k$ can then be written as
\ie \label{Binsert}
\mathcal{B}_{t^k}&= \sum_{(\mathtt{ij})}\int_{C_{\mathtt{ij}}}\left(\frac{dz_\mathtt{i}}{2\pi i}b_{z_\mathtt{i} z_\mathtt{i}}\frac{\partial z_\mathtt{i}}{\partial t^k}\bigg|_{z_\mathtt{j}}-\frac{d\bar{z}_{\mathtt{i}}}{2\pi i}b_{\bar{z}_\mathtt{i} \bar{z}_\mathtt{i}}\frac{\partial \bar{z}_\mathtt{i}}{\partial t^k}\bigg|_{z_\mathtt{j}}\right),
\fe
where the sum goes over each unordered pair $(\mathtt{ij})$ of overlapping domains $D_\mathtt{i}, D_\mathtt{j}$, and the contour segment $C_{\mathtt{ij}}$ is the segment of $\partial D_\mathtt{j}$ on which $D_\mathtt{i}, D_\mathtt{j}$ overlap. %review: is there a reason \mathtt is used here, just to differentiate from the labels of the edge segments?

% A compact expression for the string amplitude is then
% \ie
% \bigg\langle e^{\mathcal{B}}\prod_{i=1}^n \mathcal{V}_i\bigg\rangle_{\Sigma},
% \fe
% where 
% \begin{align}
% \mathcal{B}&\equiv \sum_k dt^k \mathcal{B}_{t^k},
% \end{align}
% and $\mathcal{V}_i$ is a BRST closed operator, which we will always take to be the old covariant quantization representative $c\tilde{c} V_i$, with $V_i$ is a matter CFT primary with $h=\tilde{h}=1$.

\begin{figure}
    \begin{center}
        \begin{tikzpicture}[
            >=Latex,
            line cap=round,
            boundary/.style={line width=1.1pt},
            chart boundary/.style={line width=0.9pt, dashed, blue!65!black},
            divider/.style={line width=1.1pt},
            segment label/.style={font=\small, fill=white, inner sep=1.3pt},
            chart label/.style={font=\scriptsize, fill=white, inner sep=1pt},
            chart arrow/.style={-{Latex[length=1.7mm]}, line width=0.75pt},
            endpoint filled/.style={circle, fill=black, inner sep=1.8pt},
            endpoint open/.style={circle, draw=black, fill=white, line width=0.8pt, inner sep=1.8pt},
        ]
        \def\r{2.2}
        \def\rin{1.72}
        \def\rmid{1.96}
        \def\rlabel{1.08}

        % The six annular coordinate domains.
        \path[fill=red!14]
            (0:\r) arc[start angle=0, end angle=30, radius=\r]
            -- (30:\rin) arc[start angle=30, end angle=0, radius=\rin]
            -- cycle;
        \path[fill=orange!18]
            (30:\r) arc[start angle=30, end angle=100, radius=\r]
            -- (100:\rin) arc[start angle=100, end angle=30, radius=\rin]
            -- cycle;
        \path[fill=green!14]
            (100:\r) arc[start angle=100, end angle=180, radius=\r]
            -- (180:\rin) arc[start angle=180, end angle=100, radius=\rin]
            -- cycle;
        \path[fill=cyan!16]
            (180:\r) arc[start angle=180, end angle=210, radius=\r]
            -- (210:\rin) arc[start angle=210, end angle=180, radius=\rin]
            -- cycle;
        \path[fill=blue!13]
            (210:\r) arc[start angle=210, end angle=280, radius=\r]
            -- (280:\rin) arc[start angle=280, end angle=210, radius=\rin]
            -- cycle;
        \path[fill=magenta!13]
            (280:\r) arc[start angle=280, end angle=360, radius=\r]
            -- (360:\rin) arc[start angle=360, end angle=280, radius=\rin]
            -- cycle;

        % Boundaries of D_0 and of the six boundary charts.
        \draw[chart boundary] (0:\rin) arc[start angle=0, end angle=360, radius=\rin];
        \draw[boundary] (0:\r) arc[start angle=0, end angle=360, radius=\r];
        \foreach \ang in {0,30,100,180,210,280}
            \draw[divider] (\ang:\rin) -- (\ang:\r);

        % Boundary segments, ordered counterclockwise as in the genus 1 example.
        \node[segment label] at (15:{\r+0.6}) {$S_{i_1(1)}$};
        \node[segment label] at (65:{\r+0.6}) {$S_{i_1(2)}$};
        \node[segment label] at (140:{\r+0.6}) {$S_{i_1(3)}$};
        \node[segment label] at (195:{\r+0.6}) {$S_{i_2(1)}$};
        \node[segment label] at (245:{\r+0.6}) {$S_{i_2(2)}$};
        \node[segment label] at (320:{\r+0.6}) {$S_{i_2(3)}$};

        % Labels remain inside D_0; the arrows identify the corresponding annular charts.
        \draw[chart arrow, red!75!black] (15:1.40) -- (15:\rmid);
        \draw[chart arrow, orange!85!black] (65:1.40) -- (65:\rmid);
        \draw[chart arrow, green!60!black] (140:1.40) -- (140:\rmid);
        \draw[chart arrow, cyan!65!black] (195:1.40) -- (195:\rmid);
        \draw[chart arrow, blue!75!black] (245:1.40) -- (245:\rmid);
        \draw[chart arrow, magenta!70!black] (320:1.40) -- (320:\rmid);

        \node[chart label, text=red!75!black] at (15:\rlabel) {$D_{i_1(1)}$};
        \node[chart label, text=orange!85!black] at (65:\rlabel) {$D_{i_1(2)}$};
        \node[chart label, text=green!60!black] at (140:\rlabel) {$D_{i_1(3)}$};
        \node[chart label, text=cyan!65!black] at (195:\rlabel) {$D_{i_2(1)}$};
        \node[chart label, text=blue!75!black] at (245:\rlabel) {$D_{i_2(2)}$};
        \node[chart label, text=magenta!70!black] at (320:\rlabel) {$D_{i_2(3)}$};

        \node[font=\large] at (0,0) {$D_0$};

        % Alternating endpoint markers follow the convention of figure 5.
        \node[endpoint open] at (0:\r) {};
        \node[endpoint filled] at (30:\r) {};
        \node[endpoint open] at (100:\r) {};
        \node[endpoint filled] at (180:\r) {};
        \node[endpoint open] at (210:\r) {};
        \node[endpoint filled] at (280:\r) {};
        \end{tikzpicture}
    \end{center}
    \caption{Coordinate charts $\{D_0, D_{i_1(a)}, D_{i_2(a)}\}$ used in section \ref{bcsection} on a genus 1 surface.}
    \label{charts}
\end{figure}
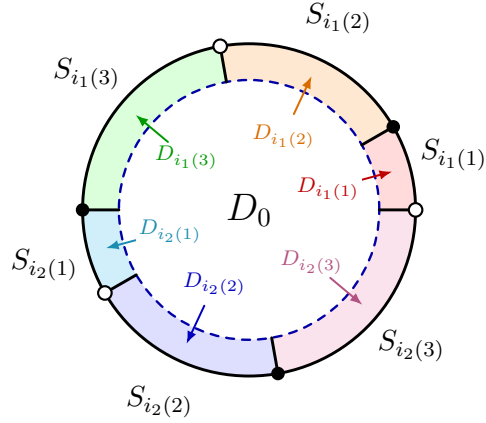

In the disc frame, we choose the set of domains $D_\mathtt{i}$ to be $\{D_0, D_i\}$, where there is a $D_i$ for each segment $S_i$. These domains are defined by
\ie
D_0&:\quad \{|z|\leq 1-\varepsilon\},\\
D_{i}&:\quad \left\{1-\varepsilon \leq |z|\leq 1,\quad \frac{2\pi}{L}\sum_{j< i}\ell_{a(j)}\leq \operatorname{arg} z \leq \frac{2\pi}{L}\sum_{j\leq i}\ell_{a(j)}\right\},
\fe
with $0<\varepsilon<1$. This configuration of charts is depicted in figure \ref{charts}. The local coordinates on each domain are defined simply by restricting the disk coordinate $z$ to them.

With this configuration of charts, the only overlapping domains with nontrivial transition functions are the pairs $(D_{i_1(a)}, D_{i_2(a)})$ for each edge $E_a$, with the transition function being 
\ie
z_{i_1(a)} = \frac{\exp\left(\frac{2\pi i}{L}\left(\sum_{j\leq i_1(a)} \ell_{a(j)}+\sum_{j< i_2(a)} \ell_{a(j)}\right)\right)}{z_{i_2(a)}}.
\fe
Among the $6g-3$ edge lengths, only $6g-4$ of them are independent, and one can take the local coordinates on the moduli space $\mathcal{M}_{g,1}$ to be $\{\ell_1,...,\ell_{6g-4}\}$. Defining
\ie
L_a(\ell_1,\cdots \ell_{6g-4}) &\equiv \left.\left(\sum_{j\leq i_1(a)}\ell_{a(j)}+ \sum_{j< i_2(a)}\ell_{a(j)}\right)\right|_{\ell_{6g-3}=\frac{L}{2}-\sum_{b=1}^{6g-4}\ell_{b}}
\fe
and inserting the transition function into \eqref{Binsert}, the final form of the $b$ ghost insertion in the disk frame is
\ie \label{Bla}
\mathcal{B}_{\ell_a} = \frac{1}{L}\sum_{\text{edges }b}\frac{\partial L_b}{\partial \ell_a}\int_{S_{i_1(b)}} dz\, z b(z) + \text{ c.c. }.
\fe

Correlators of the fermionic $(b_\lambda, c_{1-\lambda})$ system on higher genus Riemann surfaces can be computed from the holomorphic data obtained in section \ref{holodata} using the formula\cite{Verlinde:1986kw}
\ie \label{vvformula}
\left\langle\prod_{i=1}^n b_\lambda\left(z_i\right) \prod_{j=1}^m c_{1-\lambda}\left(w_j\right)\right\rangle_{\Sigma}=(2\pi)^{16(g-1)}  & Z_\text{chiral}^{(g)}\ \theta\left(\sum_i \zeta\left(z_i\right)-\sum_j \zeta\left(w_j\right)-(2 \lambda-1) \Delta \mid \Omega\right) \\
& \times \frac{\prod_{i<i^{\prime}} E\left(z_i, z_{i^{\prime}}\right) \prod_{j<j^{\prime}} E\left(w_j, w_{j^{\prime}}\right) \prod_i\left(\sigma\left(z_i\right)\right)^{2 \lambda-1}}{\prod_{i, j} E\left(z_i, w_j\right) \prod_j\left(\sigma\left(w_j\right)\right)^{2 \lambda-1}},
\fe 
where $n$ and $m$ must saturate the ghost number anomaly 
\ie \label{selection}
n-m=(2\lambda-1)(g-1).
\fe  
The factor of $(2\pi)^{16(g-1)}$ is necessary for the genus $g$ string integrand to factorize into lower genus integrands in degeneration limits.\footnote{We use the conventions for the normalization of the genus zero and genus one $bc$ ghost correlation functions specified in \cite{StringNotes}. The overall phase depends on a convention for the ordering of the holomorphic top-form, which we do not fix. } In this formula, $Z_{\text{chiral}}^{(g)}$ is the chiral boson partition function, $\zeta(z)$ is the Abel-Jacobi map, $\theta(y|\Omega)$ is the Riemann theta function, $E(z,w)$ is the prime form, and $\Delta$ is the Riemann constant vector. $\sigma(z)$ is a function that can be determined using the $\lambda = 1$, $(n,m)=(g,1)$ case of the formula \eqref{vvformula}. Their definitions and the algorithms used to compute these quantities from the holomorphic 1-forms $\omega_I$ and the period matrix $\Omega$ are summarized in Appendix \ref{holodataappendix}. Using this formula and the $b$ ghost insertion \eqref{Bla}, the string integration measure can be computed in terms of ribbon graph parameters $\{\ell_a\}$.

\section{Comparison with analytic results}

\label{numericssection}

In this section, we test the algorithms above by comparing the numerical evaluations of various observables to known analytic expressions.
 
At genus 1, since we have the exact Weyl anomaly factor \eqref{genus1Weyl}, we compare the matter partition function in the Strebel frame against the partition function in the flat torus frame. For genus 2 or higher, due to the lack of a canonical Weyl frame, we compare only quantities that are free of the Weyl anomaly: ratios of partition functions, one point functions of nontrivial operators, and correlators of the combined matter and ghost CFT. Throughout this section, we adopt the convention that $\alpha'=1$.

\subsection{Weyl anomaly at genus 1}
At genus 1, the moduli dependent component of the Weyl anomaly of the free boson partition function can be computed numerically using the prescription in Section \ref{sec:WeylAnomalyGenus1}. In particular, the ratio of free boson partition functions at different $\{\ell_a\}$ but fixed total length $L$ can be compared to the difference in Weyl anomaly as:
\ie
\label{eq:weylAnomalyComparison}
S_{\text{Weyl}}(\ell_a)-S_{\text{Weyl}}(\ell_b)=&\log Z_{\text{ren}}^{(1)}(L;\ell_a)-\log Z_{\text{ren}}^{(1)}(L;\ell_b)\\
&-\log Z_{\text{flat}}^{(1)}(L,\tau(\ell_a),\bar{\tau}(\ell_a))+\log Z_{\text{flat}}^{(1)}(L,\tau(\ell_b),\bar{\tau}(\ell_b)),
\fe
where $S_{\text{Weyl}}$ is computed using equation \eqref{genus1Weyl}, the partition function in the disc scheme $Z_{\text{ren}}^{(1)}$ is given by equation \eqref{renZ}, and the standard $\tau$ frame $Z_{\text{flat}}^{(1)}$ is given by equation \eqref{eq:Zflat}.

\begin{figure}
    \begin{center}
	 \includegraphics[width=1.0\textwidth]{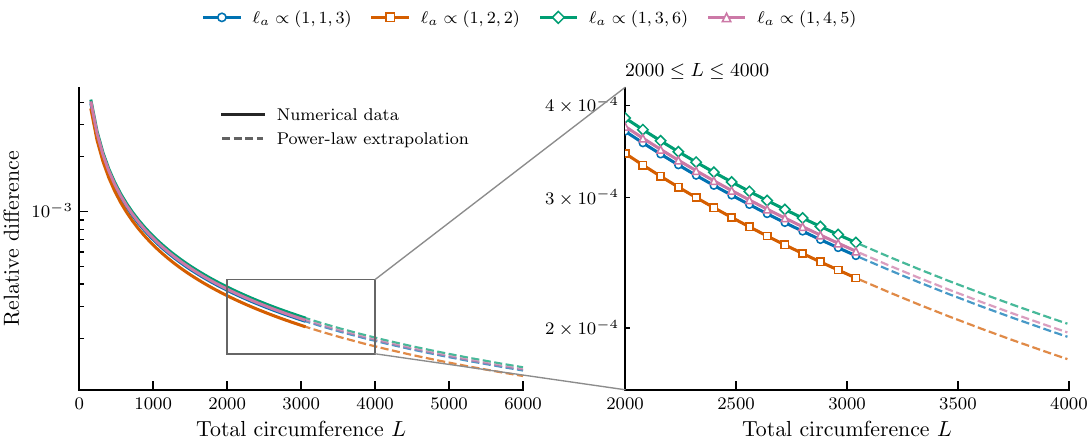}
    \end{center}
    \caption{The relative difference of the left and right hand side of equation \eqref{eq:weylAnomalyComparison}. The circles show the sampled values of $L$, the solid line interpolates between them, and the dashed line shows the extrapolation of the relative difference -- assuming a functional form of $a/L^b$ -- up to $L=6000$.}
    \label{fig:weylAnomaly}
\end{figure}

We test this formula by taking four points on the moduli space $\mathcal{M}_{1,1}^{\text{comb}}$, given by different values of $\{\ell_a\}$, and comparing the left and right hand sides of equation \eqref{eq:weylAnomalyComparison} for each of the moduli. The relative difference of the two sides, as a function of the total length $L$, is plotted in figure \ref{fig:weylAnomaly}. For total length $L=3000$, the relative difference is below $5\times 10^{-4}$ for all four moduli.

\subsection{The genus 1 integration measure}
\begin{figure}
    \begin{center}
	 \includegraphics[width=1.0\textwidth]{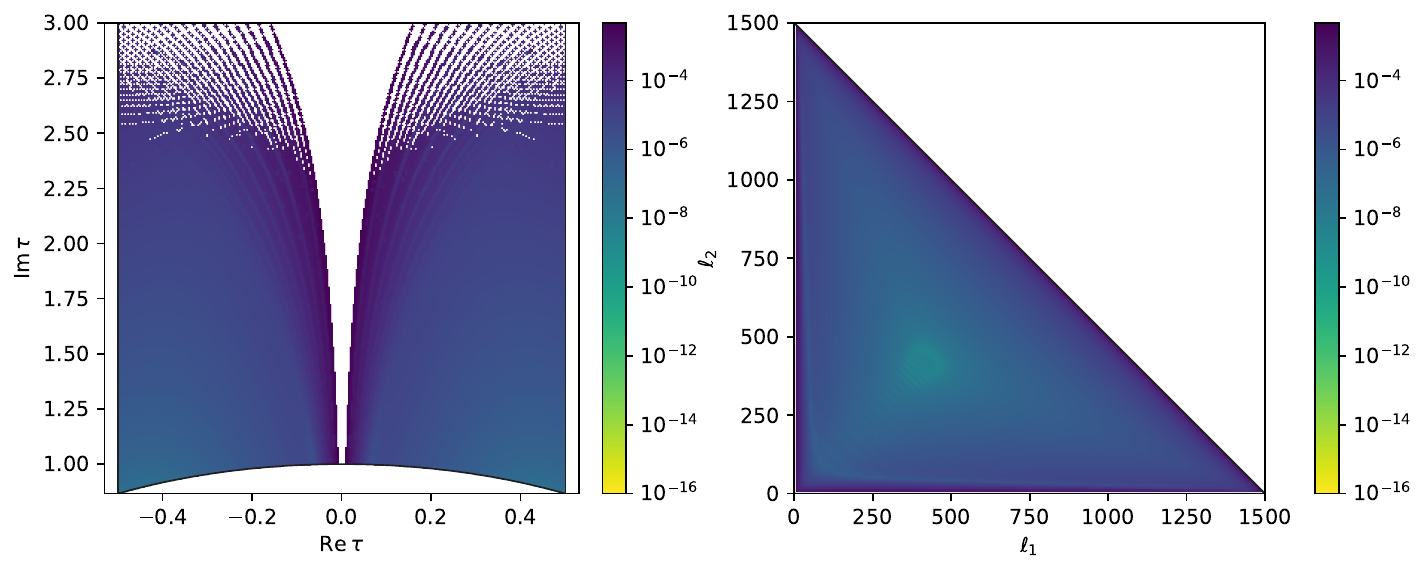}
    \end{center}
    \caption{The relative differnece between the left hand side and right hand side of equation \eqref{eq:integrationMeasure}, plotted over the fundamental domain in flat torus $\tau$ coordinates (left) and over the ribbon graph coordinates $\ell_1,\ell_2$ (right).}
    \label{fig:integrationMeasure}
\end{figure}

The integration measure of the genus 1 critical bosonic string free energy is free of Weyl anomalies, and we can compare it with the known CFT result. For genus 1 Riemann surfaces with one puncture, one can fix the puncture location using the conformal Killing group, and the moduli space $\mathcal{M}_{1,0}$ of unpunctured tori can be parameterized by the Strebel lengths $\ell_1,\ell_2$ of $\mathcal{M}_{1,1}^{\text{comb}}$. The integration measure can be computed directly from canonical quantization and is given by
\ie
\left\langle \mathcal{B}^2 \, c\tilde{c}(0)\right\rangle (Z_{\text{bos}}^{(1)})^{26}&= (2\pi)^{-24}\tau_2^{-13}|\eta(\tau)|^{-48} d\tau\wedge d\bar{\tau}
\fe
where $\mathcal{B}=\sum_{a} d\ell_a \mathcal{B}_{\ell_a}$, and the $b$ ghost insertion $\mathcal{B}_{\ell_a}$ is given by equation \eqref{Bla}. One then has
\ie \label{eq:integrationMeasure}
\left|\left\langle\mathcal{B}_{\ell_1} \wedge \mathcal{B}_{\ell_2} c \widetilde{c}(0)\right\rangle\right| (Z_{\text{bos}}^{(1)})^{26}=(2 \pi)^{-24} \tau_2^{-13}|\eta(\tau)|^{-48}\left|\frac{\partial(\tau, \bar{\tau})}{\partial\left(\ell_1, \ell_2\right)}\right|.
\fe

We test this formula by computing the LHS using the algorithm in section \ref{sec:freeBosonPartitionFunction} and section \ref{bcsection}, and computing the relation between $\tau$ and $\{\ell_a\}$ using the algorithm in section \ref{holodata}. In figure \ref{fig:integrationMeasure}, the relative and absolute difference of the left hand side and right hand side of equation \eqref{eq:integrationMeasure} are plotted as a function of $\tau$ and $\ell_1,\ell_2$, with fixed $L=3000$.

\subsection{Torus one point functions}

As a last test of our algorithms at genus 1, we compute the normalized one point functions of the normal ordered operators $:\partial X\bar\partial X:$ and $:\partial X\partial X:$. For the operators inserted using the flat torus coordinate $u\sim u+\mathbb Z+\tau\mathbb Z$ introduced in \eqref{ufromz}, the flat frame CFT results are
\ie \label{eq:flatTorusOnePoint}
\left\langle:\partial_uX\bar\partial_{\bar u}X:\right\rangle
&=\frac{1}{2\tau_2},\\
\left\langle:\partial_uX\partial_uX:\right\rangle
&=\frac{\theta_1'''(0|\tau)}{6\theta_1'(0|\tau)}
+\frac{\pi}{2\tau_2},
\fe
where the derivatives act on the first argument of the Jacobi theta function $\theta_1(u|\tau)$. Using the relation $du/dz=f(z)/\mathcal A$, the correlators in the disk frame are
\ie \label{1pttest}
\left\langle:\partial_zX\bar\partial_{\bar z}X:(0)\right\rangle &=\frac{1}{2\tau_2}\left|\frac{f(0)}{\mathcal A}\right|^2,\\
\left\langle:\partial_zX\partial_zX:(0)\right\rangle &=\left(\frac{f(0)}{\mathcal A}\right)^2 \left[ \frac{\theta_1'''(0|\tau)}{6\theta_1'(0|\tau)} +\frac{\pi}{2\tau_2}\right] -\frac{1}{12}\{u,z\}_{z=0}.
\fe
The Schwarzian derivative $\{u,z\}$ in the second line is required since
$:\partial X\partial X:$ is not a conformal primary.

\begin{figure}
    \begin{center}
	 \includegraphics[width=1.0\textwidth]{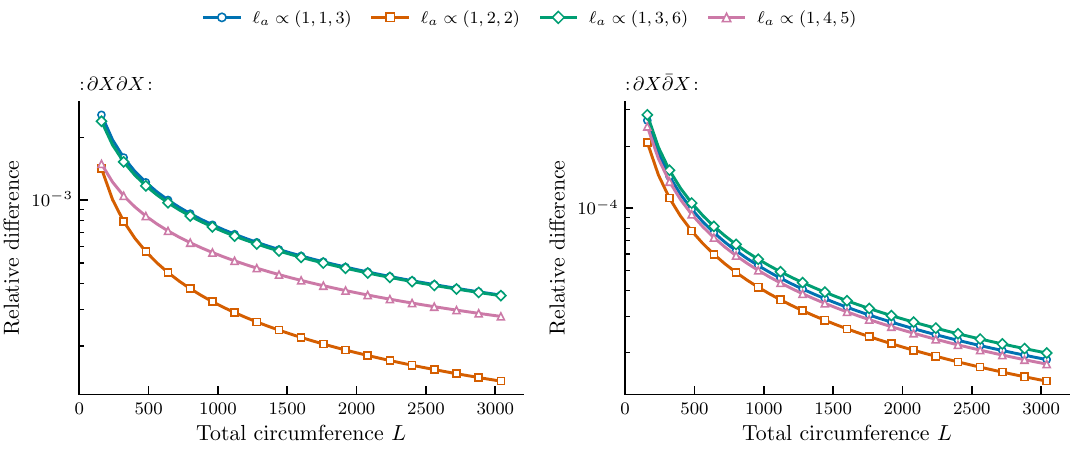}
    \end{center}
    \caption{Relative difference of the torus one point functions of $:\!\partial X\partial X\!:$ (left) and $:\!\partial X\bar\partial X\!:$ (right).}
\label{fig:onePointGenus1}
\end{figure}

The numerical one point functions are obtained from equation \eqref{1ptfuncgeneral}. In the convention of equation equation \eqref{psionefx}, the wavefunctionals used in the calculation are
\ie \label{eq:onePointWavefunctions}
\Psi_{:\partial X\partial X:}[X(\sigma)]
&=X_1^2\Psi_1[X(\sigma)],\\
\Psi_{:\partial X\bar\partial X:}[X(\sigma)]
&=\frac{1}{\pi}\left(\frac{1}{2}-X_1X_{-1}\right)
\Psi_1[X(\sigma)].
\fe
They are discretized and sewn in the same way as the identity wavefunctional in section \ref{sec:freeBosonPartitionFunction}, and the resulting normalized one point functions are compared directly with \eqref{1pttest}. The results for this test are shown in figure \ref{fig:onePointGenus1}, where both sides are transformed to the flat coordinate $u$ before forming the relative difference. At total length $L=3040$, the largest relative differences among these moduli are $7.29\times10^{-4}$ for $:\partial X\partial X:$ and $3.27\times10^{-5}$ for $:\partial X\bar\partial X:$.

\subsection{Compact boson at higher genus}

We also test the algorithms for Riemann surfaces with genus $g\geq 2$. The algorithm we use to generate the set genus $g$ ribbon graphs with one face is described in appendix \ref{app:ribbonGraphGeneration}. In the disk frame, the genus $g$ compact boson partition function is given by
\ie
\label{eq:compactBosonPartFunc}
Z^{(g)}_\text{comp}(R,\ell_a)\propto \frac{R}{\sqrt{\det \mathbf{A}'}}\Theta(4 i R^2 \mathbf{T}'),
\fe
where $R$ is the target space radius $X\sim X+2\pi R$, $\Theta$ denotes the Siegel theta function, and the matrices $\mathbf{A}', \mathbf{T}'$ are defined in equations \eqref{Aprime} and \eqref{WTTprime}, respectively.

We test the compact boson partition function against the conventional period matrix expression,
\ie
Z_{\mathrm{comp}}^{(g)}(R, \Omega)=2\pi R\,Z_{\mathrm{bos}}^{(g)}(\Omega) \sum_{m,n\in\mathbb Z^g}\exp\left[-\pi R^2(m+\Omega n)^T(\operatorname{Im}\Omega)^{-1}(m+\bar\Omega n)\right],
\fe
where $Z_{\mathrm{bos}}^{(g)}(\Omega)$ is the genus $g$ noncompact boson partition function, which is independent of $R$.

\begin{figure}[H]
    \centering
    \includegraphics[width=1.0\textwidth]{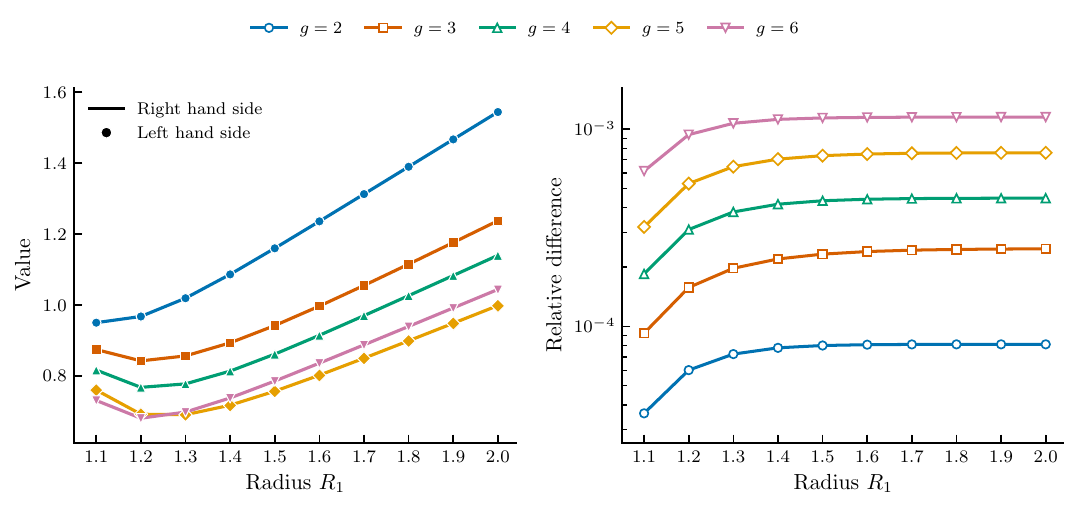}
    \caption{The values (left) and relative difference (right) of the two sides of \eqref{classicalthetatest} as a function of $R_1$ at fixed $R_2=1$ and ribbon graph topology for genus $2\leq g\leq 6$. All edge lengths are equal for a given genus, with $\ell_a=200$ for $g=2$, $\ell_a=120$ for $g=3$, $\ell_a=86$ for $g=4$, $\ell_a=67$ for $g=5$, and $\ell_a=55$ for $g=6$.}
    \label{testcompactratio}
\end{figure}

The ratio of two $Z^{(g)}(R, \Omega)$'s with different radii is free from Weyl anomaly, and we test the equation\footnote{In other words, this equation tests, for $\Omega=X+iY$ and
\ie
Q(\Omega)\equiv\begin{pmatrix}
Y^{-1}&Y^{-1}X\\
X^TY^{-1}&Y+X^TY^{-1}X
\end{pmatrix},
\fe 
that $4T'=U^TQ(\Omega)U$ for some $U\in GL(2g,\mathbb{Z})$.
}
\ie \label{classicalthetatest}
\frac{R_1\Theta(4iR_1^2\mathbf T')}{R_2\Theta(4iR_2^2\mathbf T')}=\frac{R_1\displaystyle\sum_{m,n\in\mathbb Z^g}\exp\left[-\pi R_1^2(m+\Omega n)^T(\operatorname{Im}\Omega)^{-1}(m+\bar\Omega n)\right]}{R_2\displaystyle\sum_{m,n\in\mathbb Z^g}\exp\left[-\pi R_2^2(m+\Omega n)^T(\operatorname{Im}\Omega)^{-1}(m+\bar\Omega n)\right]}.
\fe
We fix the denominator of both sides of \eqref{classicalthetatest} at $R_2=1$ and vary $R_1$. Figure \ref{testcompactratio} shows the relative difference between the two sides of equation \ref{classicalthetatest} for genus $2\leq g\leq 6$ and fixed ribbon graph topology and $\ell_i$. The largest relative difference for genus 2 is $8.11\times 10^{-5}$ and for genus 6 is $1.15\times 10^{-3}$.

%\begin{figure}[H]
%    \centering
%    \includegraphics[width=0.78\textwidth]{figures/genus3_genus4_genus5_genus6_tduality.pdf}
%    \caption{The relative difference of the two sides of \eqref{Tdualityresum} for one ribbon graph topology at each genus. The genus 3 test uses the theta function cutoff $N=4$, the genus 4 and genus 5 tests use $N=3$, and the genus 6 test uses $N=2$. In every case, each edge length is $\ell_a=200$. The corresponding total lengths are $L=6000$, $L=8400$, $L=10800$, and $L=13200$, respectively.}
%    \label{fig:compactG36}
%\end{figure}

\subsection{The genus 2 Igusa cusp form}

The next object we compute is the moduli integrand of the genus 2 vacuum amplitude for the critical bosonic string. Given two points $\Omega_a, \Omega_b$ on $\mathcal{M}_{2,0}$, which can be specified by two sets of edge lengths, we test the ratio of the $bc$ ghost + free boson correlators at 3 arbitrary points on the genus 2 Riemann surface:
\ie \label{g2ratiocheck}
\frac{\left|\left\langle b(z_1)b(z_2)b(z_3)\right\rangle_{\Omega_a}\right|^2}{\left|\left\langle b(z_1)b(z_2)b(z_3)\right\rangle_{\Omega_b}\right|^2}\frac{\left|Z_{\text {chiral}}^{(2)}\left(\Omega_a\right)\right|^{52}}{\left|Z_{\text {chiral}}^{(2)}\left(\Omega_b\right)\right|^{52}} =\frac{\left|\operatorname{det} S_i\left(z_j ; \Omega_a\right)\right|^2}{\left|\operatorname{det} S_i\left(z_j ; \Omega_b\right)\right|^2} \frac{\left|\chi_{10}\left(\Omega_b\right)\right|^2}{\left|\chi_{10}\left(\Omega_a\right)\right|^2}.
\fe

Let $K$ be the canonical bundle of the genus 2 Riemann surface $\Sigma$. The quantities $S_i(z;\Omega)$ are the basis of the space $H^0(\Sigma, K^2)$ of holomorphic quadratic differentials. In terms of the two ${\mathcal A}$-normalized
holomorphic 1-forms $\hat{\omega}_1(z),\, \hat{\omega}_2(z)$ (cf. \eqref{anormalize}), the basis $S_i(z;\Omega)$ can be taken to be
\ie
S_1(z;\Omega)=\hat{\omega}_1(z)^2,\quad S_2(z;\Omega)=\hat{\omega}_2(z)^2,\quad S_3(z;\Omega)=\hat{\omega}_1(z)\hat{\omega}_2(z),
\fe
and the determinant $|\det S_i(z_j;\Omega)|^2$ arises in the zero mode sector of the $bc$ ghost path integral.

The quantity $\chi_{10}(\Omega)$ is known as the Igusa cusp form \cite{Igusa:1962}, and this factor in \eqref{g2ratiocheck} is pinned down from its analytic properties and its modular transformation under the $Sp(4,\mathbb{Z})$ modular group \cite{DHokerPhong:1988, BelavinKnizhnik:1986}. It can be written in terms of the Riemann theta functions introduced in equation \eqref{eq:RiemannThetaFunction} by
\ie
\chi_{10}(\Omega) = \prod_{\delta \text{ even}} \theta[\delta](0|\Omega)^2,
\fe
where the product is over the 10 even spin structures on $\Sigma$.\footnote{Our convention for the Igusa cusp form differs from that of \cite{StringNotes} by a factor of $2^{12}$ but agrees with that of \cite{DHokerPhong:2002IV}.} 

\begin{figure}[H]
    \centering
    \includegraphics[width=1.0\textwidth]{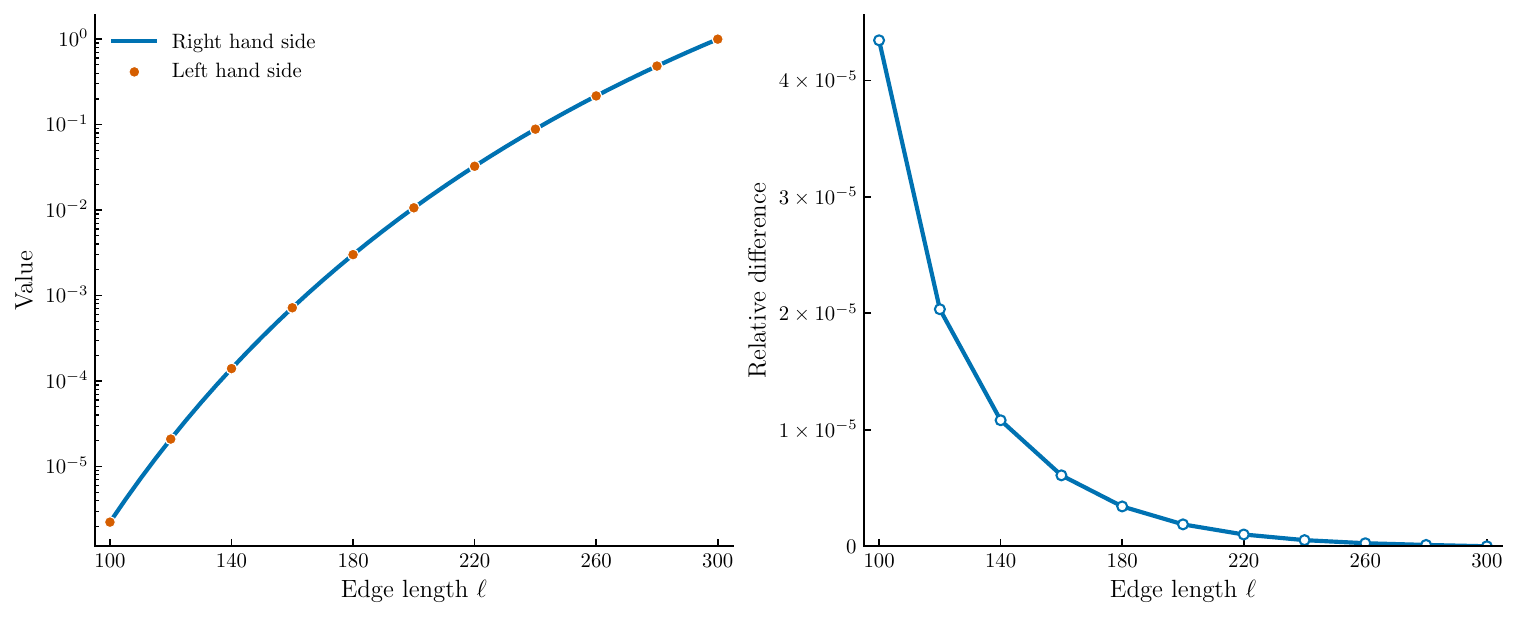}
    \caption{The values (left) and relative difference (right) of the two sides of \eqref{g2ratiocheck}. The parameterization of the edges is $\ell_{1},...,\ell_8=\ell,\ell_9=2700-8\ell$.}
    \label{igusaFixedLFamily}
\end{figure}

The $bc$ ghost correlators can be computed using the formula \eqref{vvformula}. Although our convention for $Z_{\text{chiral}}$ is specified in \eqref{Zchiral}, the combination in \eqref{g2ratiocheck} is independent of this choice: the $Z_{\text{chiral}}$ dependence of the $bc$ ghostcorrelator cancels the explicit factor of $\lvert Z_{\text{chiral}}\rvert^{52}$. The holomorphic 1-forms $\omega_I$, the period matrix $\Omega$, and the Igusa cusp form on the right hand side can be computed using the algorithm in section \ref{holodata}. We numerically check this relation for the moduli points $\Omega_a$ obtained from a fixed ribbon graph topology with edge lengths $\ell_{1,\cdots, 8} = \ell, \quad \ell_9 = 2700-8\ell$, and $\Omega_b$ given by the same topology and $\ell_{1,\cdots, 9} = 300$. The details and the results of this calculation are shown in figure \ref{igusaFixedLFamily}. Therefore this test only serves as a check of the algorithm in section \ref{holodata} for computing holomorphic data, but not the algorithm for computing free boson correlators in section \ref{sec:freeBosonPartitionFunction}. 

As a genus 2 test to the formula \eqref{Bla} of $b$ ghost insertions, we consider the moduli space integrand for the genus 2 one point function of the nonphysical vertex operator $c\tilde{c}$. Namely, we consider the top form component of the quantity $\langle e^{\mathcal{B}} c\tilde{c}(0)\rangle$ on the moduli space $\mathcal{M}_{2,1}$ of genus 2 Riemann surfaces with 1 puncture. 

Let us first compute this quantity analytically. The vertex operator $c\tilde{c}$ is not of conformal weight $(h,\tilde{h})=(0,0)$, hence the quantity $\langle e^{\mathcal{B}} c\tilde{c}(0)\rangle$ depends on the local coordinate where $c\tilde{c}$ is inserted. In the ribbon graph disk frame, the puncture is located at the origin $z=0$. The coordinates on the moduli space $\mathcal{M}_{2,1}$ can be taken as the 3 complex components of the period matrix $\Omega_{IJ}$, together with the puncture coordinates $u$, defined as
\ie
u = \int_q^0 \hat{\omega}_1 = \int_q^0 \hat{f}_1(z)dz ,
\fe
where $\hat{\omega}_1$ is the first $\mathcal{A}$-normalized holomorphic 1-form as in \eqref{anormalize}. The reference point $q$ is in principle arbitrary, and we take it to be a zero of $\hat{\omega}_1$ in our calculation. Using the coordinates $(\Omega_{IJ}, u)$, the quantity $\langle e^{\mathcal{B}} c\tilde{c}(0)\rangle$ can be related to the integrand of the genus 2 vacuum amplitude by
\ie 
\langle e^{\mathcal{B}} c\tilde{c}(0)\rangle &=  |\hat{\omega_1}(0)|^{-2} \left\langle \prod_{1\leq I\leq J\leq 2}\mathcal{B}_{\Omega_{IJ}}\wedge \mathcal{B}_{\bar{\Omega}_{IJ}}\right\rangle du\wedge d\bar{u}\\
& = \frac{2^{24}(2\pi)^{-50}}{(\det \operatorname{Im}\Omega)^{13}|\hat{f}_1(0)|^2|\chi_{10}(\Omega)|^2} du\wedge d\bar{u}\wedge \prod_{1\leq I\leq J\leq 2} d\Omega_{IJ}\wedge d\bar{\Omega}_{IJ}.
\fe

On the ribbon graph side, the 8 real moduli of $\mathcal{M}_{2,1}$ are given by the 8 independent edge lengths $\ell_1, \cdots \ell_8$. Using the $b$ ghost insertion \eqref{Bla} and the formula \eqref{vvformula}, the quantity $\langle e^{\mathcal{B}} c\tilde{c}(0)\rangle$ can be computed and tested against
\ie \label{eq:genus2Comparison}
\left\langle \mathcal{B}_{\ell_1}\wedge \cdots \mathcal{B}_{\ell_8} c\tilde{c}(0)\right\rangle (Z_\text{bos}^{(2)})^{26} = \frac{2^{24}(2\pi)^{-50}}{(\det \operatorname{Im}\Omega)^{13}|\hat{f}_1(0)|^{2}|\chi_{10}(\Omega)|^{2}} \left|\frac{\partial(\Omega_{IJ}, \bar{\Omega}_{IJ}, u, \bar{u})}{\partial(\ell_1,\cdots,\ell_8)}\right|.
\fe

\begin{figure}
    \begin{center}
	 \includegraphics[width=1.0\textwidth]{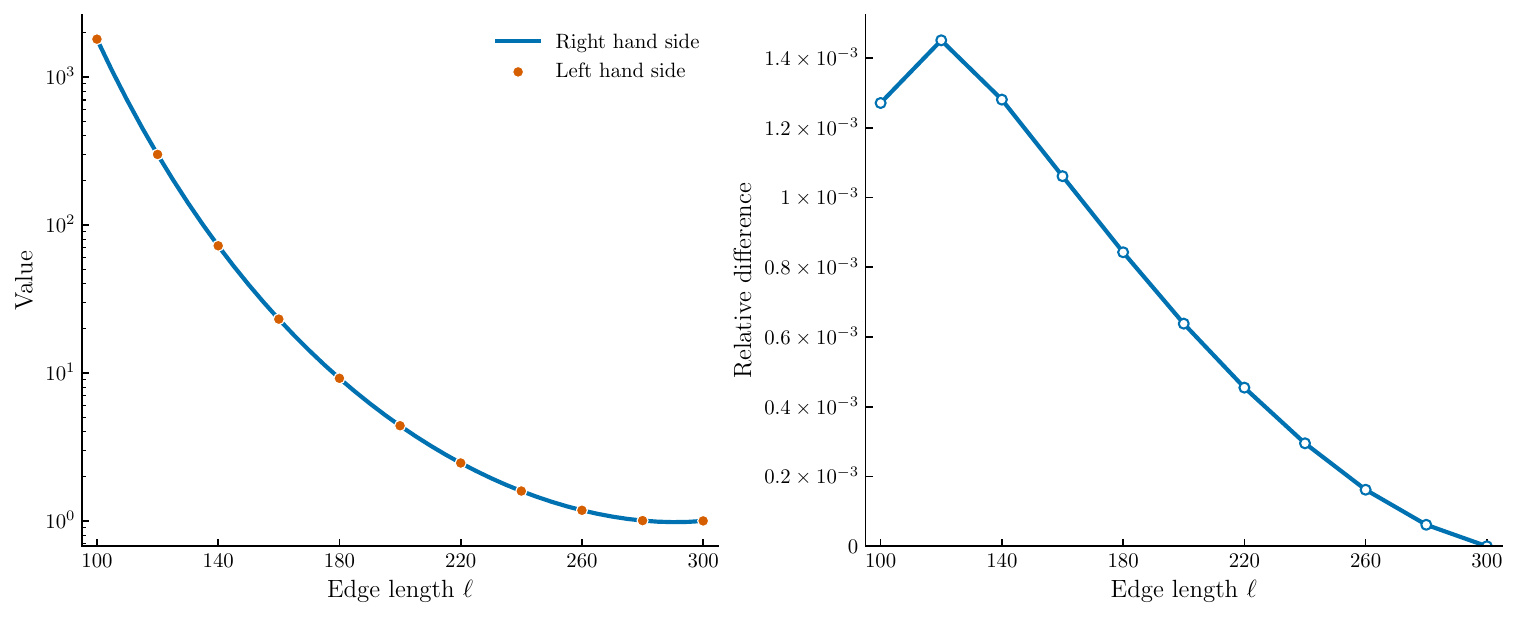}
    \end{center}
    \caption{The values of the two sides of equation \eqref{eq:genus2Comparison} are shown on the left. The corresponding relative difference is shown on the right.  The parameterization of the edges is $\ell_{1},...,\ell_8=\ell,\ell_9=2700-8\ell$.}
    \label{fig:onePointGenus2}
\end{figure}

The comparison of the left hand side and right hand side of equation \eqref{eq:genus2Comparison} are plotted in figure \ref{fig:onePointGenus2} for a single genus 2 topology. For the $\ell_i$ tested, the largest relative disagreement between the two sides of equation \eqref{eq:genus2Comparison} is $1.45\times 10^{-3}$.

\subsection{The genus 3 Igusa cusp form}

For nonhyperelliptic genus 3 surfaces, the six holomorphic quadratic differentials $\hat{\omega}_I \hat{\omega}_J$ span the space $H^0(\Sigma, K^2)$, and the period matrix $\Omega_{IJ}$ is a good coordinate on the moduli space $\mathcal{M}_{3,0}$. Similar to the genus 2 relation \eqref{g2ratiocheck}, at genus 3
\ie \label{corrgenus3}
\left|\left\langle \prod_{i=1}^6 b(z_i) \right\rangle_\Omega \right|^2|Z_\text{chiral}^{(3)}(\Omega)|^{52} \propto |\det S_i(z_j;\Omega)|^2 |\chi_{18}(\Omega)|^{-1}.
\fe
Here, $\chi_{18}$ stands for the genus 3 Igusa cusp form, which is given by a product of Riemann theta functions \cite{Igusa:1967}
\ie
\chi_{18}(\Omega)&=\prod_{\delta \text{ even}}\theta[\delta](0|\Omega),
\fe
where the product is over the 36 even spin structures. 

\begin{figure}[H]
    \centering
    \includegraphics[width=1.0\textwidth]{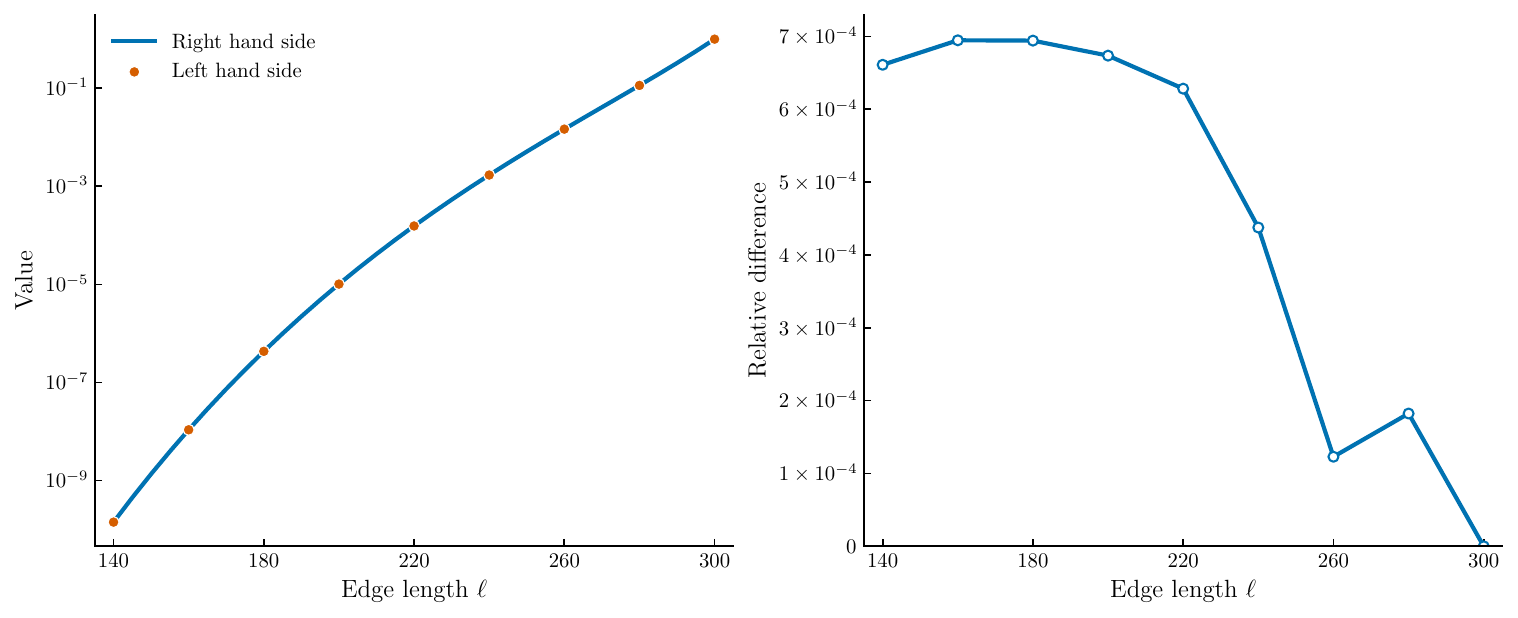}
    \caption{The values of the two sides of \eqref{g3ratiocheck} are shown on the left, with
    the analytical right hand side shown as a
    solid curve, and the numerical left hand side shown as points.  The
    vertical axis is logarithmic.  The corresponding relative difference is
    shown on the right.  The genus 3 fixed $L$ family is
    $\ell_1=\cdots=\ell_{14}=\ell$, $\ell_{15}=4500-14\ell$, shown for
    $140\leq\ell\leq300$.  Theta functions are evaluated with cutoff $N=4$.}
    \label{fig:g3ratio}
\end{figure}

We test this relation by checking
\ie \label{g3ratiocheck}
\frac{\left|\left\langle \prod_{i=1}^6 b(z_i)\right\rangle_{\Omega_a}\right|^2}{\left|\left\langle \prod_{i=1}^6 b(z_i)\right\rangle_{\Omega_b}\right|^2}\frac{\left|Z_{\text {chiral}}^{(3)}\left(\Omega_a\right)\right|^{52}}{\left|Z_{\text {chiral}}^{(3)}\left(\Omega_b\right)\right|^{52}} =\frac{\left|\operatorname{det} S_i\left(z_j ; \Omega_a\right)\right|^2}{\left|\operatorname{det} S_i\left(z_j ; \Omega_b\right)\right|^2} \frac{\left|\chi_{18}\left(\Omega_b\right)\right|}{\left|\chi_{18}\left(\Omega_a\right)\right|} 
\fe
on a specific ribbon graph topology. %review: see earlier comment about topology labels, maybe call it topology [3,1]
Specifically, we take $\Omega_a$ along the family $\ell_1=\cdots =\ell_{14}=\ell$, $\ell_{15}=4500-14\ell$, and $\Omega_b$ to be the moduli corresponding to $\ell = 300$. The 6 $b$ ghost insertions are taken to be the following points in the disk frame 
\ie\begin{gathered}
    z_1 = 0.08+0.12i,\quad z_2 = -0.14+0.16i,\quad z_3=0.19-0.09i,\\
    z_4=0.21+0.09i,\quad z_5=-0.16+0.12i,\quad z_6 = 0.05-0.18i.
\end{gathered}
\fe
The relative difference between two sides of \eqref{g3ratiocheck} are plotted in figure \ref{fig:g3ratio}. For $140\leq \ell \leq 280$, the relative differences are below $7\times 10^{-4}$.

Again, similar to the genus 2 test, this test only checks our algorithm for holomorphic data, as the chiral boson correlator dependence cancels in the left hand side of \eqref{corrgenus3}. The test above concerns the local matter ghost correlator. We have not carried out the genus 3 analogue of the full $\mathcal M_{3,1}$ Jacobian comparison in \eqref{eq:genus2Comparison}.\footnote{
We should emphasize that the relation \eqref{corrgenus3} cannot be used for hyperelliptic genus 3 Riemann surfaces. Such a hyperelliptic surface can be described by
\ie
y^2 = \prod_{a=1}^8 (x-e_a).
\fe
After fixing three branch points $e_a$ at $0, 1,\infty$ using a $PSL(2,\mathbb{C})$ Mobius transformation on $x$, the remaining 5 $e_a$'s parametrize a complex locus of dimension 5 in the 6 dimensional complex moduli space $\mathcal{M}_{3,0}$. The basis of $H^0(\Sigma, K)$ can be taken to be $dx/y, xdx/y$ and $x^2dx/y$, and their quadratic products are subject to the linear relation 
\ie
\frac{dx}{y} \frac{x^2dx}{y} - \left(\frac{xdx}{y}\right)^2 = 0.
\fe
Therefore, there is a linear relation among the products $\hat{\omega}_I \hat{\omega}_J$ of ${\mathcal A}$-normalized holomorphic 1-forms, and the determinant $\det S_i(z_j;\Omega)$ vanishes. This phenomenon indicates that $\Omega_{IJ}$ fail to be good local coordinates on the moduli space $\mathcal{M}_{3,0}$ on the hyperelliptic locus.
}

%Lastly, Let $\{t^k\}$ be a set of local complex coordinates on $\mathcal{M}_{3,0}$, and let $\mu_k$ be the Beltrami differential associated with the modulus $t^k$. The Rauch variation formula then relates the quadratic products $\hat{\omega}_I \hat{\omega}_J$ with the variation of the period matrix $\Omega_{IJ}$ by \cite{Rauch:1959}
%\ie
%\frac{\partial \Omega_{IJ}}{\partial t^k} = \int_{\Sigma} \hat{\omega}_{I,z} \hat{\omega}_{J,z}(\mu_k)_{\bar{z}}{}^{z} dz\wedge d\bar{z},
%\fe
%and thus the linear relation among $\hat{\omega}_I \hat{\omega}_J$ leads to a vanishing Jacobian $\det \frac{\partial \Omega_{IJ}}{\partial t^k}$ on the hyperelliptic locus, which indicates that $\Omega_{IJ}$ do not give local coordinates on $\mathcal{M}_{3,0}$.

\section{Discussion}
\label{discussionsection}

In this work, we have developed a numerical framework for computing higher genus CFT correlators and string amplitude integrands. Using Kontsevich's isomorphism between punctured Riemann surfaces and metric ribbon graphs \cite{Kontsevich:1992ti,Strebel:1967,Strebel:1984}, we reconstructed holomorphic data directly from ribbon graph parameters and evaluated the free boson path integral at arbitrary genus. We have tested the method against a number of known results. At genus 1, it reproduces the expected partition functions and one point functions. At higher genus, the matter ghost correlators at genera 2 and 3 reproduce the expected Mumford form dependence \cite{BelavinKnizhnik:1986,DHokerPhong:1988,Igusa:1962,Igusa:1967}, while the compact boson partition functions have the expected radius dependence for genera 2 through 6. Note that the genus 3 test of the Mumford form concerns the local matter + ghost correlator; we have not performed the corresponding comparison that includes the full Jacobian on $\mathcal M_{3,1}$.

The computation of string observables further requires integration over moduli space. The ribbon graph parametrization appears to be well suited for this purpose: distinct ribbon graph topologies label cells in the moduli space $\mathcal{M}_{g,1}$ with disjoint interiors, and the cells are glued together along loci where one or more edges collapse. The moduli integral can then be organized as a sum of ordinary integrals over positive edge lengths. This may offer a practical advantage at higher genus, where the period matrix does not provide local coordinates on moduli space, although the rapidly growing number of cells becomes a challenge for $g \geq 4$.

At the moment, the main technical obstacle to performing the moduli space integral is the loss of numerical accuracy near the boundaries of the ribbon graph cells. Our present discretization works well when all edge lengths are of the same order. When an edge becomes short, however, only a small number of discretization points lie on the corresponding boundary segment, and the errors in the reconstructed holomorphic data and in the path integral become large. Since these regions are unavoidable in a global moduli integral, a more efficient discretization scheme will be needed. We leave the implementation of such refinements to future work. 

It would also be interesting to generalize the free boson construction to interacting CFTs with a path integral description. For a general CFT, the identity wavefunctional is no longer Gaussian. One must therefore compute this wavefunctional numerically and subsequently perform the functional integral on the circle with glued segments. The first step may be approached through Hamiltonian truncation \cite{RychkovVitale:2015}, neural network variational states \cite{CarleoTroyer:2017,MartynNajafiLuo:2023}, or a direct lattice Monte Carlo evaluation of the path integral inside the disk \cite{BrowerFlemingNeuberger:2013}. Given a sufficiently accurate representation of the disk wavefunctional, the remaining gluing integral could in principle also be carried out by Monte Carlo. Whether these methods can be implemented with sufficient precision for higher genus correlators is an interesting question for future work.

Finally, let us mention that the algorithm for reconstructing holomorphic data can be generalized beyond the disk frame, in particular to the plumbing frame, where a Riemann surface is constructed by sewing pairs of punctures on 3-punctured spheres. One may construct the holomorphic 1-forms by making a suitable ansatz on each 3-punctured sphere and imposing matching conditions across the identified points. This provides an algorithm for computing the period matrix from the plumbing parameters. Since arbitrary genus Virasoro conformal blocks can be evaluated recursively in the same plumbing parameters \cite{Cho-Collier-Yin}, this would give a direct method for constructing moduli integrands in interacting CFTs and expressing them in terms of the period matrix. In particular, it should make possible a direct numerical study of higher loop observables in $c=1$ string theory, which we are investigating in ongoing work \cite{c1paper}.

\section*{Acknowledgements}

We would like to thank Alexander Michel and Jaroslav Scheinpflug for discussions. BM, YW and XY thank Benasque Science Center and the organizers of the workshop ``String field theory and flux compactification" for their hospitality during the course of this work. GPT-5.4 through GPT-5.6 were used for coding and algorithm optimization, while Claude Opus 4.6 through Claude Opus 4.8 were used during the early stages of this work. All derivations in this work were carried out without using AI. SC is supported in part by an NSF GRFP fellowship. This work is supported by DOE grant DE-SC0007870.

\appendix

\section{Ribbon graph generation}
\label{app:ribbonGraphGeneration}

In order to make use of the gluing map for general genus in equation \eqref{gluinggenusg}, we need to determine the gluing data for all ribbon graphs of a given genus and number of faces. In this section, we describe the algorithm to generate such data.

A candidate ribbon graph can be formulated as a set of vertices $V$, a set of edges $E$ connecting the vertices (with the graph being connected and with only three incident edges at each vertex), and a fixed cyclic ordering of the incident edges at each vertex, which we will denote as the ``rotation system'' $R$. The number of vertices is equal to $2(2g-2+F)$ and the number of edges is equal to $3(2g-2+F)$, with $F$ the number of faces.

The specification of these three pieces of data uniquely specifies the ribbon graph embedding onto a closed oriented surface (up to an orientation preserving homeomorphism).

Our algorithm first constructs all possible sets of edges that can connect a given set of vertices. Edges are encoded in the adjacency matrix $A$, with $A_{ij}$ being the number of edges between vertices $i,j$. Therefore, we construct all possible $A$ with

\ie
&A_{ij}=A_{ji},~~A_{ij}\in \{0,1,2\} \quad &\forall i,j\\
&A_{ii}=0,~~\sum_j A_{ij}=3 \quad &\forall i.
\fe

For genus 1, one can also have $A_{ij}= 3$. Below, we will assume that the genus is greater than 1, as the unique genus 1 graph with one face can be found by hand.

We discard candidate $A$ which correspond to disconnected graphs. 

We want to then retain only one representative of each graph under isomorphism. Performing an exact test of isomorphism between a pair of graphs is expensive, so we first apply the Weisfeiler-Leman (WL) test \cite{WeisfeilerLeman:1968, Shervashidze:2011WL}. The WL test has computational complexity $\mathcal{O}( V\log V)$ for $V$ vertices, while in its worse case, the exact isomorphism test we use can have computational complexity $\mathcal{O}(V!)$. If the WL test returns false, then the two graphs are not isomorphic, but the algorithm is not able to conclude if two graphs are isomorphic.

For the remaining candidates, we test if they are exactly isomorphic using the isomorphism test in the $\texttt{NetworkX}$ python package. Given a pair of isomorphic graphs, we keep one. This step is empirically the most computationally expensive of our algorithm when the number of vertices is high (e.g. genus 3 or higher).

For each remaining candidate, we generate all possible rotation systems $R$ of edges incident at a given vertex. This is all the data of a candidate ribbon graph, except we must now determine the number of faces. Begin with a given edge $e_1$ connecting vertices $v_0,v_1$ in the ribbon graph and traverse in an arbitrary direction to vertex $v_1$. If the local rotation of $v_1$ is $\{e_1, e_2,e_3\}$, traverse to the vertex on the other end of $e_2$ next. Continue this procedure until one returns to the original edge $e_1$, traversing from $v_0$ to $v_1$ (it is not sufficient to merely return to the original vertex, as one face can return to the same vertex multiple times, nor is it sufficient to return to the original edge traversing from $v_1$ to $v_0$. If there is a single face, each edge is included in the face twice.). For every edge not visited during this traversal, repeat until all edges have been visited. The number of disjoint cycles is equal to the number of faces. We discard all candidates that do not have the desired number of faces.

All remaining candidates are ribbon graphs, but there can exist isomorphic ribbon graphs amongst them. For a given ribbon graph, we first enumerate all possible automorphisms of the vertices and edges (i.e. momentarily dropping the local rotation system associated with each vertex). A given permutation $\pi(u)=u'$ of the vertices correponds to an automorphism of the collection of vertices + edges if $A_{\pi(i)\pi(j)}=A_{ij}$, with $A$ the adjacency matrix defined above. If two edges connect the same vertices, the permutation of the vertices does not fully specify the permutation of the edges. In this case, we further specify the two distinct permutations of the edges connecting the same vertices. Finally, the map that $\pi$ induces on the edges is applied to the rotation system $R$ associated with each vertex to produce a ribbon graph isomorphism.

For a given candidate ribbon graph $\Gamma$, we now have the full set of ribbon graphs isomorphic to $\Gamma$, which we denote as $K_\Gamma$. In order to decide if two sets $K_\Gamma,K_{\Gamma'}$ are the same, we pick one canonical representative from each set. A natural numbering of the vertices is associated with its index $i$ in the adjacency matrix $A$. Meanwhile, edges are assigned numbers based on increasing label of their first endpoint and, for a fixed first endpoint label, by increasing label of their second endpoint. Two edges that have the same endpoints are assigned consecutive numbers. We order each rotation system $\{e_1,e_2,e_3\}$ such that $(e_1 \leq e_2,e_1\leq e_3)$ (while respecting the original cyclic ordering). We then compare all pairs of entries of $K_\Gamma$. Denote one pair as $\Gamma_1,\Gamma_2$. Begin with vertex $1$ of $\Gamma_1,\Gamma_2$. Compare the two rotation systems associated with this vertex: $\{e_{1,1},e_{2,1},e_{3,1}\},\{e_{1,2},e_{2,2},e_{3,2}\}$. If $e_{1,1}< e_{1,2}$, keep $\Gamma_1$ but delete $\Gamma_2$, and vice versa. If both are equal, apply the same test to $e_{2,1}$. Proceed to apply the same test inductively for each vertex $i$ until only one of $\Gamma_1,\Gamma_2$ is kept. Repeat the same procedure for all entries in $K_\Gamma$ until one canonical entry is kept, $\Gamma_{c}\in K_\Gamma$

Now, one can easily test if two $\Gamma_c,\Gamma_{c}'$ are equivalent. What we are left with is the entire set of nonisomorphic ribbon graphs of genus $g$ with $F$ faces, as well as the ordering of edges around each face. %review: this would be the place to mention canonical ordering of the set of topologies by their representatives.

%description of 

\section{Details of the compact boson partition function}

\label{compactappendix}

In this appendix, we explain the detailed conditions for the physical shifts $\{s_a\}$ appearing in the compact boson partition function, and we explicitly perform the Gaussian integral in \eqref{compactPI}.

For each Strebel vertex $V_A$ ($A=1,\cdots, 4g-2$), the condition that the total shift vanishes when going around a contractible cycle can be written as
\ie \label{nowindingeq}
\sum_{a=1}^{6g-3} D_{A,a}s_a = 0,
\fe
where $D_{A,a}$ is defined as the incidence matrix of the ribbon graph $\Gamma$:
\ie
D_{A,a}=\begin{cases}
 1 & \text{if edge } E_a \text{ comes into vertex } V_A,\\
 -1 & \text{if edge } E_a \text{ goes out of vertex } V_A, \\
 0 & \text{if edge } E_a \text{ is not connected to vertex }V_A.
\end{cases}
\fe
The matrix equation \eqref{nowindingeq} then tells us that the independent shifts live in the kernel $\operatorname{ker}_{\mathbb{Z}}(D)$, which is a $2g$ dimensional lattice that corresponds exactly to the shifts around the $2g$ independent cycles on the Riemann surface $\Sigma$.

\begin{figure}
    \begin{center}
    	\subfloat[]{
    		\includegraphics[width=0.5\textwidth]{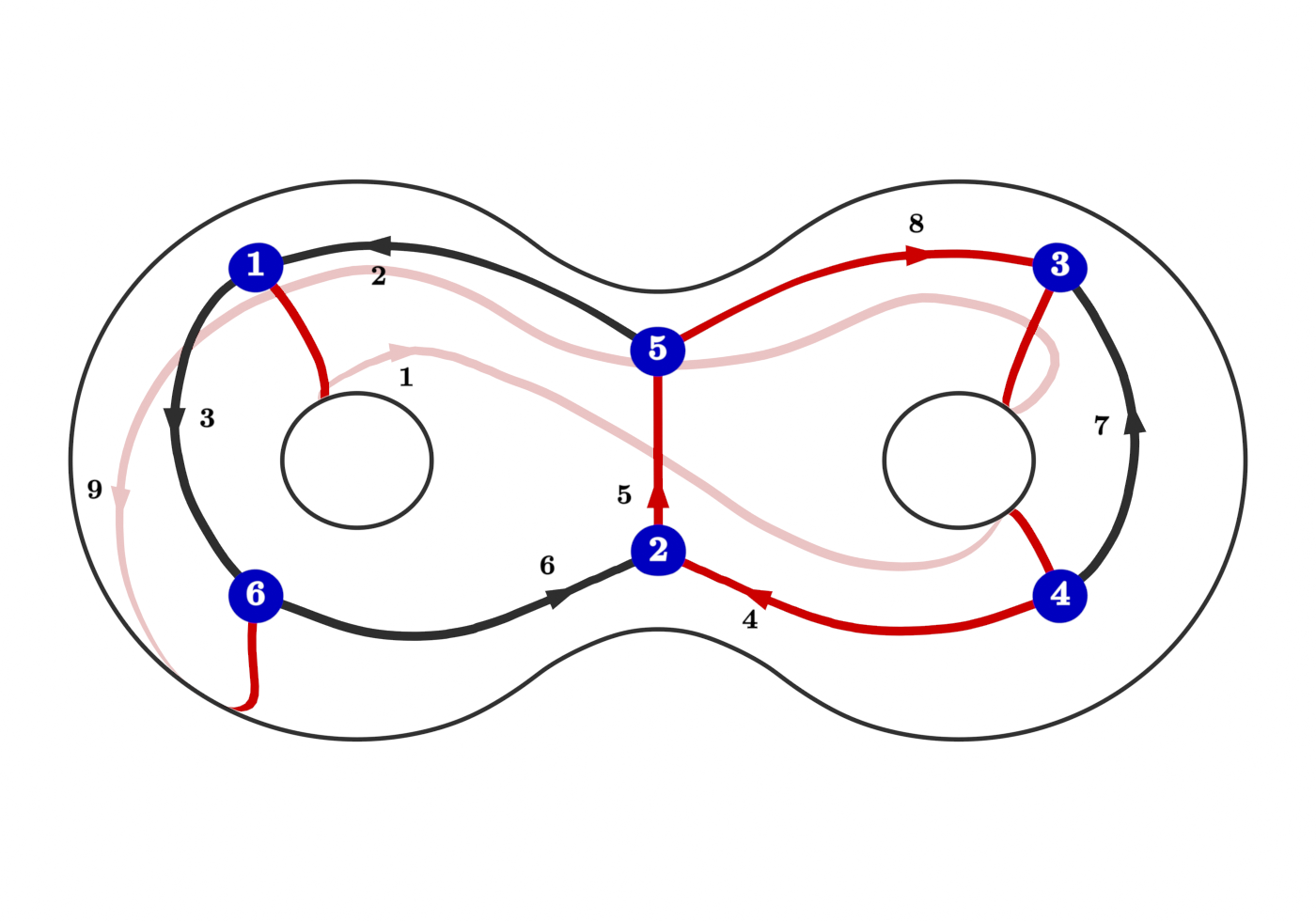}
    	}
    	\subfloat[]{
    		\includegraphics[width=0.5\textwidth]{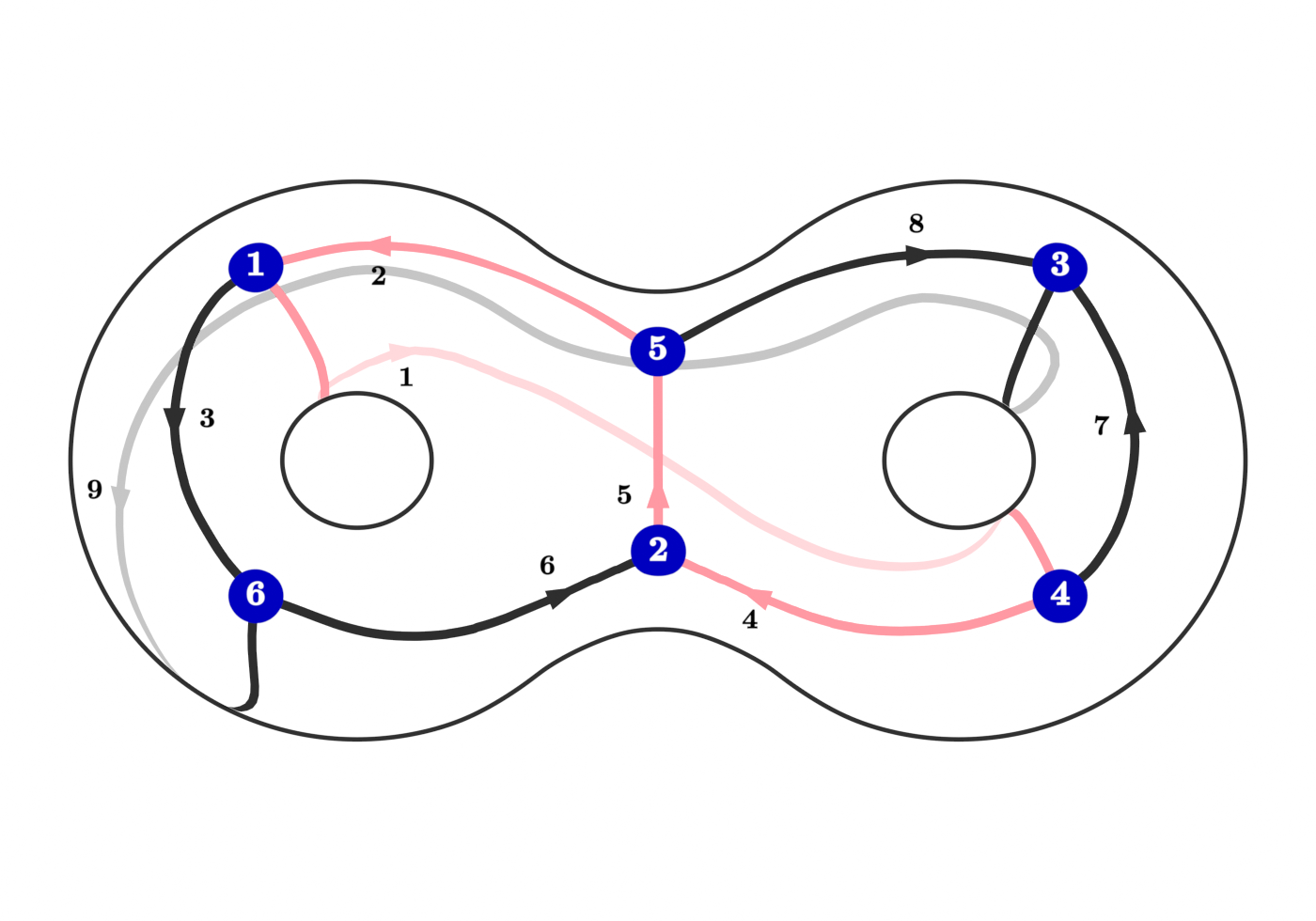}
    	}
    \end{center}
    \caption{The oriented graph $\hat{\Gamma}$ associated with a genus 2 ribbon graph $\Gamma$ is shown on the left, with a spanning tree highlighted in red. The cycle $C_e$ in $\hat{\Gamma}$ for the nontree edge $e=2$ is shown on the right in pink.}
    \label{cycleCe}
\end{figure}

A basis of $\operatorname{ker}_{\mathbb{Z}}(D)$ can be built in the following way. One first chooses a spanning tree of the ribbon graph $\Gamma$, which contains $4g-3$ edges. For each of the $2g$ edges $E_e$ that are not in the spanning tree, we denote its start point and end point by $V_{A_1(e)}$ and $V_{A_2(e)}$, respectively. Given such a nontree edge $E_e$, one can always find a cycle $C_e$ in $\hat{\Gamma}$ by starting from $V_{A_1(e)}$, going through $E_e$ to get to $V_{A_2(e)}$, and then taking the unique path in the spanning tree from $V_{A_2(e)}$ back to $V_{A_1(e)}$, as shown in figure \ref{cycleCe}. From this cycle $C_e$, one then defines a lattice vector $b_e\in \mathbb{Z}^{6g-3}$ by
\ie
(b_e)_a=\begin{cases}
 1 & \text{if edge } E_a \text{ is traversed with the same orientation in } C_e,\\
 -1 & \text{if edge } E_a \text{ is traversed with the opposite orientation in } C_e,\\
 0 & \text{if edge } E_a \text{ is not in } C_e,
\end{cases}
\fe
and the $2g$ vectors $\{b_e\}$ will form an integral basis of $\text{ker}_{\mathbb{Z}}(D)$. If we define the matrix $\mathbf{B}\in \mathbb{Z}^{(6g-3)\times 2g}$ by 
\ie \label{Bmatrix}
\mathbf{B}_{a,e}=(b_e)_a,
\fe 
then an arbitrary lattice vector $s$ in $\text{ker}_{\mathbb{Z}}(D)$ can be represented by 
\ie \label{physicalshifts} s=\mathbf{B}n,\fe
where $n\in \mathbb{Z}^{2g}$ is a lattice vector in the $2g$ dimensional integral lattice. Namely, the physical shifts $s_a$ all take the form \eqref{physicalshifts}.

The discretized compact boson partition function is then obtained by summing over all physical shifts
\ie
Z_{\text{comp}}^{(g)}(R;L;\ell_a)= \sum_{n\in \mathbb{Z}^{2g}} \int \prod_{k=1}^{L} dX(k)\prod_{a=1}^{6g-3} \prod_{k\in S_{i_1(a)}} \delta\left(X(k')-X(k)-2\pi R s_a(n)\right)\,\Psi_1[X],
\fe
where $s_a(n)$ is given by \eqref{physicalshifts} in terms of the lattice vector $n$. After replacing the variables $X(k)$ with $k\in \{1,\cdots, L\}\backslash \mathcal{K}$ using the delta functions and factoring out the zero mode, one obtains
\ie
Z_{\text{comp}}^{(g)}(R;L;\ell_a) = 2\pi R\sum_{n\in \mathbb{Z}^{2g}}\int_{k\in \mathcal{K}'} \prod_{k\in \mathcal{K}}dX(k) \exp\left[-\pi^{-1}\left(\sum_{k,\ell\in\mathcal{K}'}\mathbf{A}'_{k\ell}X(k)X(\ell)+2\sum_{k\in \mathcal{K}'}\mathbf{v}'_k X(k) +\mathbf{C}\right)\right],
\fe
where the set $\mathcal{K}'$ is defined to be $\mathcal{K}\backslash\{1\}$. The matrix $\mathbf{A}'$, the vector $\mathbf{v}'$, and the constant $\mathbf{C}$ are defined as
\ie
&\mathbf{M}_{k\ell}=\frac{1}{2 \alpha' L}\frac{\sin \frac{\pi}{L}}{\cos \frac{2\pi (k-\ell)}{L}-\cos \frac{\pi}{L}},\\
&\mathbf{A}_{k\ell} = \mathbf{M}_{k\ell}+ \mathbf{M}_{k'\ell} + \mathbf{M}_{k \ell'} +\mathbf{M}_{k'\ell'},\quad \mathbf{A}'_{k\ell}= \mathbf{A}_{k\ell} - \mathbf{A}_{1,\ell} - \mathbf{A}_{k,1} + \mathbf{A}_{1,1},\\
&\mathbf{v}_k = 2\pi R\sum_{a=1}^{6g-3} s_a(n) \sum_{\ell\in S_{i_1(a)}}(\mathbf{M}_{k\ell'}+\mathbf{M}_{k'\ell'}),\quad \mathbf{v}'_k = \mathbf{v}_k-\mathbf{v}_1,\\
&\mathbf{C}= (2\pi R)^2 \sum_{a,b=1}^{6g-3} s_a(n) s_b(n) \sum_{k\in S_{i_1(a)}} \sum_{\ell \in S_{i_1(b)}} \mathbf{M}_{k'\ell'}.
\fe
Performing the Gaussian integral then gives
\ie
Z_{\text{comp}}^{(g)}(R;L;\ell_a) &\propto \frac{R}{\sqrt{\det \mathbf{A}'}}\sum_{n\in \mathbb{Z}^{2g}}\exp\left(-4\pi R^2 n^\mathrm{T}\mathbf{T}' n\right)=\frac{R}{\sqrt{\det \mathbf{A}'}}\Theta(4 i R^2 \mathbf{T}'),
\fe
where $\Theta$ is the Siegel theta function, and the $2g\times 2g$ matrix $\mathbf{T}'$ is given by
\ie \label{WTTprime}
&\mathbf{W}_{k\ell} = \mathbf{M}_{k\ell'} + \mathbf{M}_{k'\ell'} -\mathbf{M}_{1,\ell'} - \mathbf{M}_{1',\ell'},\\
&\mathbf{T}_{ab} = \sum_{k\in S_{i_1(a)}}\sum_{\ell \in S_{i_1(b)}} \left(\mathbf{M}_{k'\ell'}-\sum_{m,n\in \mathcal{K}'}(\mathbf{A}')^{-1}_{mn} \mathbf{W}_{mk}\mathbf{W}_{n\ell}\right),\\
& \mathbf{T}' = \mathbf{B}^\mathrm{T}\mathbf{T} \mathbf{B},
\fe
and the matrix $\mathbf{B}$ is defined in \eqref{Bmatrix}.

\section{Details of the formula \eqref{vvformula}}

\label{holodataappendix}

In this appendix, we summarize the definitions and the algorithms used to compute the quantities appearing in \eqref{vvformula} in details. 

\subsubsection*{The chiral boson partition function}

The chiral boson partition function $Z_{\text{chiral}}$ is the holomorphic conformal block of the $U(1)$ current algebra generated by $\partial X$ in the free boson CFT. It can be related to the free boson partition function by
\ie
\label{eq:boschiralrelation}
Z_\text{bos}^{(g)}(\Omega) = V_X\int \frac{d^gk}{(2\pi)^g} e^{-\pi k^I k^J\mathrm{Im}(\Omega_{IJ})}|Z_\text{chiral}|^2 \propto \left[\det \mathrm{Im}(\Omega)\right]^{-1/2}|Z_\text{chiral}^{(g)}|^2.
\fe

Both $Z_\text{bos}^{(g)}$ and $Z_\text{chiral}^{(g)}$ depend on the renormalization scheme. Moreover, the quantity $Z_{\text{chiral}}^{(g)}$ depends on the Weyl frame and is subject to a gravitational anomaly due to the unbalanced central charge $(1,0)$. In principle, to specify $Z_{\text{chiral}}^{(g)}$ fully, one must choose a Hermitian metric on the worldsheet $\Sigma$ as well as a conformal frame that specifies the plumbing fixtures used to construct $\Sigma$ by sewing along the $\alpha^I$ cycles. The quantity $|Z_\text{chiral}^{(g)}|^2$ depends only on the Weyl frame, and one must take that Weyl frame to be the same one used for free boson correlators in bosonic string amplitudes.

In practice, we can choose a worldsheet metric in which $Z_\text{chiral}^{(g)}$ is given by
\ie \label{Zchiral}
Z_\text{chiral}^{(g)} = \left[\frac{\det \mathrm{Im}(\Omega)}{\left(\det \mathbf{A}'(\Omega)\right)_\text{ren}}\right]^{\frac{1}{4}},
\fe
where the branch of the fourth root is chosen such that $Z_\text{chiral}^{(g)}$ is always positive, and the renormalized determinant $\left(\det \mathbf{A}'(\Omega)\right)_\text{ren}$ is defined similarly to \eqref{renZ}, namely $\log\left(\det \mathbf{A}'(\Omega)\right)_\text{ren}$ is taken to be the finite part of $\log \det \mathbf{A}'(L;\ell_a)$ after stripping off the large $L$ divergence of the form $\gamma L+\alpha\log L$. This corresponds to a specific choice of worldsheet metric due to the gravitational anomaly. It does not affect the physical correlators, since we always have both holomorphic and antiholomorphic $bc$ ghosts in the physical amplitudes, and such amplitudes depend only on $|Z_\text{chiral}^{(g)}|^2$. The Weyl frame of \eqref{Zchiral} automatically agrees with that of the free boson correlators computed in section \ref{bosonsection}.

\subsubsection*{The Abel-Jacobi map}

The Abel-Jacobi map $\zeta(z)=(\zeta_1(z),...,\zeta_g(z))$ is defined as
\ie \label{AJmap}
\zeta_I(z) = \int_{z_0}^z \omega_I,
\fe
where $z_0$ is a fixed point on $\Sigma$ and $\omega_I$ are the holomorphic 1-forms normalized along $\alpha$ cycles:
\ie
\oint_{\alpha^J}\omega_I=\tensor{\delta}{_I^J}.
\fe Note that \eqref{AJmap} depends on the integration contour from $z_0$ to $z$. Moreover, $\zeta(z)$ is defined modulo shifts by periods
\ie \label{AJshift}
\zeta_I(z) \rightarrow \zeta_I(z)+ m_I + \Omega_{IJ}n^J,
\fe
where $m_I$ and $n^I$ are lattice vectors in $\mathbb{Z}^g$. In practice, we compute $\zeta(z)$ using \eqref{AJmap} by taking $z_0$ to be the origin of the unit disk and the contour to be the radial ray from $0$ to the point $z$ on the disk.

\subsubsection*{Spin structures and the Riemann theta function}

The Riemann theta function $\theta(y|\Omega)$ that enters \eqref{vvformula} directly is defined as
\ie
\label{eq:RiemannThetaFunction}
\theta(y|\Omega)=\sum_{n \in \mathbb{Z}^g} \exp \left(\pi in^I \Omega_{I J}n^J+2 \pi in^I y_I\right).
\fe

Given a spin structure on $\Sigma$, one can generalize this definition as follows. A spin structure on $\Sigma$ can be parameterized by the characteristic
\ie \label{spinstr}
\delta=(\delta^I, \delta_I')\in \left\{0,\tfrac{1}{2}\right\}^{2g},
\fe
where the periodicity of a spinor field along the $\alpha^I$ cycle is given by $\epsilon^I=(-1)^{2\delta^I-1}$ in the conformal frame in which the cycle $\alpha^I$ can be represented by the circumference of a flat cylinder. The spinor periodicity along $\beta_I$ is given by $\epsilon_I'=(-1)^{2\delta'_I-1}$ in a similar conformal frame. The Arf invariant of the spin structure $\delta$ is given by
\ie \label{arf}
\mathrm{Arf}(\delta)=4\sum_{I}\delta^I \delta_I'\ (\text{mod }2).
\fe 
It is clear that $\mathrm{Arf}(\delta)$ can only be $0$ or $1$. The spin structures $\delta$ with $\mathrm{Arf}(\delta) = 0$ are called even, and those with $\mathrm{Arf}(\delta) = 1$ are called odd. The Riemann theta function with characteristic is then defined by 
\ie
\theta[\delta](y |\Omega) = \sum_{n \in \mathbb{Z}^g} \exp \left[\pi i (n^I+\delta^I) \Omega_{I J}(n^J+\delta^J)+2 \pi i(n^I+\delta^I)(y_I+\delta_I^{\prime})\right].
\fe

\subsubsection*{The prime form}

The prime form $E(z,w)$ is the unique holomorphic differential of weight $(-\frac{1}{2})$ in both $z$ and $w$ that vanishes only at $z=w$. It can be computed explicitly using
\ie \label{EZW}
E(z,w) = \frac{\theta[\delta](\zeta(z)-\zeta(w)|\Omega)}{\sqrt{\omega[\delta](z)\omega[\delta](w)}},
\fe
where $\delta$ parametrizes an odd spin structure on $\Sigma$. It can be proved that \eqref{EZW} is independent of the odd spin structure $\delta$. The quantity $\omega[\delta](z)$ is a holomorphic 1-form defined as
\ie \label{omegadeltaz}
\omega[\delta](z)=\omega_I(z)\frac{\partial}{\partial y_I}\theta[\delta](y|\Omega)|_{y=0}.
\fe
In practice, the Riemann theta function is computed by truncating the lattice sum, and the prime form follows directly from the definitions \eqref{EZW} and \eqref{omegadeltaz}.

\subsubsection*{The Riemann constant vector}

The Riemann constant vector $\Delta=(\Delta_1,\cdots, \Delta_g)$ is the vector that appears in the Riemann vanishing theorem, whose statement is that $\theta(y|\Omega)$ vanishes if and only if $y$ can be written as
\ie
y = \sum_{i=1}^{g-1} \zeta(z_i) - \Delta,
\fe
where $\{z_i\}$ is a set of $g-1$ points on the Riemann surface $\Sigma$. We adopt the algorithm in \cite{deconinck2015computing} for computing $\Delta$, which is based on the following theorem.\footnote{A proof of this theorem can be found in \cite{farkas1980riemann}, section VI.3.6.}
\begin{theorem}
Let $\omega$ be a nonzero holomorphic 1-form on the Riemann surface $\Sigma$ with distinct zeros $\{c_r\}$, where the zero at $c_r$ has degree $m_r$. Then the Riemann constant vector $\Delta$ satisfies
\ie \label{canonicaldiv}
2\Delta = \sum_r m_r \zeta(c_r)
\fe
modulo the period ambiguity \eqref{AJshift}.
\end{theorem}

In the code, we take this holomorphic 1-form to be the first $\alpha$ normalized holomorphic 1-form $\hat{\omega}_1$, defined as
\ie \label{anormalize}
\hat{\omega}_I = (\mathcal{A}^{-1})_I{}^J \omega_J,
\fe
where the matrix $\mathcal{A}$ is defined in \eqref{periodmatrix}. One first computes the coefficients $\{a_n\}_{n=1}^{L/2}$ for $\hat{\omega}_1$ in \eqref{fzdz} from the known coefficients of $\omega_I$ and the matrix $\mathcal{A}$. The zeros $c_r$ of $\hat{\omega}_1$ are the roots of the polynomial $\sum_{n=1}^{L/2}a_n z^{n-1}$, with corresponding degrees $m_r$. 

Since the Abel-Jacobi map $\zeta(z)$ is subject to the ambiguity \eqref{AJshift}, the equation \eqref{canonicaldiv} only determines $\Delta$ up to a half period. Namely, the actual Riemann constant vector is among the $2^{2g}$ candidates
\ie
\Delta_\text{cand}=\frac{1}{2}\sum_{\text{zeros }r}m_r\,\zeta(c_r)+m+\Omega n,\qquad m,n\in\{0,\tfrac12\}^g.
\fe
To select the correct $\Delta$ among these candidates, we apply the Riemann vanishing theorem to a small fixed set of points $\{p_k\}$ in the interior of the disk, and choose the candidate $\Delta_\text{cand}$ that minimizes
\ie
\sum_k \left|\theta\Bigl((g-1)\zeta(p_k)-\Delta_{\text{cand}}\,\Big|\,\Omega\Bigr)\right|^2,
\fe
and take it to be the final answer for the Riemann constant vector.

\subsubsection*{The function $\sigma(z)$}

Finally, the quantity $\sigma(z)$ can be determined by taking the $\lambda = 1, (n,m)=(g,1)$ case of \eqref{vvformula}, whose left hand side can be computed directly from the path integral:
\ie \label{lambda1eq}
(Z_\text{chiral}^{(g)})^{-2} \det(\omega_I(z_i)) = Z_\text{chiral}^{(g)}\ \theta\Big(\sum_{i=1}^g \zeta(z_i)-\zeta(w)-\Delta\Big|\Omega\Big) \frac{\prod_{1\leq i< i'\leq g}E(z_i, z_{i'})\prod_{i=1}^g\sigma(z_i)}{\prod_{i=1}^g E(z_i, w)\sigma(w)}.
\fe

From this equation, one can directly read off the ratio of $\sigma(z)$ to $\sigma(w)$ at different points $z$ and $w$:
\ie
\frac{\sigma(z)}{\sigma(w)} = \frac{\theta(\sum_{i=1}^g \zeta(z_i)-\zeta(z)-\Delta|\Omega)}{\theta(\sum_{i=1}^g \zeta(z_i)-\zeta(w)-\Delta|\Omega)} \frac{\prod_{i=1}^g E(z_i,w)}{\prod_{i=1}^g E(z_i,z)},
\fe
where $\{z_i\}_{i=1}^g$ is an arbitrary set of auxiliary points on the Riemann surface. Given this relation, we can normalize $\sigma(0)$ at the origin to be $\sigma(0)=\mathcal{N}_\sigma$. We then have 
\ie
\sigma(z) = \mathcal{N}_\sigma\, \frac{\theta(\sum_{i=1}^g \zeta(z_i)-\zeta(z)-\Delta|\Omega)}{\theta(\sum_{i=1}^g \zeta(z_i)-\zeta(0)-\Delta|\Omega)} \frac{\prod_{i=1}^g E(z_i,0)}{\prod_{i=1}^g E(z_i,z)}.
\fe
Plugging this equation into \eqref{lambda1eq}, one can solve for the normalization factor $\mathcal{N}_\sigma$, which determines the function $\sigma(z)$ completely. Note that this equation does not fix the normalization of $\sigma(z)$ for genus 1, but the normalization factor $\mathcal{N}_\sigma$ does not enter physical correlators anyway, since the selection rule $n-m=(2\lambda-1)(g-1)$ forces $\mathcal{N}_\sigma$ to cancel out.

For a genus $g$ Riemann surface, the function $\sigma(z)$ obtained in this way is proportional to $(Z_\text{chiral}^{(g)})^{-\frac{3}{g-1}}$ as far as the Weyl anomaly is concerned. The correlator \eqref{vvformula} for the usual $\lambda = 2$ $bc$ system with $(3g-3)$ $b$ ghost insertions then carries the same Weyl anomaly as $(Z_\text{chiral}^{(g)})^{-26}$, which cancels the Weyl anomaly from the $c=26$ matter system, as expected.

\printbibliography

@article{Kontsevich:1992ti,
    author = "Kontsevich, M.",
    title = "{Intersection theory on the moduli space of curves and the matrix Airy function}",
    doi = "10.1007/BF02099526",
    journal = "Commun. Math. Phys.",
    volume = "147",
    pages = "1--23",
    year = "1992"
}

@book{Strebel:1984,
    author = "Strebel, K.",
    title = "{Quadratic Differentials}",
    doi = "10.1007/978-3-662-02414-0",
    year = "1984"
}

@article{Strebel:1967,
	author = {Strebel, Kurt},
	journal = {Journal d'Analyse Math{\'e}matique},
	number = {1},
	pages = {373--382},
	title = {On quadratic differentials with closed trajectories and second order poles},
	volume = {19},
	year = {1967}}

@article{BACHER200213,
title = {Counting 1-vertex triangulations of oriented surfaces},
journal = {Discrete Mathematics},
volume = {246},
number = {1},
pages = {13-27},
year = {2002},
note = {Formal Power Series and Algebraic Combinatorics 1999},
issn = {0012-365X},
doi = {https://doi.org/10.1016/S0012-365X(01)00249-7},
url = {https://www.sciencedirect.com/science/article/pii/S0012365X01002497},
author = {Roland Bacher and Alina Vdovina}
}

@article{Verlinde:1986kw,
    author = "Verlinde, Erik P. and Verlinde, Herman L.",
    editor = "Stone, M.",
    title = "{Chiral Bosonization, Determinants and the String Partition Function}",
    reportNumber = "Print-86-1416 (UTRECHT)",
    doi = "10.1016/0550-3213(87)90219-7",
    journal = "Nucl. Phys. B",
    volume = "288",
    pages = "357",
    year = "1987"
}

@article{deconinck2015computing,
  title={Computing the Riemann constant vector},
  author={Deconinck, Bernard and Patterson, Matthew S and Swierczewski, Christopber},
  journal={preprint, available online at https://depts. washington. edu/bdecon/papers/pdfs/rcv. pdf},
  year={2015}
}

@book{farkas1980riemann,
  author = {Farkas, Hershel M. and Kra, Irwin},
  title = {Riemann Surfaces},
  series = {Graduate Texts in Mathematics},
  volume = {71},
  publisher = {Springer},
  address = {New York},
  year = {1980},
  doi = {10.1007/978-1-4684-9930-8}
}

@online{StringNotes,
  author = {Yin, Xi},
  title = {Foundations of String Theory},
  url = {https://github.com/xiyin137/stringbook/},
  urldate = {2026-07-23},
  note = {Online lecture notes}
}

@article{Igusa:1962,
  author = {Igusa, Jun-Ichi},
  title = {On Siegel Modular Forms of Genus Two},
  journal = {American Journal of Mathematics},
  volume = {84},
  number = {1},
  pages = {175--200},
  year = {1962},
  doi = {10.2307/2372812}
}

@article{Igusa:1967,
  author = {Igusa, Jun-ichi},
  title = {Modular Forms and Projective Invariants},
  journal = {American Journal of Mathematics},
  volume = {89},
  number = {3},
  pages = {817--855},
  year = {1967},
  doi = {10.2307/2373243}
}

@article{DHokerPhong:1988,
  author = {D'Hoker, Eric and Phong, D. H.},
  title = {The geometry of string perturbation theory},
  journal = {Reviews of Modern Physics},
  volume = {60},
  number = {4},
  pages = {917--1065},
  year = {1988},
  doi = {10.1103/RevModPhys.60.917}
}

@article{BelavinKnizhnik:1986,
  author = {Belavin, A. A. and Knizhnik, V. G.},
  title = {Algebraic geometry and the geometry of quantum strings},
  journal = {Physics Letters B},
  volume = {168},
  number = {3},
  pages = {201--206},
  year = {1986},
  doi = {10.1016/0370-2693(86)90963-9}
}

@article{Rauch:1959,
  author = {Rauch, H. E.},
  title = {Weierstrass points, branch points, and moduli of Riemann surfaces},
  journal = {Communications on Pure and Applied Mathematics},
  volume = {12},
  number = {3},
  pages = {543--560},
  year = {1959},
  doi = {10.1002/cpa.3160120310}
}

@article{RychkovVitale:2015,
  author = {Rychkov, Slava and Vitale, Lorenzo G.},
  title = {Hamiltonian Truncation Study of the $\phi^4$ Theory in Two Dimensions},
  journal = {Physical Review D},
  volume = {91},
  number = {8},
  pages = {085011},
  year = {2015},
  doi = {10.1103/PhysRevD.91.085011},
  eprint = {1412.3460},
  archivePrefix = {arXiv},
  primaryClass = {hep-th}
}

@article{CarleoTroyer:2017,
  author = {Carleo, Giuseppe and Troyer, Matthias},
  title = {Solving the Quantum Many-Body Problem with Artificial Neural Networks},
  journal = {Science},
  volume = {355},
  number = {6325},
  pages = {602--606},
  year = {2017},
  doi = {10.1126/science.aag2302},
  eprint = {1606.02318},
  archivePrefix = {arXiv},
  primaryClass = {cond-mat.dis-nn}
}

@article{MartynNajafiLuo:2023,
  author = {Martyn, John M. and Najafi, Khadijeh and Luo, Di},
  title = {Variational Neural-Network Ansatz for Continuum Quantum Field Theory},
  journal = {Physical Review Letters},
  volume = {131},
  number = {8},
  pages = {081601},
  year = {2023},
  doi = {10.1103/PhysRevLett.131.081601},
  eprint = {2212.00782},
  archivePrefix = {arXiv}
}

@article{BrowerFlemingNeuberger:2013,
  author = {Brower, Richard C. and Fleming, George T. and Neuberger, Herbert},
  title = {Lattice Radial Quantization: 3D Ising},
  journal = {Physics Letters B},
  volume = {721},
  number = {4--5},
  pages = {299--305},
  year = {2013},
  doi = {10.1016/j.physletb.2013.03.009},
  eprint = {1212.1757},
  archivePrefix = {arXiv},
  primaryClass = {hep-lat}
}

@article{Cho-Collier-Yin,
  author = {Cho, Minjae and Collier, Scott and Yin, Xi},
  title = {Recursive Representations of Arbitrary Virasoro Conformal Blocks},
  journal = {JHEP},
  volume = {04},
  pages = {018},
  year = {2019},
  doi = {10.1007/JHEP04(2019)018},
  eprint = {1703.09805},
  archivePrefix = {arXiv},
  primaryClass = {hep-th}
}

@article{Zamolodchikov,
  author = {Zamolodchikov, Al. B.},
  title = {Conformal Symmetry in Two-Dimensional Space: Recursion Representation of Conformal Block},
  journal = {Theoretical and Mathematical Physics},
  volume = {73},
  number = {1},
  pages = {1088--1093},
  year = {1987},
  doi = {10.1007/BF01022967}
}

@incollection{Penner,
  author = {Penner, Robert C.},
  title = {Cell Decomposition and Compactification of Riemann's Moduli Space in Decorated Teichmuller Theory},
  booktitle = {Woods Hole Mathematics: Perspectives in Mathematics and Physics},
  editor = {Tongring, Nils and Penner, Robert C.},
  series = {Series on Knots and Everything},
  volume = {34},
  pages = {263--301},
  publisher = {World Scientific},
  year = {2004},
  doi = {10.1142/9789812701398_0006},
  eprint = {math/0306190},
  archivePrefix = {arXiv}
}

@misc{c1paper,
  author = {Christian, Sam and Mazel, Ben and Wang, Yuchen and Yin, Xi and Zhang, Yutai},
  note = {Work in progress}
}

@article{Polyakov:1981,
  author = {Polyakov, Alexander M.},
  title = {Quantum Geometry of Bosonic Strings},
  journal = {Physics Letters B},
  volume = {103},
  pages = {207--210},
  year = {1981},
  doi = {10.1016/0370-2693(81)90743-7}
}

@article{DHokerPhong:1989Chiral,
  author = {D'Hoker, Eric and Phong, D. H.},
  title = {Conformal Scalar Fields and Chiral Splitting on Super Riemann Surfaces},
  journal = {Communications in Mathematical Physics},
  volume = {125},
  pages = {469--513},
  year = {1989},
  doi = {10.1007/BF01218413}
}

@article{DHokerPhong:2002IV,
  author = {D'Hoker, Eric and Phong, D. H.},
  title = {Two-Loop Superstrings. IV. The Cosmological Constant and Modular Forms},
  journal = {Nuclear Physics B},
  volume = {639},
  pages = {129--181},
  year = {2002},
  doi = {10.1016/S0550-3213(02)00516-3},
  eprint = {hep-th/0111040},
  archivePrefix = {arXiv}
}

@article{DHokerGutperlePhong:2005,
  author = {D'Hoker, Eric and Gutperle, Michael and Phong, D. H.},
  title = {Two-Loop Superstrings and {S}-Duality},
  journal = {Nuclear Physics B},
  volume = {722},
  pages = {81--118},
  year = {2005},
  doi = {10.1016/j.nuclphysb.2005.06.010},
  eprint = {hep-th/0503180},
  archivePrefix = {arXiv}
}

@article{DHokerGreenPiolineRusso:2015,
  author = {D'Hoker, Eric and Green, Michael B. and Pioline, Boris and Russo, Rodolfo},
  title = {Matching the {$D^6R^4$} Interaction at Two Loops},
  journal = {JHEP},
  volume = {01},
  pages = {031},
  year = {2015},
  doi = {10.1007/JHEP01(2015)031},
  eprint = {1405.6226},
  archivePrefix = {arXiv},
  primaryClass = {hep-th}
}

@article{GomezMafra:2013,
  author = {Gomez, Humberto and Mafra, Carlos R.},
  title = {The Closed-String Three-Loop Amplitude and {S}-Duality},
  journal = {JHEP},
  volume = {10},
  pages = {217},
  year = {2013},
  doi = {10.1007/JHEP10(2013)217},
  eprint = {1308.6567},
  archivePrefix = {arXiv},
  primaryClass = {hep-th}
}

@article{Witten:1991Intersection,
  author = {Witten, Edward},
  title = {Two-Dimensional Gravity and Intersection Theory on Moduli Space},
  journal = {Surveys in Differential Geometry},
  volume = {1},
  pages = {243--310},
  year = {1991}
}

@article{BCOV:1993,
  author = {Bershadsky, Michael and Cecotti, Sergio and Ooguri, Hirosi and Vafa, Cumrun},
  title = {Kodaira--Spencer Theory of Gravity and Exact Results for Quantum String Amplitudes},
  journal = {Communications in Mathematical Physics},
  volume = {165},
  pages = {311--427},
  year = {1994},
  doi = {10.1007/BF02099774},
  eprint = {hep-th/9309140},
  archivePrefix = {arXiv}
}

@article{Mirzakhani:2007,
  author = {Mirzakhani, Maryam},
  title = {Simple Geodesics and Weil--Petersson Volumes of Moduli Spaces of Bordered Riemann Surfaces},
  journal = {Inventiones Mathematicae},
  volume = {167},
  pages = {179--222},
  year = {2007},
  doi = {10.1007/s00222-006-0013-2}
}

@article{EynardOrantin:2007,
  author = {Eynard, Bertrand and Orantin, Nicolas},
  title = {Weil--Petersson Volume of Moduli Spaces, Mirzakhani's Recursion and Matrix Models},
  year = {2007},
  eprint = {0705.3600},
  archivePrefix = {arXiv},
  primaryClass = {math-ph}
}

@article{CollierEtAl:2024,
  author = {Collier, Scott and Eberhardt, Lorenz and M{\"u}hlmann, Beatrix and Rodriguez, Victor A.},
  title = {The Virasoro Minimal String},
  journal = {SciPost Physics},
  volume = {16},
  number = {2},
  pages = {057},
  year = {2024},
  doi = {10.21468/SciPostPhys.16.2.057},
  eprint = {2309.10846},
  archivePrefix = {arXiv},
  primaryClass = {hep-th}
}

@article{Berkovits:2004Multiloop,
  author = {Berkovits, Nathan},
  title = {Multiloop Amplitudes and Vanishing Theorems Using the Pure Spinor Formalism for the Superstring},
  journal = {JHEP},
  volume = {09},
  pages = {047},
  year = {2004},
  doi = {10.1088/1126-6708/2004/09/047},
  eprint = {hep-th/0406055},
  archivePrefix = {arXiv}
}

@article{Berkovits:2006HigherDerivative,
  author = {Berkovits, Nathan},
  title = {New Higher-Derivative {$R^4$} Theorems},
  journal = {Physical Review Letters},
  volume = {98},
  pages = {211601},
  year = {2007},
  doi = {10.1103/PhysRevLett.98.211601},
  eprint = {hep-th/0609006},
  archivePrefix = {arXiv}
}

@article{Polchinski:1988Factorization,
  author = {Polchinski, Joseph},
  title = {Factorization of Bosonic String Amplitudes},
  journal = {Nuclear Physics B},
  volume = {307},
  pages = {61--92},
  year = {1988},
  doi = {10.1016/0550-3213(88)90522-6}
}

@article{Shervashidze:2011WL,
    author  = {Shervashidze, Nino and Schweitzer, Pascal and
               van Leeuwen, Erik Jan and Mehlhorn, Kurt and
               Borgwardt, Karsten M.},
    title   = {Weisfeiler--Lehman Graph Kernels},
    journal = {Journal of Machine Learning Research},
    volume  = {12},
    number  = {77},
    pages   = {2539--2561},
    year    = {2011},
    url     = {https://www.jmlr.org/papers/v12/shervashidze11a.html}
}

@article{WeisfeilerLeman:1968,
    author  = {Weisfeiler, Boris Yu. and Leman, Andrei A.},
    title   = {The Reduction of a Graph to Canonical Form and the
               Algebra Which Appears Therein},
    journal = {Nauchno-Technicheskaya Informatsia, Series 2},
    number  = {9},
    pages   = {12--16},
    year    = {1968},
    note    = {In Russian; English translation available at
               \url{https://www.iti.zcu.cz/wl2018/pdf/wl_paper_translation.pdf}}
}
\end{document}